\documentclass[prd,twocolumn,showpacs,preprintnumbers,amsmath,amssymb,nofootinbib]{revtex4}
\usepackage{graphicx}
\usepackage{dcolumn}
\usepackage{bm}
\usepackage[english]{babel}
\usepackage{latexsym}
\usepackage{amsmath}
\usepackage{amsfonts}
\usepackage{amssymb}
\usepackage{textcomp}
\usepackage{graphicx}
\usepackage{enumerate}
\usepackage{hhline}
\usepackage{array}
\usepackage{multirow}
\usepackage[utf8]{inputenc}

\def\lsim{\mathrel{\lower2.5pt\vbox{\lineskip=0pt\baselineskip=0pt
\hbox{$<$}\hbox{$\sim$}}}}
\def\gsim{\mathrel{\lower2.5pt\vbox{\lineskip=0pt\baselineskip=0pt
\hbox{$>$}\hbox{$\sim$}}}}

\newcommand{\be}{\begin{eqnarray}}
\newcommand{\ee}{\end{eqnarray}}
\newcommand{\sbe}{\begin{eqnarray*}}
\newcommand{\see}{\end{eqnarray*}}
\newcommand{\Nc}{N_{\mathrm{c}}\,}
\newcommand{\qq}{${\overline{q}}q\;$}

\newlength{\arrayrulewidthOriginal}
\newcommand{\Cline}[2]{%
  \noalign{\global\setlength{\arrayrulewidthOriginal}{\arrayrulewidth}}%
  \noalign{\global\setlength{\arrayrulewidth}{#1}}\cline{#2}%
  \noalign{\global\setlength{\arrayrulewidth}{\arrayrulewidthOriginal}}}

\begin{document}

\title{Chiral Perturbation Theory, the $\mathbf{1/\Nc}$ expansion\\ and Regge behavior determine the structure of the lightest scalar meson}
\author{J. R. Pel\'aez$^a$}
\author{M.R. Pennington$^{b}$\footnote{Notice: Authored by Jefferson Science Associates, LLC under U.S. DOE Contract No. DE-AC05-06OR23177. The U.S. Government retains a non-exclusive, paid-up, irrevocable, world-wide license to publish or reproduce this manuscript for U.S. Government purposes.}\noindent} 
\author{J. Ruiz de Elvira$^{a}$}
\author{ D.J. Wilson$^{c}$}

\affiliation{$^a$ Departamento de F{\'\i}sica Te{\'o}rica II,
  Universidad Complutense de Madrid, 28040   Madrid,\ \ Spain\\
$^b$ Theory Center, Thomas Jefferson National Accelerator Facility, 12000 Jefferson Avenue, Newport News, VA 23606, U.S.A.\\
$^c$ Physics Division, Argonne National Laboratory, Argonne, IL 60439, U.S.A.}
\begin{abstract} The leading $1/\Nc$ behaviour of Unitarised 
Chiral Perturbation Theory distinguishes
the nature of the $\rho$ and the $\sigma$: 
The $\rho$ is a \qq meson, while
 the $\sigma$ is not.
 However, semi-local duality 
between resonances and Regge behaviour cannot
 be satisfied for larger $N_c$, if such a distinction 
holds. While the
 $\sigma$ at $\Nc=\,3$ is inevitably 
dominated by its di-pion component,
 Unitarised Chiral Perturbation Theory also suggests that as
 $\Nc$ increases above 6-8, the $\sigma$ may have a sub-dominant \qq fraction up at
 1.2 GeV. Remarkably this ensures semi-local duality is fulfilled for the
 range of $\Nc \lesssim 15-30$, where the unitarisation procedure adopted applies
\end{abstract}
\maketitle

\section{Introduction}

Long ago Jaffe \cite{Jaffe:1976ig} identified the distinct nature of mesons: those built simply of a quark and an antiquark, and those with additional ${\overline q}q$ pairs. Of course,
even well established ${\overline q}q$ resonances, like the $\rho$ and $\omega$, spend part of their time in four and six quark configurations as this is how they decay to $\pi\pi$ and $3\pi$ respectively. However, the $1/\Nc$ expansion \cite{'tHooft:1973jz} provides a method of clarifying such differences. If we could tune $\Nc$ up from 3, we would see that an intrinsically ${\overline q}q$ state would become narrower and narrower. As $\Nc$ increases, the underlying pole, which defines the resonant state, moves along the unphysical sheet(s) towards the real axis. In contrast a tetraquark state would become wider and wider and its pole would effectively disappear from \lq\lq physical'' effect: if only we could tune $\Nc$.

A long recognised feature of the world with $\Nc=\,3$ is that of ``local duality''\cite{dhs,schmid,Donnachie:2002en}. In a scattering process, as the energy increases from threshold, distinct resonant structures give way to a smooth Regge behaviour. At low energy the scattering amplitude is well represented by a sum of resonances (with a background), but as the energy increases the resonances (having more phase space for decay) become wider and increasingly overlap. This overlap generates a smooth behaviour of the cross-section most readily described not by a sum of a large number of resonances in the direct channel, but the contribution of a small number of crossed channel Regge exchanges. Indeed, detailed studies \cite{schmid,piN} of meson-baryon scattering processes show that the sum of resonance contributions at all energies ``averages'' (in a well-defined sense to be recalled below) the higher energy Regge behaviour. Indeed, these early studies\cite{dhs,schmid} revealed 
 how
  this property starts right from $\pi N$ threshold, so that this ``local duality'' holds across the whole energy regime. Thus resonances in the $s$-channel know about Regge exchanges in the $t$-channel. Indeed, these resonance and Regge components are not to be added like Feynman diagram contributions, but are ``dual'' to each other: one uses one {\bf or} the other. Indeed, the wonderful formula discovered by Veneziano \cite{Veneziano:1968yb}
is an explicit realisation of this remarkable property. This has allowed the idea of ``duality'' first found in meson-nucleon reactions to be extended to baryon-antibaryon reactions, as well as to the  simpler meson-meson scattering channel we consider here\cite{pipiduality}.  Unlike the idealised Veneziano model with its exact local duality, the real world, with finite width resonances, has  a ``semi-local duality'' quantified by averaging over the typical spacing of resonance towers defined by the inverse of the slope of relevant Regge trajectory.

Regge exchanges too are built from ${\overline q}q$ and multiquark contributions. In a channel like that with isospin 2 in $\pi\pi$ scattering, or isospin 3/2 in $K\pi$ scattering, there are no ${\overline q}q$ resonances, and so the Regge exchanges with these quantum numbers must involve multi-quark components. Data teach us that even at $\Nc=\,3$ these components are suppressed compared to the dominant ${\overline q}q$ exchanges. Semi-local duality means that in $\pi^+\pi^-\to\pi^-\pi^+$ scattering, the low energy resonances must have contributions to the cross-section that ``on the average'' cancel, since this process is purely isospin 2 in the $t-$channel. The meaning of semi-local duality is that this cancellation happens right from $\pi\pi$ threshold. 

Now in $\pi\pi$ scattering below 900 MeV, there are just two low energy resonances: the $\rho$ with $I=J=1$ and the $\sigma$ with $I=J=0$. In the model of Veneziano, where resonances contribute as delta-functions, exact local duality is achieved by the $\sigma$ and $\rho$ having exactly the same mass, and the coupling squared of the $\sigma$ is 9/2 times that of the $\rho$. Of course, the Veneziano amplitude is too simplistic and does not respect two body unitarity. Yet nevertheless, in the real world with $\Nc=\,3$ with finite width resonances ``semi-local'' duality is at play right from threshold.
There is a cancellation between the $\rho$ with a width of 150 MeV, which is believed to be predominantly a ${\overline q}q$ state, and the $\sigma$, which is very broad, at least 500 MeV wide, with a shape that is not Breit-Wigner-like, and might well be a tetraquark, molecular~\cite{Pennington:2010dc} or gluonic state~\cite{mink,narison}, or possibly a mixture of all of these. Its short-lived nature certainly means it spends most of its existence in a di-pion configuration. The contribution of these two resonances to the $\pi^+\pi^-$ cross-section do indeed ``on average'' cancel in keeping with $I=2$ exchange in the $t-$channel. However, such a distinct nature for the $\rho$ and $\sigma$ would prove a difficulty if we could increase $\Nc$. A tetraquark $\sigma$ would become still broader and its contribution to the cross-section less and less, while its companion the $\rho$ would become more delta-function-like and have nothing to cancel. Semi-local duality would fail. The 
 correct Regge behaviour would not be generated
 . It would just be a feature of the world with $\Nc=\,3$ and not for higher values. Yet our theoretical expectation is quite the contrary, the multiquark Regge exchange should be even better suppressed as $\Nc$ increases above 3. This paradox clearly poses a problem for the description of the $\sigma$ as a non-\qq state. The aim of this paper is to show how unitarised chiral perturbation theory provides a picture of how this paradox is resolved.

Chiral Perturbation Theory ($\chi$PT) \cite{chpt1} provides a systematic procedure for computing processes involving the Goldstone bosons of chiral symmetry breaking, particularly pions. The domain of applicability is naturally restricted to low energies where the pion momenta $p$ and the pion mass $m_\pi$ are much less than the natural scale of the theory specified by the pion decay constant $f_\pi$ scaled by $4\pi$, {\it i.e.} 1 GeV. The presence at low energies of  elastic resonances, like the $\rho$ and $\sigma$, means that the unitarity limit is reached at well below this scale of 1 GeV. Consequently, the fact that $\chi$PT satisfies unitarity order-by-order is not sufficiently fast for these key low energy resonances to be described beyond their near threshold tails. Much effort has been devoted to accelerating the process of unitarisation 
\cite{Truong:1988zp,Dobado:1996ps,Dobado3,Oller:1997ti,Oller:1998zr,GomezNicola:2001as,Nieves:2001de}. Low orders in $\chi$PT must already contain information about key components at all orders for unitarisation to be achieved. It surely pays to sum these known contributions up even when working ostensibly at low orders in perturbation theory. One method for achieving a Unitarised Chiral Perturbation Theory (UChPT) is the {\it Inverse Amplitude Method}
\cite{Truong:1988zp,Dobado:1996ps,Dobado3,GomezNicola:2007qj}. This is based on the very simple idea that in the region of elastic unitarity, the imaginary part of the inverse of each partial wave amplitude is determined by phase space --- dynamics resides in the real part of the inverse amplitude. This procedure leads naturally to resonant effects in the strongly attractive $I=0$ and $I=1$ channels. At tree level $\chi$PT involves just one parameter, the pion decay constant. However, at higher orders new Low Energy Constants (LECs) enter in the pion-pion scattering amplitudes~: 
4 at one loop order \cite{chpt1}, 6 more at two loops \cite{Bijnens:1997vq}, etc. These have to be fixed from experiment. Clearly, the predictive power of the theory, so apparent at tree level, where every pion process just depends on the scale set by $f_\pi$, becomes clouded as higher loops become significant with the LECs poorly known. While the elastic Inverse Amplitude Method delays the onset of these new terms with their additional LECs, this is still restricted to the region below 1 GeV (or 1.2 GeV if the IAM is used within a coupled channel formalism, although this has other problems not present in the elastic treatment -- see \cite{GomezNicola:2001as}).

A beauty of Chiral Lagrangians is that the $\Nc$ dependence of the parameters is determined. 
Every LEC, starting with $f_{\pi}$ has a well-defined leading $\Nc$ behaviour \cite{chpt1,Peris:1994dh}, for instance, $f_\pi \sim\,\sqrt{\Nc}$. At one loop order with central values for the LECs, one of us~(JRP) has studied unitarised low energy $\pi\pi$ 
scattering as $\Nc$ increases \cite{Pelaez:2003dy}, showing how the $\rho$ does indeed become narrower (as expected of a ${\overline q}q$ resonance). In contrast, at least for not too large $N_c$, the $\sigma$ pole became wider and moved away from the 
400 to 600 MeV region of the real energy axis,
 as anticipated by a largely ${\overline{qq}}qq$ nature. As we shall discuss, and as already introduced, 
this means that for the central values and most parameter space, 
the semi-local duality implicit in Finite Energy Sum Rules (FESRs) is not satisfied as 
$\Nc$ increases.

Subsequently, one of us (JRP) together with Rios showed \cite{Pelaez:2006nj}
that the $\Nc$ behaviour becomes more subtle when two loop $\chi$PT effects are included. 
In particular, for the best fits of the unitarised two loop $\chi$PT,
there is
a \qq component of the $\sigma$, which while sub-dominant at $\Nc=\,3$, becomes increasingly 
important as $\Nc$ increases. 
The $\sigma$ pole still moves away from the 400-600 MeV region of the real 
axis, but the pole trajectory turns around moving back towards 
the real axis above 1 GeV as $\Nc$ becomes 
larger than 10 or so. This occurs rather naturally in the two-loop results
but was only hinted in some part of the one-loop parameter space.
Such a behaviour would indicate that while the $\sigma$ is predominantly non-${\overline q}q$  at $\Nc=\,3$, it does have a ${\overline q}q$ component. As we show here, it is this component that ensures FESRs are satisfied. Regge expectations then hold at all $\Nc$. Indeed, imposing this as a physical requirement places a constraint on the second order LECS: a constraint readily satisfied with LECs in fair agreement with current crude estimates.

Thus chiral dynamics already contains the resolution of the paradox that was the motivation for this study: namely how does the suppression of $I=2$ Regge exchanges happen if resonances like the $\rho$ and $\sigma$ are intrinsically different. We will see that the $\sigma$ 
may naturally contain a small but all important ${\overline q}q$ component. At large $\Nc$ this would be the seed of this state. As $\Nc$ is lowered this state will have an increased coupling to pions, and it is these that dominate its existence when $\Nc=\,3$.
We will, of course, discuss the range of $\Nc$ for which the IAM applies and where replacing the LECs (at $N_{\mathrm{c}}=3$) with their leading $\Nc$ form is appropriate.

\section {Semi-local duality and finite energy sum rules}

\subsection{Regge theory and semi-local duality}

Regge considerations lead us to study $s$-channel $\pi\pi$ scattering amplitudes with definite isospin
in the $t$-channel, labeled $A^{tI}(s,t)$. These can, of course, be written in terms of amplitudes with definite isospin in the direct channel, $A^{sI}(s,t)$, using the well-known crossing relationships, so that
\be\nonumber\label{At}
A^{t0}(s,t)&=&\frac{1}{3}\,A^{s0}(s,t) +  \;\;\;A^{s1}(s,t) + \frac{5}{3}\,A^{s2}(s,t)\\\nonumber
A^{t1}(s,t)&=&\frac{1}{3}\,A^{s0}(s,t) + \frac{1}{2}\,A^{s1}(s,t) - \frac{5}{6}\,A^{s2}(s,t)\\
A^{t2}(s,t)&=&\frac{1}{3}\,A^{s0}(s,t) - \frac{1}{2}\,A^{s1}(s,t) + \frac{1}{6}\,A^{s2}(s,t) .
\ee
It is convenient to  denote the common channel threshold by $s_{\mathrm{th}}\equiv t_{\mathrm{th}}\equiv 4m_\pi^2$.
The amplitudes of Eq.~(1) have definite symmetry under $s\to u$ and this will be reflected in writing them as functions of $\nu\,=\,(s-u)/2$, a variable for which $\nu=s=-u$ along the line $t=t_{\mathrm{th}}$. To check semi-local duality, we need to continue the well-known Regge asymptotics at fixed $t$ down to threshold. To do this we follow~\cite{barger-cline} with:
\be\nonumber
{\mathrm{Im}}\,A^{tI}_{\mathrm{Regge}}(\nu,t) = \sum_R \beta_R(t) \Theta(\nu) \left[\alpha'^2\,(\nu^2-\nu_{\mathrm{th}}^2)\right]^{\alpha_R(t)/2}\!\!\! ,\\
{\hspace{-10mm}}
\ee
where as usual the $\alpha_R(t)$ denote the Regge trajectories with the appropriate $t$-channel quantum numbers, $\beta_R(t)$ their Regge couplings and $\alpha'$ is the universal slope of the ${\overline q}q$ meson trajectories ($\sim 0.9$ GeV$^{-2}$).  The crossing function
$\Theta(\nu)\,=\,\left[1\,-{\nu^2_{\mathrm{th}}}/{\nu^2}\right]^{(1+\gamma)}$
having $\gamma=0$ for $s-u$-even amplitudes, and $\gamma=1/2$ if they are crossing-odd, ensures the imaginary parts of the amplitudes vanish at threshold, while being unity when $\nu$ is large.
$\nu_{\mathrm{th}}$ is the value of $\nu$ at threshold, {\it viz.} $\nu_{\mathrm{th}}\,=\,(s_{\mathrm{th}}+t)/2$. For the amplitude with $I=1$ in the $t$-channel, for which $\gamma=1$, the sum in Eq.~(2) will be dominated by $\rho$-exchange with a trajectory $\alpha(t)\,=\alpha_0\,+\,\alpha'\,t$ that has the value 1 at $t=m_\rho^2$ and 3 at $t=m_{\rho_3}^2$~\cite{PDG}, {\it i.e.}
$\alpha_0=0.467$ and $\alpha' = 0.889$ GeV$^{-2}$. 
For isoscalar exchange the dominant trajectories are the Pomeron with $\alpha_P(t)= 1.083 + 0.25 t$ (with $t$ in GeV$^2$ units)~\cite{Donnachie:1985iz}{\footnote{Because of the rapid convergence of the sum rules we consider, the fact the Pomeron form used violates the Froissart bound is of no consequence. This has been explicitly checked by also using the parametrization of Cudell {\it et al.}~\cite{Cudell:2001pn}.}  and the $f_2$-trajectory which is almost degenerate with that of the $\rho$. For the exotic $I=2$ channel with its leading Regge exchange being a $\rho-\rho$ cut, we expect $\alpha(0) << \alpha_\rho(0)$, and its couplings to be correspondingly smaller.

Semi-local duality between Regge and resonance contributions teaches us that
\be
\begin{split}
\int_{\nu_1}^{\nu_2}\;d\nu\;\nu^{-n}\; {\mathrm{Im}}\,{A^{tI}_{\mathrm{resonance}}}(\nu,t) \simeq\\
\int_{\nu_1}^{\nu_2}\;d\nu\;\nu^{-n}\; {\mathrm{Im}}\,{A^{tI}_{\mathrm{Regge}}}(\nu,t)\quad ,
\end{split}
\ee
the \lq\lq averaging'' should take place over at least one resonance tower. Thus
the integration region $\nu_2\,-\nu_1$ should be a multiple of $1/\alpha'$, typically 1 GeV$^2$. We will consider two ranges from threshold to 1 GeV$^2$ and up to 2 GeV$^2$.

This duality should hold for values of $t$ close to the forward scattering direction, and so we consider both $t=0$ and $t=t_{\mathrm{th}}$. The difference in results between these two gives us a measure of the accuracy of semi-local duality,
as expressed in Eq.~(3). Since we are interested in the resonance integrals being saturated by the lightest states, we consider values of $n=0$ to $n=3$. We will find that with $n=1,2,3$ the low mass resonances do indeed control these Finite Energy Sum Rules.

\subsection{Finite Energy Sum Rules from data\\ ({\it i.e.} $\Nc=\,3$)}
\begin{center}
    \begin{table*}
\begin{center}
    \begin{tabular}{|c|c||c|c||c|c|}\cline{3-6}
      \multicolumn{2}{c}{}
      &\multicolumn{2}{|c||}{$\mathbf{I_t=0}$}
      &\multicolumn{2}{c||}{$\mathbf {I_t=1}$}
      \\
\hline
      \multicolumn{1}{|c|}{} & $n$ &t =t$_{th}$&t=0&t=t$_{th}$&t=0\\
\hline
      \multirow{4}{0.4in}{~\\R\\E\\G\\G\\E} 
&&&&&\\[-2.mm]
       & 0 & 0.225 & 0.233 & 0.325 & 0.353 \\[1.mm]
       & 1 & 0.425 & 0.452 & 0.578 & 0.642 \\[1.mm]
       & 2 & 0.705 & 0.765 & 0.839 & 0.908 \\[1.mm]
       & 3 & 0.916 & 0.958 & 0.966 & 0.990 \\[3.mm]\hline\hline
      \multirow{4}{0.4in}{\\~\\KPY\\~\\$S,P,D$}
&&&&&\\[-1.mm] 
       & 0 & 0.337 $\pm$ 0.093 & 0.342 $\pm$ 0.083 & 0.479 $\pm$ 0.213 & 0.492 $\pm$ 0.191\\[1.mm]
       & 1 & 0.567 $\pm$ 0.095 & 0.582 $\pm$ 0.082 & 0.725 $\pm$ 0.157 & 0.741 $\pm$ 0.131\\[1.mm]
       & 2 & 0.788 $\pm$ 0.061 & 0.815 $\pm$ 0.047 & 0.894 $\pm$ 0.072 & 0.911 $\pm$ 0.052\\[1.mm]
       & 3 & 0.927 $\pm$ 0.023 & 0.953 $\pm$ 0.013 & 0.971 $\pm$ 0.022 & 0.982 $\pm$ 0.011
          \\[3.mm]\hline\hline
      \multirow{4}{0.4in}{\\~\\KPY\\~\\$S,P$}
&&&&&\\[-1.mm] 
       & 0 & 0.615 $\pm$ 0.169 & 0.572 $\pm$ 0.133 & 0.743 $\pm$ 0.187 & 0.709 $\pm$ 0.103\\[1.mm]
       & 1 & 0.796 $\pm$ 0.145 & 0.771 $\pm$ 0.120 & 0.874 $\pm$ 0.123 & 0.861 $\pm$ 0.064\\[1.mm]
       & 2 & 0.912 $\pm$ 0.088 & 0.909 $\pm$ 0.068 & 0.950 $\pm$ 0.062 & 0.950 $\pm$ 0.026\\[1.mm]
       & 3 & 0.971 $\pm$ 0.038 & 0.977 $\pm$ 0.021 & 0.984 $\pm$ 0.023 & 0.989 $\pm$ 0.006
           \\[3.mm]\hline\hline
\end{tabular}
\end{center}
    \caption{ $R_n^I$ ratios defined in Eq.~(4) evaluated using the Regge model  of Eq.~(2) and the KPY $\pi\pi$ parameterization~\cite{Kaminski:2006qe} with and without $D$-waves. }
    \centering\label{Table: Regge_KPY}
  \end{table*}
\end{center}

Let us first look at $\pi\pi$ scattering data and see how well it approximates this relationship, before we consider the various resonances contributions that make up the ``data'' and in turn how these might change with $\Nc$.
To do this it is useful to define the following ratio
\begin{eqnarray}
\nonumber\\
R_n^I\;=\frac{\;\int_{\nu_1}^{\nu_2}\;d\nu\;\nu^{-n}\; {\mathrm{Im}}\,A^{tI}(\nu,t)\,\,}{
\int_{\nu_1}^{\nu_3}\;d\nu\;\nu^{-n}\; {\mathrm{Im}}\,A^{tI}(\nu,t)}\quad . 
\\\nonumber
\end{eqnarray}
The behaviour of such a ratio which tests the way the low energy amplitudes average the expected leading Regge energy dependence of Eq.~(2) --- the leading Regge behaviour because only then does the Regge coupling $\beta_R(t)$ cancel out in the ratio.
We will consider these ratios with $\nu_1$ at its threshold value, $\nu_2=1$ GeV$^2$ and $\nu_3=2$ GeV$^2$.
In evaluating the amplitudes in Eq.~(1), we represent them by a sum of $s$-channel partial waves, so that
\be
{\mathrm{Im}} A^{sI}(s,t)\;=\;\sum_{\ell}\;(2\ell+1)\, {\mathrm{Im}} {\cal A}^I_{\ell}(s)\;P_{\ell}(z_s)\quad ,
\ee
where the sum involves only even $\ell$ for $I=0,\, 2$ and odd $\ell$ for $I=1$. $P_{\ell}(z_s)$ are the usual Legendre polynomials, with $z_s$ the cosine of the $s$-channel c.m. scattering angle related to the Mandelstam variables by $z_s=1\,+\,2t/(s-s_{\mathrm{th}})$. It is useful to note that the partial wave amplitudes behave towards threshold like 
${\cal A}_{\ell}\,\sim (s-s_{\mathrm{th}})^{\ell}$, so that the imaginary parts that appear in Eqs.~(5,3,4) behave like $(s-s_{\mathrm{th}})^{2\ell+1}$ from unitarity.

 We use the partial wave parametrization from Kami\'nski, Pel\'aez and Yndurain (KPY)~\cite{Kaminski:2006qe} to represent the data. The partial wave sum is performed in two ways: first including partial waves up to and including $\ell=2$, and second with just the $S$ and $P$ waves. We compare each of these in Table~I with the evaluation of the ratios in Eq.~(4) using the leading Regge pole contribution.
 This serves as a guide as to 
\begin{itemize}
\item[(i)] how well semi-local duality of Eq.~(3) works from experimental data in the world of $\Nc=\,3$ by comparing the Regge ``prediction'' with the KPY representation of experiment, and 
\item[(ii)] by comparing  how well the integrals are dominated by just the lowest partial waves $\ell \le 1$ with $\ell \le 2$.
\end{itemize}

This will be needed  to address how the duality relation of Eq.~(3) puts constraints on the nature of the $\rho$ and $\sigma$ resonances.
We present these results  in Table~\ref{Table: Regge_KPY}. The $n=1$ integral would with $t=0$ be closest to averaging the total cross-section. The  table shows that the data follow the expectations of semi-local duality from the dominant Pomeron and $\rho$ Regge exchange immediately above threshold to 1 and 2 GeV$^2$. 
 As expected this works best for $n \ge 1$ when the low energy regime dominates. We see that including just $S$ and $P$-waves is not sufficient for this agreement.  For the $n=0$ sum rule even higher waves than $D$ are crucial in integrating up to 2 GeV$^2$. In contrast for $n=3$ of course just $S$ and $P$ are naturally sufficient.
 Higher values of $n$ would weight the near threshold behaviour of all waves even more and this region is less directly controlled by resonance contributions alone but their tails down to threshold, where Regge averaging is less likely to be valid. Thus we restrict attention to our Finite Energy Sum Rules with $n=1-3$.  It is important to note that all we require is the fact that the $I_t=2$ exchange is lower lying than those with  $I_t=0,\ 1$. That the continuation of Regge behaviour for the absorptive parts of the amplitude actually does average resonance-dominated low energy data even with sum-rules with $n=2,\ 3$ is proved by considering the $P$ and $D$-wave scattering lengths. With scattering lengths defined by being the limit of the real part of the appropriate partial waves, Eq.~(5), as the momentum tends to zero:
\be
a_\ell^I\;=\;\lim_{p\to 0}\,{\cal A}_\ell^I(s)/(p/m_\pi)^{2\ell}
\ee
where $p = \frac{1}{2}\, \sqrt{s-s_{\mathrm{th}}}$. Then by using the Froissart-Gribov representation for the partial wave amplitudes, we have
\be
a^1_1&=&\frac{4}{3\pi}\;\int_{s_{\mathrm{th}}}^\infty\,\frac{ds}{s^2}\;\mathrm{Im}\,A^{t1}(s,t_{\mathrm{th}})\\[2.5mm]
a^0_2&=&\frac{16}{15\pi}\;\int_{s_{\mathrm{th}}}^\infty\,\frac{ds}{s^3}\;\mathrm{Im}\,A^{t0}(s,t_{\mathrm{th}})\;.
\ee
If we evaluate these integrals using just the Regge representation from threshold up, we find the following result
\be
\nonumber
m_\pi^2\,a^1_1&=&\frac{1}{12\pi}\,\beta_\rho(t_{\mathrm{th}})\,(\alpha'\,s_{\mathrm{th}})^{\alpha_\rho}\\
&&\hspace{1cm}\cdot\; \Gamma\left(\frac{5}{2}+\frac{\alpha_\rho}{2}\right)\,\Gamma\left(\frac{1}{2}-\frac{\alpha_\rho}{2}\right),\\
\nonumber m_\pi^4\,a^0_2&=&\frac{1}{120\pi}\,\sum_{R=P,f_2}\,\beta_R(t_{\mathrm{th}})\,(\alpha'\,s_{\mathrm{th}})^{\alpha_R}\\
&&\hspace{1cm}\cdot\; \Gamma\left(2+\frac{\alpha_R}{2}\right)\,\Gamma\left(1-\frac{\alpha_R}{2}\right)\;, 
\ee
where each $\alpha_R$ is to be evaluated at $t=t_{\mathrm{th}}$. Analysis of high energy $NN$ and $\pi N$ scattering~\cite{rarita,PY} determines the couplings $\beta_R$ of the contributing Regge poles to $\pi\pi$ scattering through factorization~\cite{barger-cline}. In the case of the $\rho$ the value of the residue is known to be almost proportional to $\alpha_\rho(t)$ putting a zero close to $t\simeq -0.5$ (GeV$^2$) and reproducing the correct $\rho\pi\pi$ coupling at $t = m_\rho^2$. This is more like the shape shown in Ref.~\cite{pennannphys} than that proposed earlier by Rarita {\it et al.}~\cite{rarita,PY}. This fixes $\beta_\rho(t=t_{\mathrm{th}}) = 0.84 \pm 0.13$ from the ``best value'' of the analysis of Ref.~\cite{PY}\footnote{note that the amplitudes defined in~\cite{PY} are $\pi/4$ times those used here.}. The suppression of $I=2$ $s$-channel amplitudes that is basic to our assumptions here requires an exchange degeneracy between the $\rho$ and $f_2$ trajectories, 
 so that $\beta_{f_2}\,=\,3 \beta_\rho/2$, as in the ``best value'' fit of Ref.~\cite{PY}. With the Pomeron contribution proportional to a $\pi\pi$ cross-section of $16 \pm 2$ mb for $s \simeq 5-8$ GeV$^2$. This gives
\be
\nonumber m_\pi^2\,a^1_1&=& (3.4 \pm 0.5)\cdot 10^{-2}\quad,\\
m_\pi^4\,a^0_2 &=& (1.67 \pm 0.19)\cdot 10^{-3}
\ee
to be compared with the precise values found by Colangelo, Gasser and Leutwyler~\cite{cgl}
from a dispersive analysis of  $\pi\pi$ amplitudes combining Roy Eqs. and 
$\chi$PT predictions
\be
\nonumber
m_\pi^2\,a^1_1&=& (3.79 \pm 0.06)\cdot 10^{-2}\quad,\\
 m_\pi^4\,a^0_2&=& (1.75 \pm 0.03)\cdot 10^{-3}\quad ,
\ee
or the recent dispersive analysis by two of us and other collaborators in \cite{GarciaMartin:2011cn}, which includes the latest NA48/2 $K_{e4}$ decay results \cite{Batley:2010zza} and no $\chi$PT
\be
\nonumber
m_\pi^2\,a^1_1&=& (3.81 \pm 0.09)\cdot 10^{-2}\quad,\\
 m_\pi^4\,a^0_2&=& (1.78 \pm 0.03)\cdot 10^{-3}\quad .
\ee

We see that the presumption that Regge parametrization averages the low energy scattering in terms of sum-rules with $n=2,3$ is borne out with remarkable accuracy: far greater accuracy than underlies our fundamental assumption that $I=2$ $s$-channel resonances and $t$-channel exchanges are suppressed relative to those with $I=0$ and 1. This is further supported by the fact that the $I=2$ $D$-wave scattering length as determined in \cite{cgl,GarciaMartin:2011cn}  is indeed a factor of 10 smaller than that for $I=0$. The required cancellation between the $\rho$ and the $\sigma$ contributions that is the subject of this paper requires a less stringent relation than nature imposes at $\Nc=3$.

\section{$\boldsymbol{N_\mathrm{c}}$ dependence of $\boldsymbol{\pi\pi}$ scattering to one loop 
  UChPT:  Dominant non-\qq behaviour of the $\sigma$}

Having confirmed that semi-local duality between resonances and Regge behaviour works for $\Nc=\,3$, we turn to the description of amplitudes within Chiral Perturbation Theory and the Inverse Amplitude Method (IAM).
For orientation we recapitulate first the central results of Ref.~\cite{Pelaez:2003dy} and we will discuss the uncertainties
at the end.
We plot in Fig.~\ref{fig:TIJsu3}, the imaginary part of the $\pi\pi$
scattering partial waves, $T^I_J$, with $I=J=0$ and $I=J=1$ from unitarised one loop $SU(3)$ $\chi$PT, which fits the experimental data very well for $\Nc=\,3$.
The virtue of $\chi$PT is the fact that the constants all have a dependence on $\Nc$ that is well-defined at leading order.

\begin{figure*}
  \begin{center}
    \includegraphics[scale=0.28]{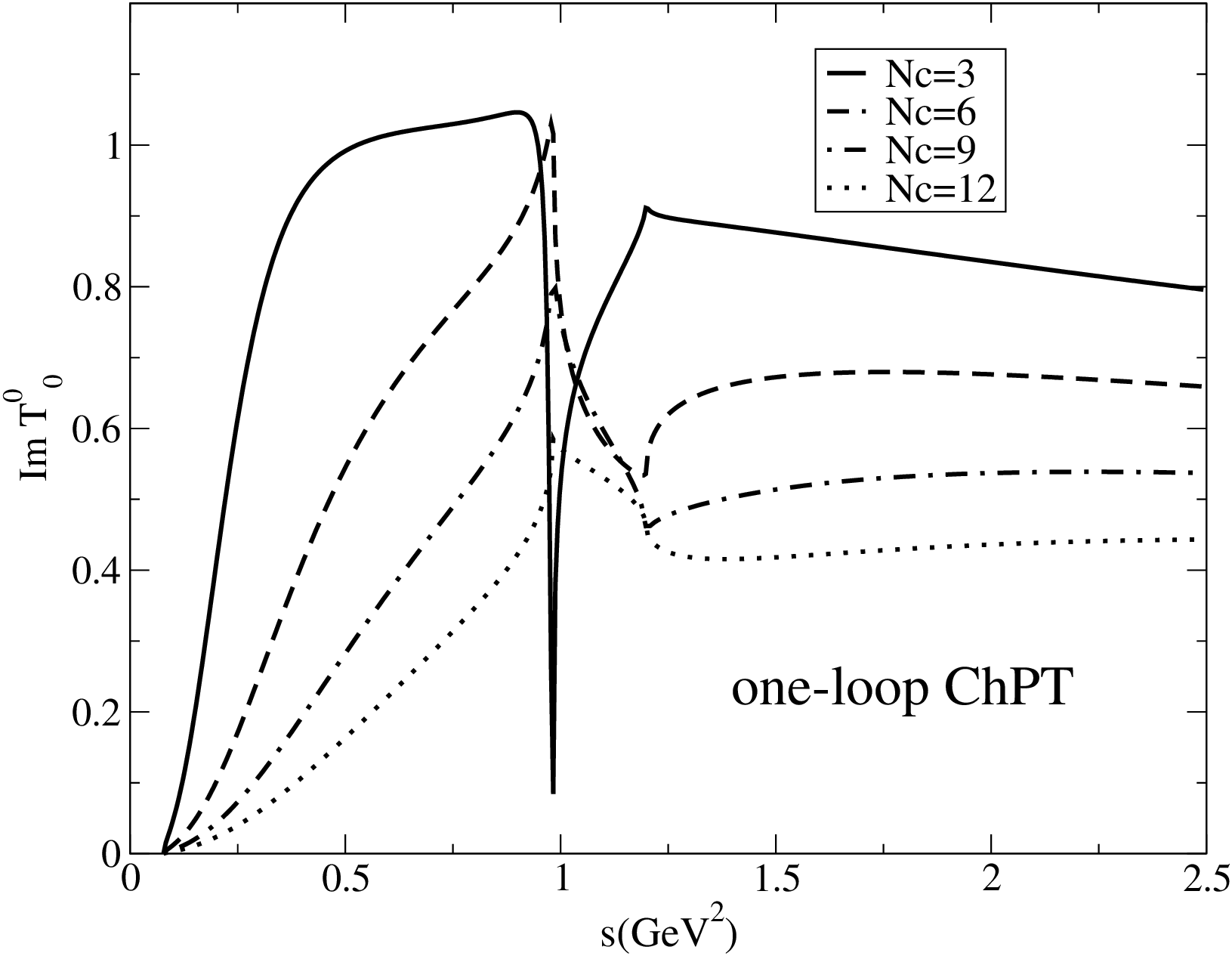}\hspace{4mm}\includegraphics[scale=0.28]{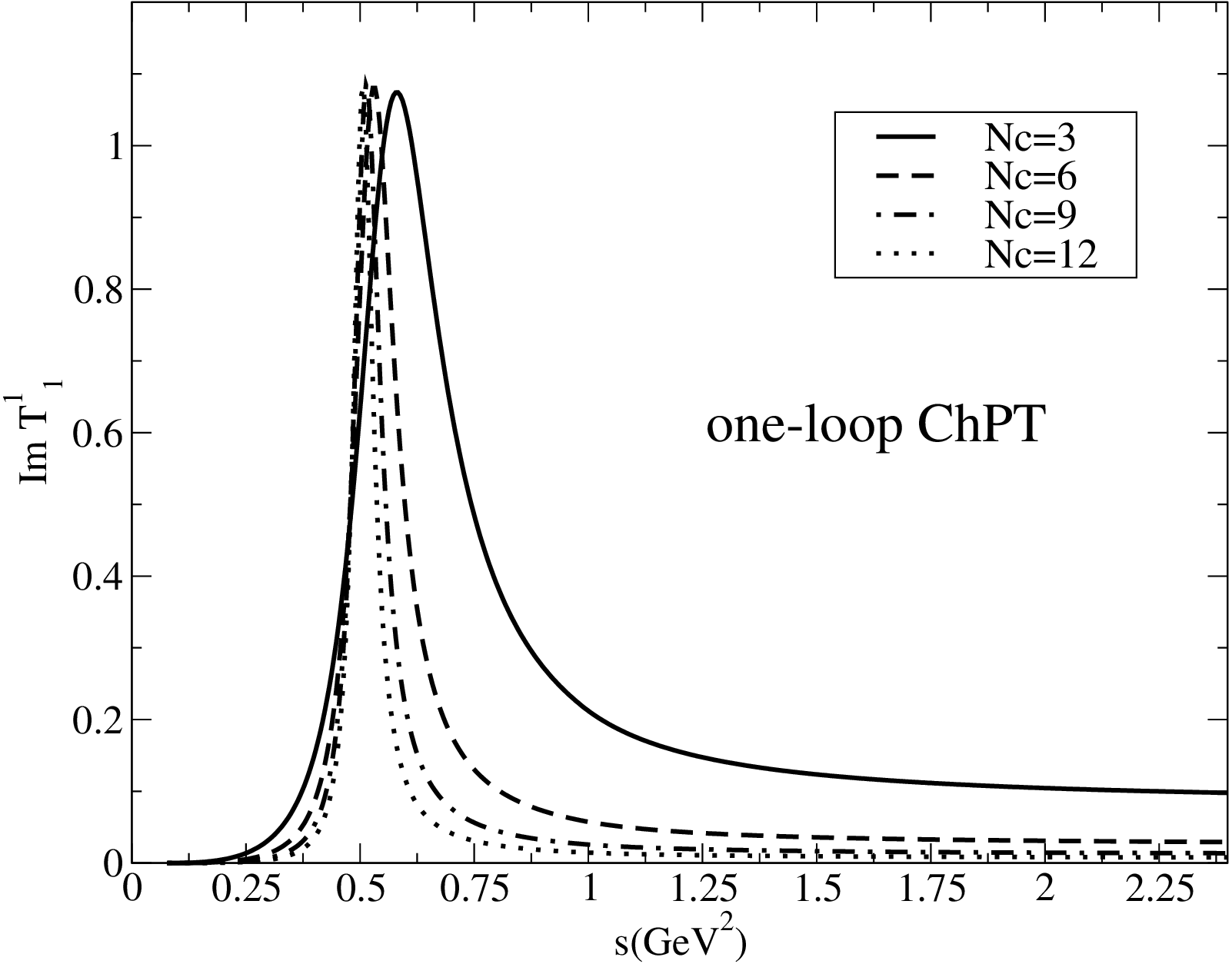}
    \caption{Absorptive parts of key partial wave amplitudes, ${\mathrm{Im}T^I_J}(s)$ with $I=J=0$ and $I=J=1$. Parameters are fixed from a coupled channel $SU(3)$ chiral fit at $\Nc=\,3$ to data.\label{fig:TIJsu3}}
  \end{center}
\end{figure*}

As anticipated by the work of one of us (JRP)~\cite{Pelaez:2003dy}, Fig.~\ref{fig:TIJsu3} shows how the $\rho$ 
peak narrows as $\Nc$ increases  and how its mass barely moves (for the LECs used here
the mass decreases slightly, whereas for those in \cite{Pelaez:2003dy}, with a coupled channel IAM,
it increases, but again by very little).
In contrast, any scalar resonance contribution to the isoscalar amplitude becomes smaller and flatter below 1 GeV. Indeed, the positions of the $\rho$ and $\sigma$ poles move along the unphysical sheet as $\Nc$ increases from 3. It is useful to replicate these results here, as shown in Fig.~\ref{fig:polessu3}. We see the $\rho$-pole move towards the real axis, while that for the $\sigma$ moves away from the real axis region below 1 GeV. 
This is, of course, reflected in the behaviour of the amplitudes with definite $t$-channel isospin, Eq.~(\ref{At}).

\begin{table}
  \begin{center}
    \begin{tabular}{c|c}
      LECs(x $10^3$) & One-loop IAM\\\hline
      $L_1^r$ & ~0.60 ~$\pm$ ~0.09\\
      $L_2^r$ & ~1.22 ~$\pm$ ~0.08\\
      $L_3^r$ & -3.02 ~$\pm$ ~0.06\\
      $L_4^r$ & 0(fixed)\\
      $L_5^r$ & ~1.90 ~$\pm$ ~0.03\\
      $L_6^r$ & -0.07 ~$\pm$ ~0.20\\
      $L_7^r$ & -0.25 ~$\pm$ ~0.18\\
      $L_8^r$ & ~0.84 ~$\pm$ ~0.23
    \end{tabular}\caption{One-loop IAM LECs we have used~\cite{Pelaez:2004xp}.}
  \end{center}
\label{tab:LECs}
\end{table}

\begin{figure}
  \begin{center}
    \includegraphics[scale=0.27]{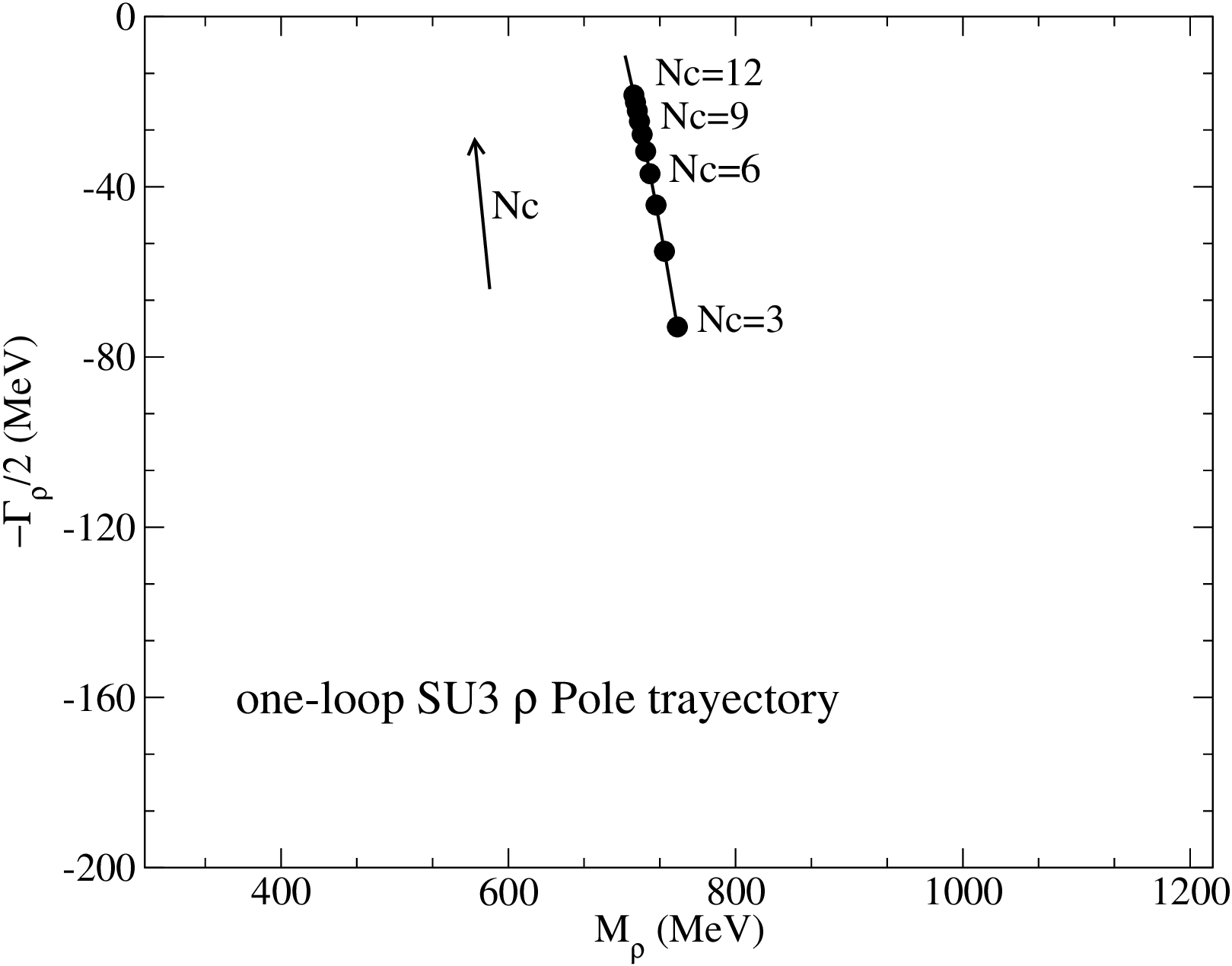}\hspace{4mm}\includegraphics[scale=0.27]{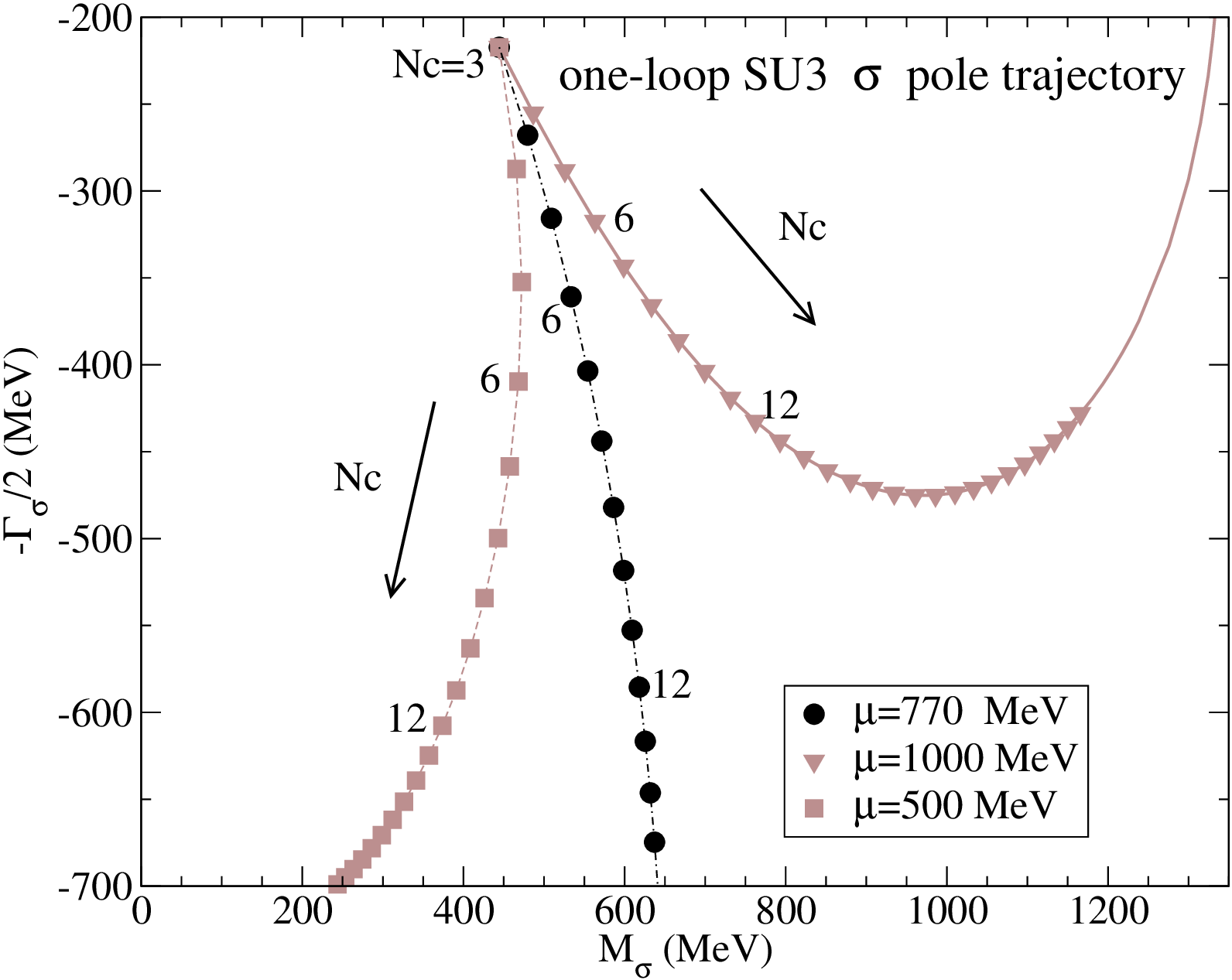}
    \caption{Position of the $\rho$ and $\sigma$ poles in the complex energy plane as a function of $\Nc$ in one loop $\chi$PT.  Black lines 
correspond to  the fit described in the text~\cite{Pelaez:2004xp}
imposing the leading $1/N_c$ behaviour of the LECs at 
the usual renormalization scale $\mu=770\,$MeV.  Note the different vertical scales for the $\rho$ and $\sigma$ poles. 
The lighter points delineate the estimated uncertainty from the choice of $\mu$.
This range is not plotted for the $\rho$, 
since  it is so very close to the central line.
\label{fig:polessu3}
}
  \end{center}
\end{figure}

The one-loop LECs we have used are those from Ref.~\cite{Pelaez:2004xp}.
These are listed in Table~II.
Constructing the IAM analysis of ~\cite{Pelaez:2006nj} using these LECs, we show in Fig.~\ref{fig:Atsu3} the imaginary parts of the resulting amplitudes as functions of $s$. We see for instance in looking at Im$A^{t2}(\nu,t_{\mathrm{th}})/\nu^2$ that at $\Nc=3$ the positive $\sigma$ and negative $\rho$ components cancel. This is not the case as $\Nc$ increases to 12.

\begin{figure*}
  \begin{center}
          \includegraphics[scale=0.28]{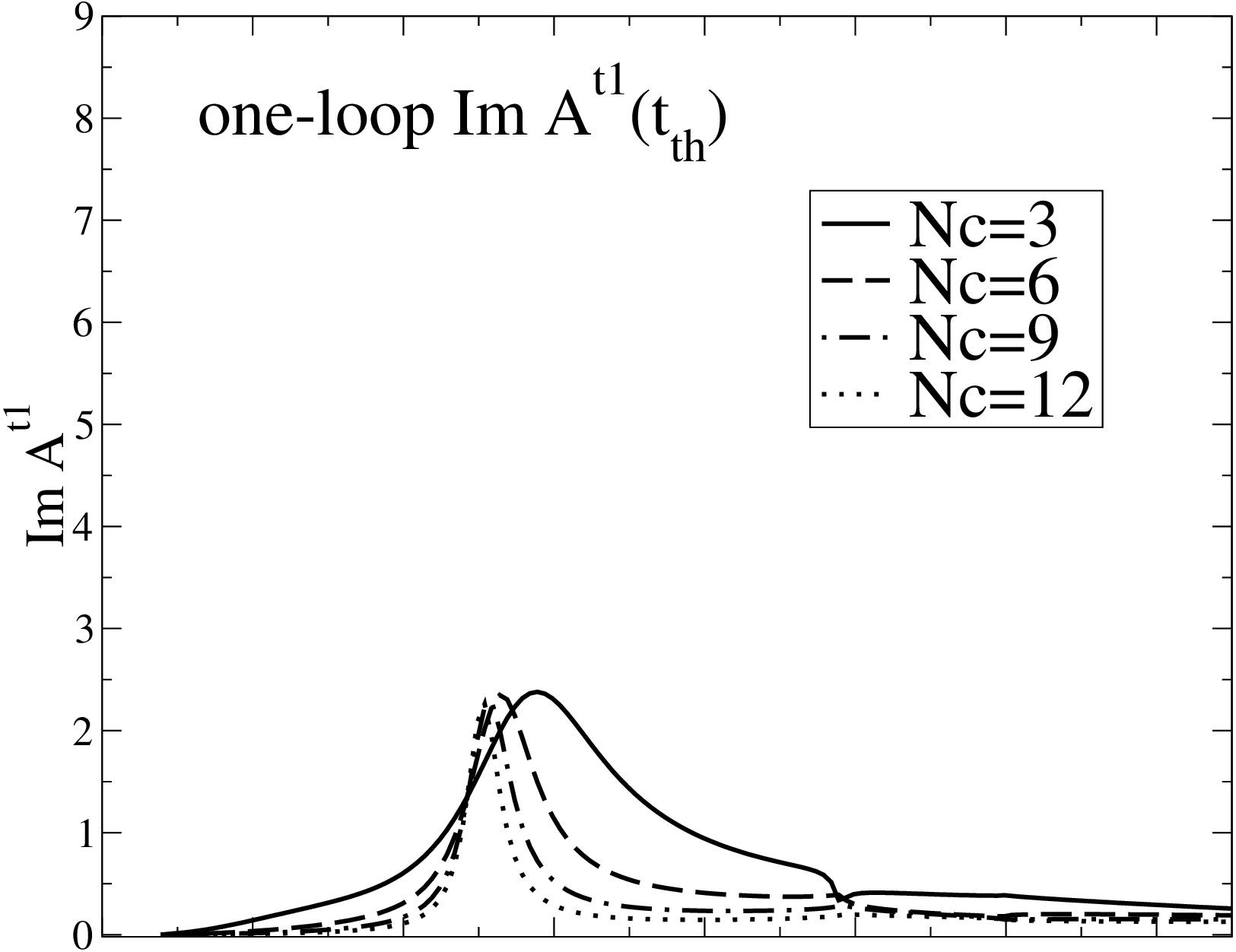}\hspace{4mm}\includegraphics[scale=0.28]{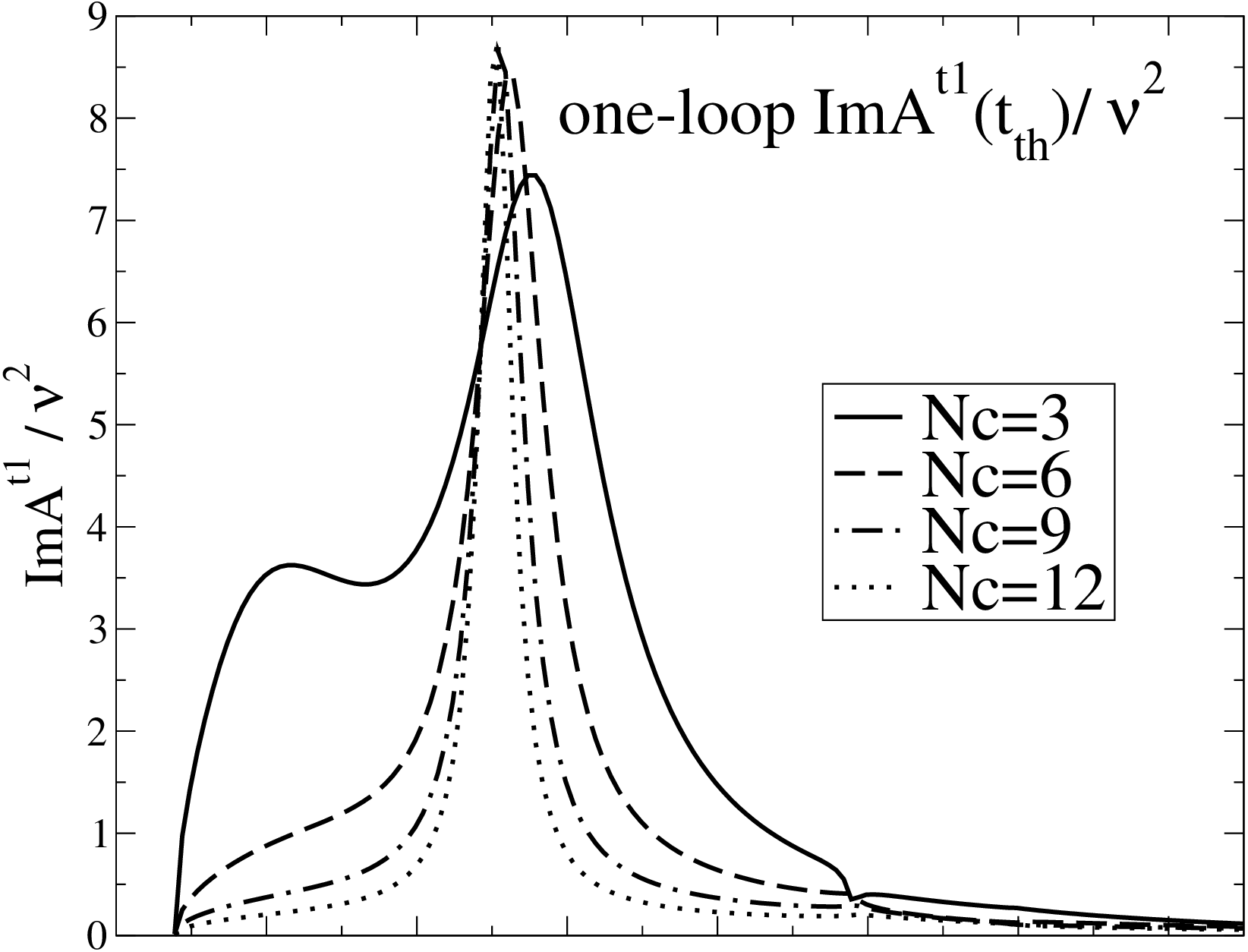}\\
          \includegraphics[scale=0.28]{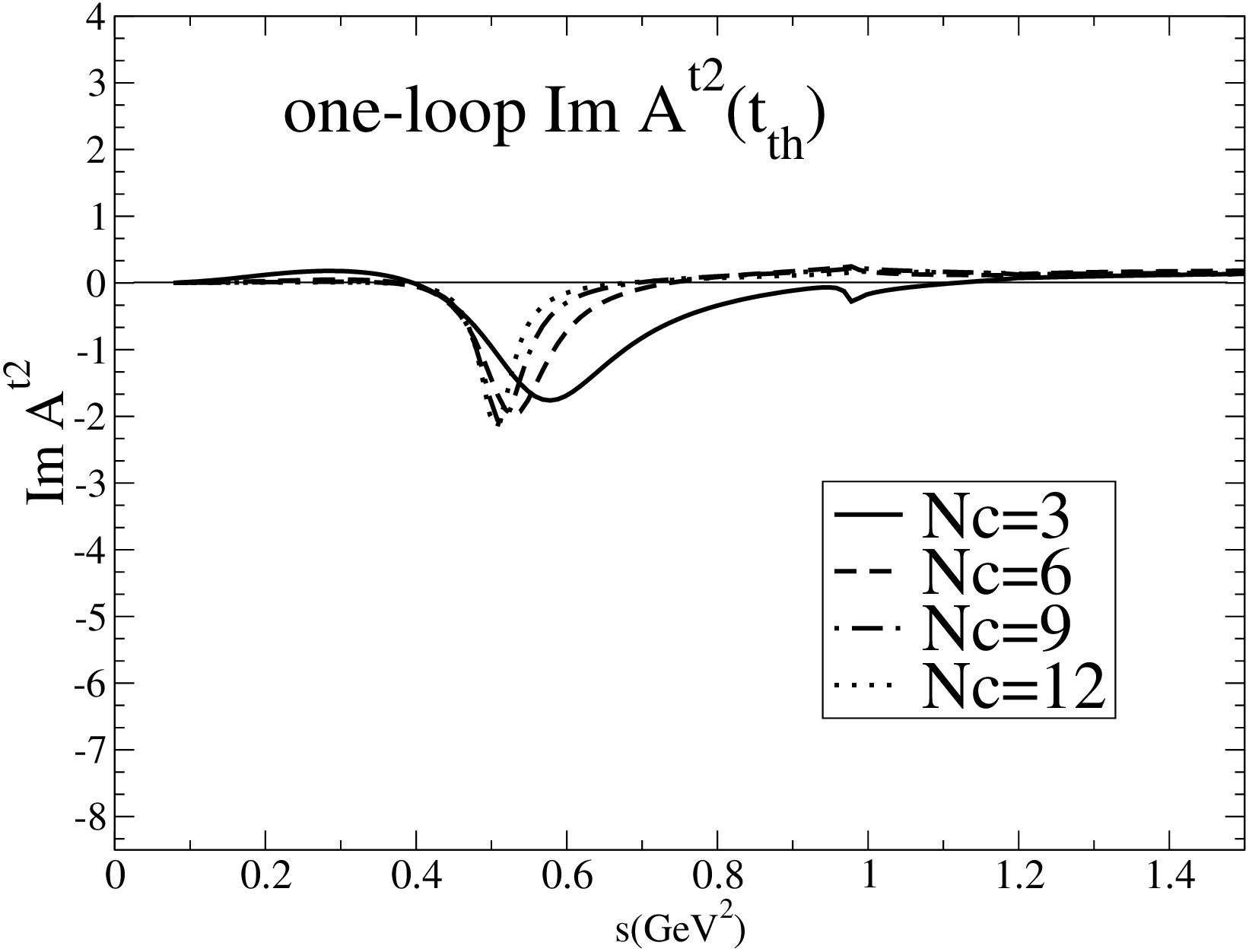}\hspace{4mm}\includegraphics[scale=0.28]{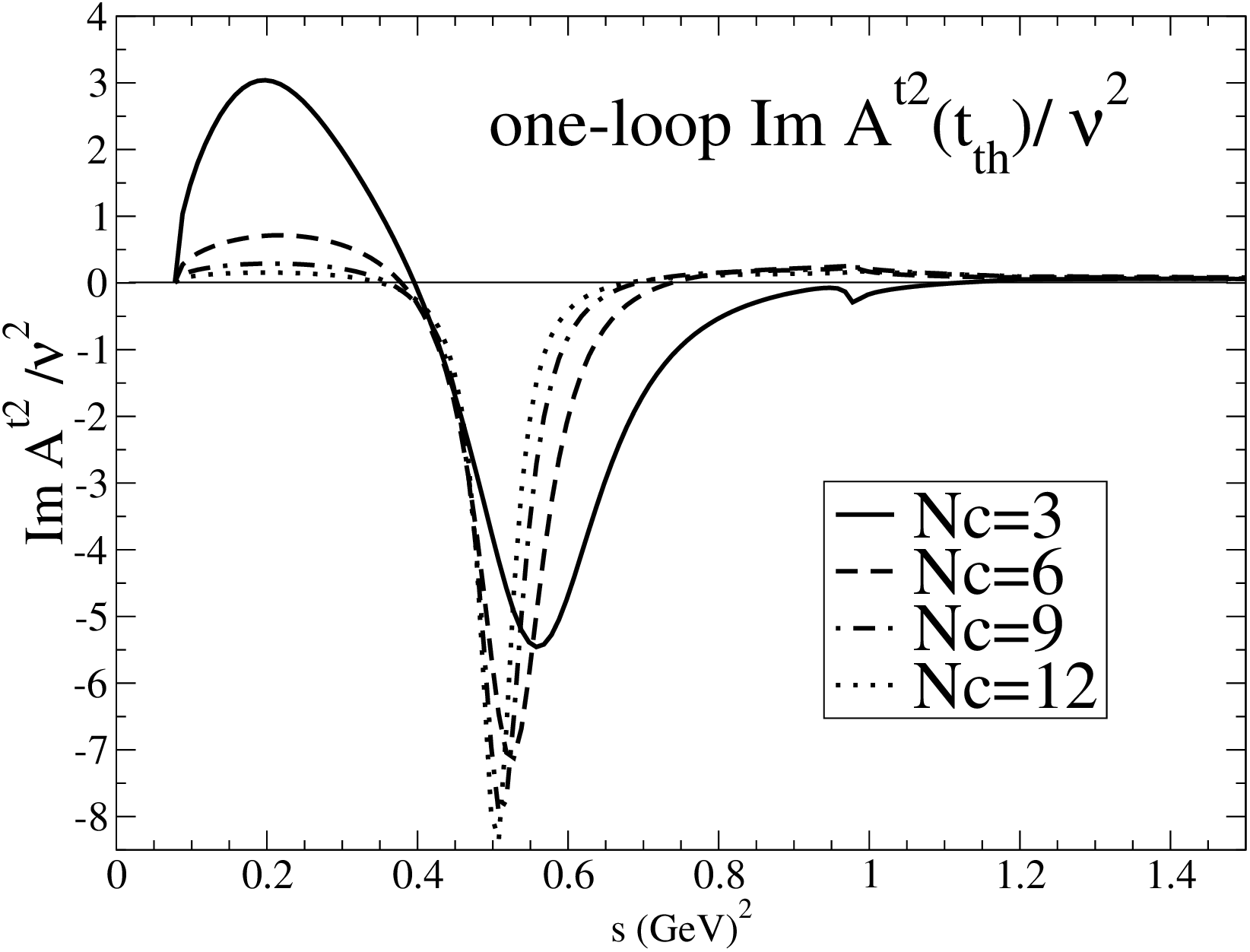}
      \caption{Absorptive part of amplitudes with definite $t$-channel isospin, ${\mathrm{Im}A^{tI}(s,t_{\mathrm{th}})}/\nu^n$. The top pair of graphs have $I=1$ and the lower with $I=2$, and on the left hand $n=0$ and right hand $n=2$.
        Parameters have been fixed from a coupled channel $SU(3)$ chiral fit at $\Nc=\,3$ to data.}\label{fig:Atsu3}
    \end{center}
  \end{figure*}

To quantify the $\Nc$ dependence at different orders in $\chi$PT and with different choices of LECs, 
we calculate the value of Finite Energy Sum Rules (FESR) ratios: 
\begin{eqnarray}
\nonumber\\
F_n^{\;I\:I'}(t)= \frac{\int_{\nu_{\mathrm{th}}}^{\nu_{\mathrm{max}}}{d\nu\; {\mathrm {Im}}\ A^{tI}(\nu,t,N_\mathrm{c})/\nu^n}}{\int_{\nu_{\mathrm{th}}}^{\nu_{\mathrm{max}}}{d\nu\;{\mathrm {Im}}\ A^{tI'}(\nu,t,N_\mathrm{c})/\nu^n}},
\label{ratios}
\end{eqnarray}
for different values of $n=0-3$,  and $N_c$, $t$, $\nu_\mathrm{max}$, and isospin
$t$-channels $I, I'$. The ratio $F^{10}$ compares the amplitude given by $\rho$ Regge-exchange with that controlled by the Pomeron, while the ratio $F^{21}$ compares the ``exotic'' four quark exchange with \qq $\rho$-exchange.

\begin{figure*}
  \begin{center}
      \includegraphics[scale=0.24]{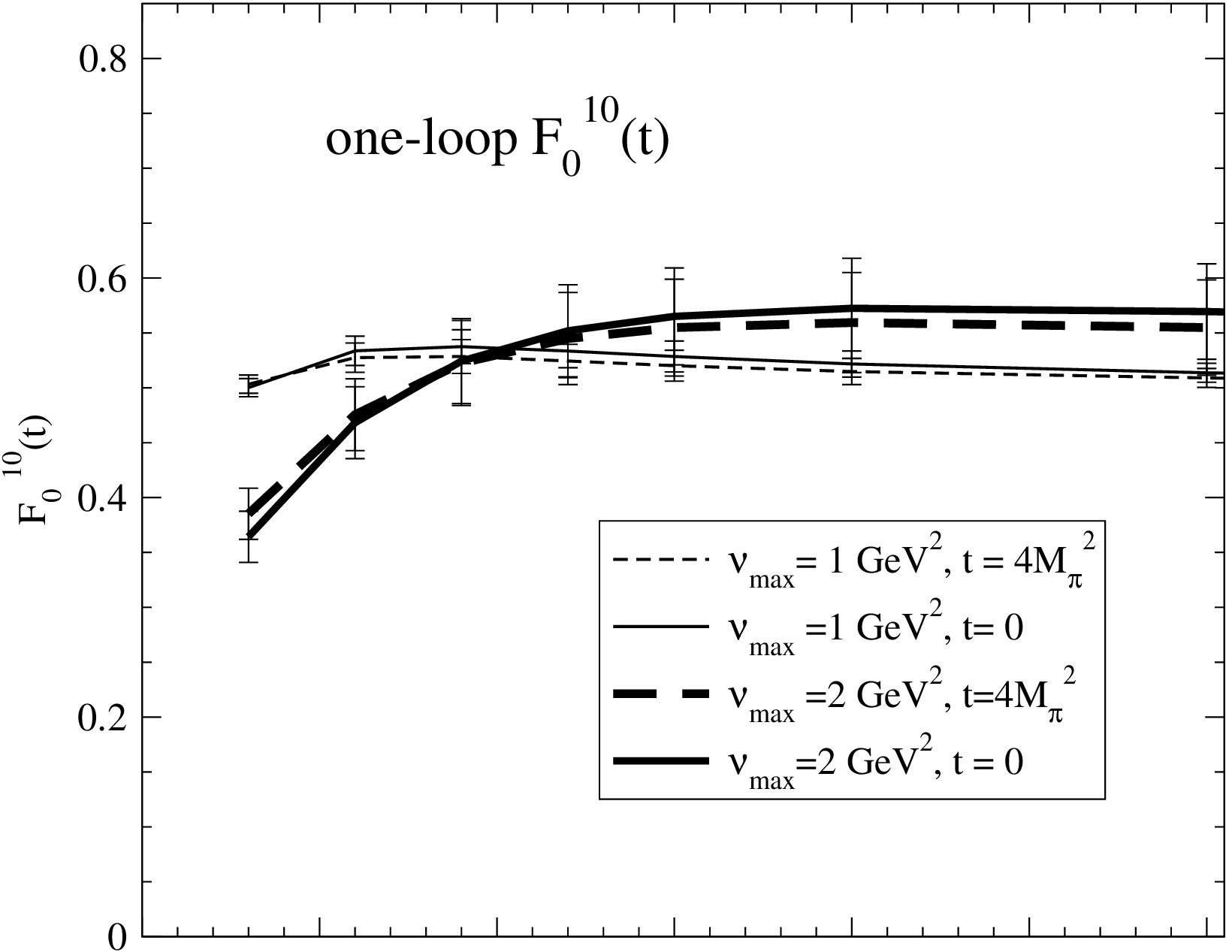}\hspace{4mm}\includegraphics[scale=0.24]{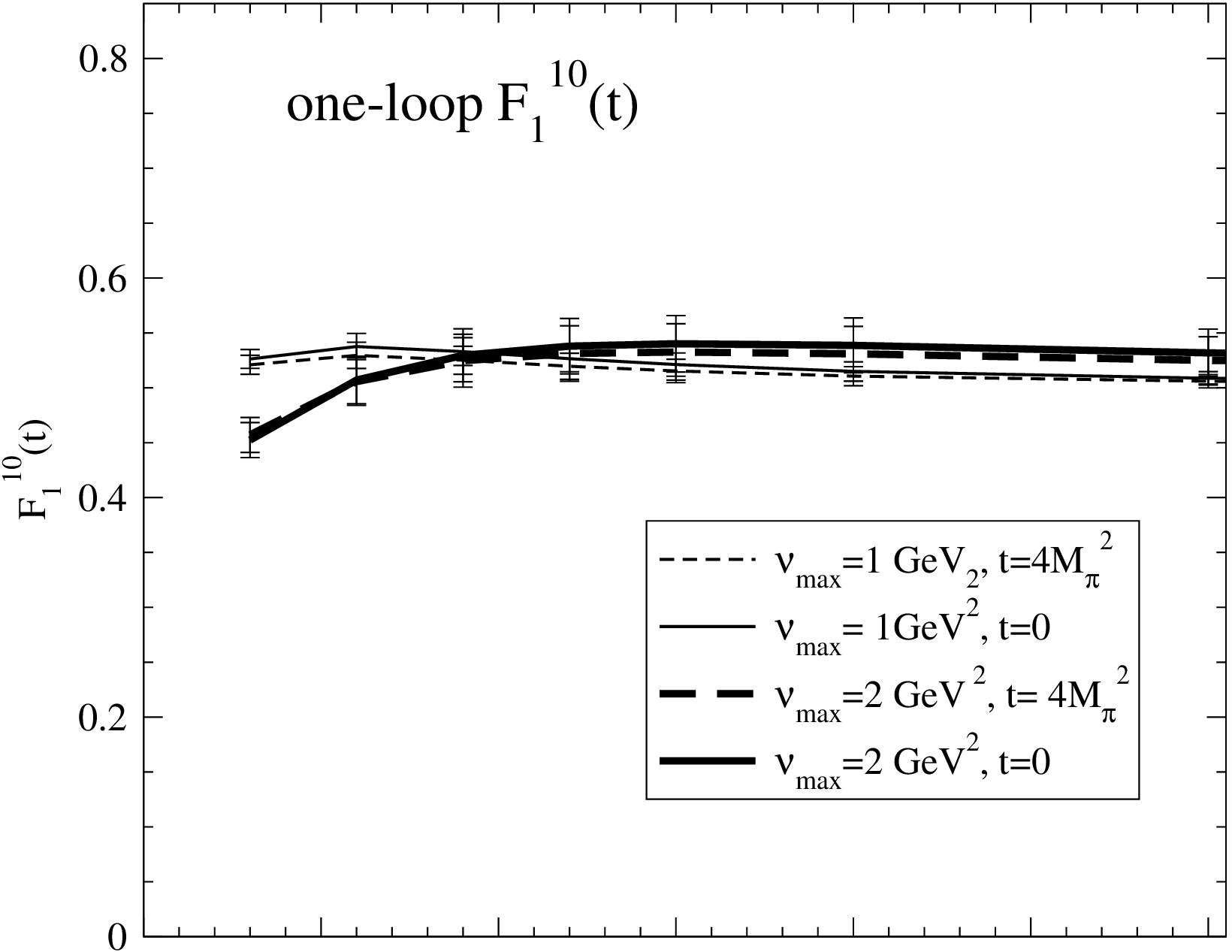}\\
      \includegraphics[scale=0.24]{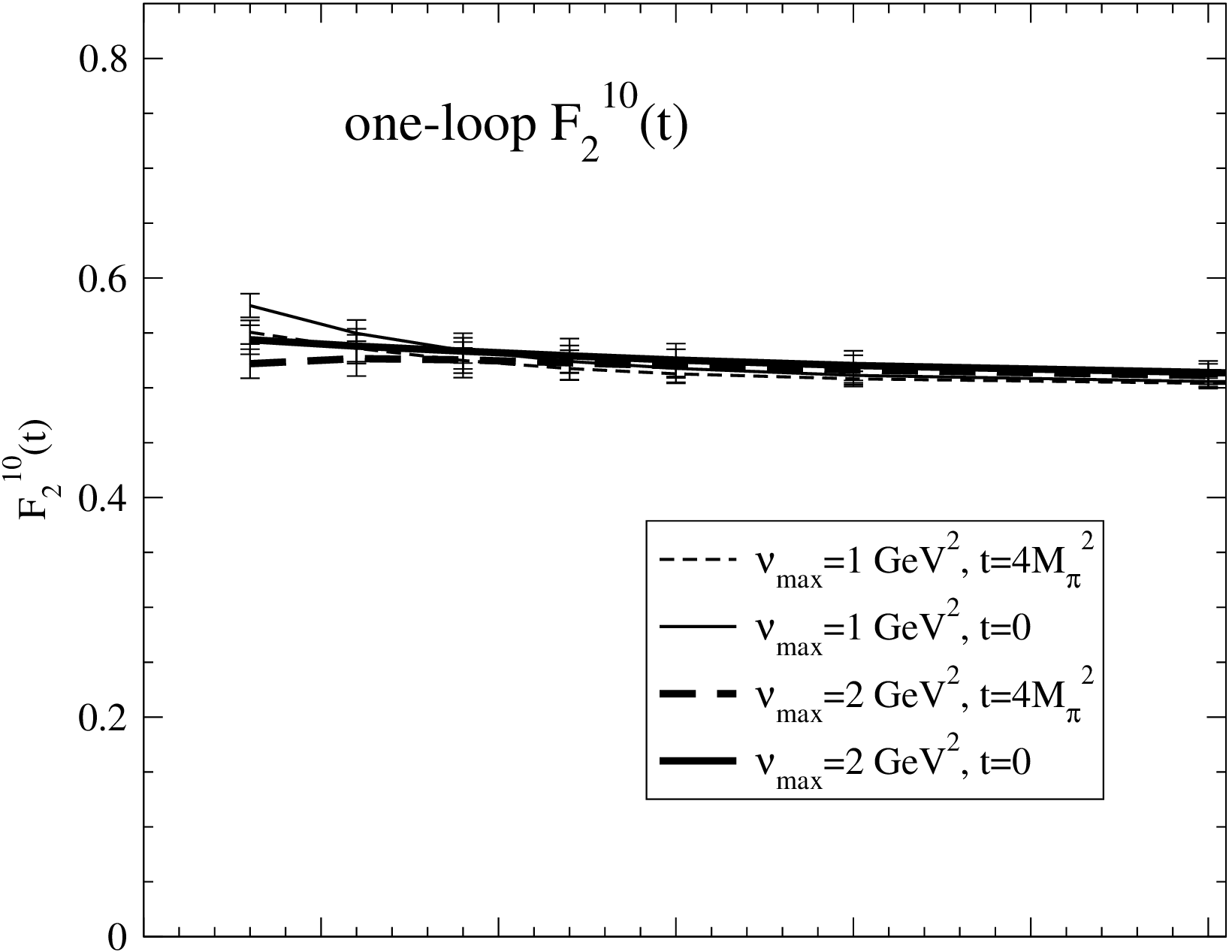}\hspace{4mm}\includegraphics[scale=0.24]{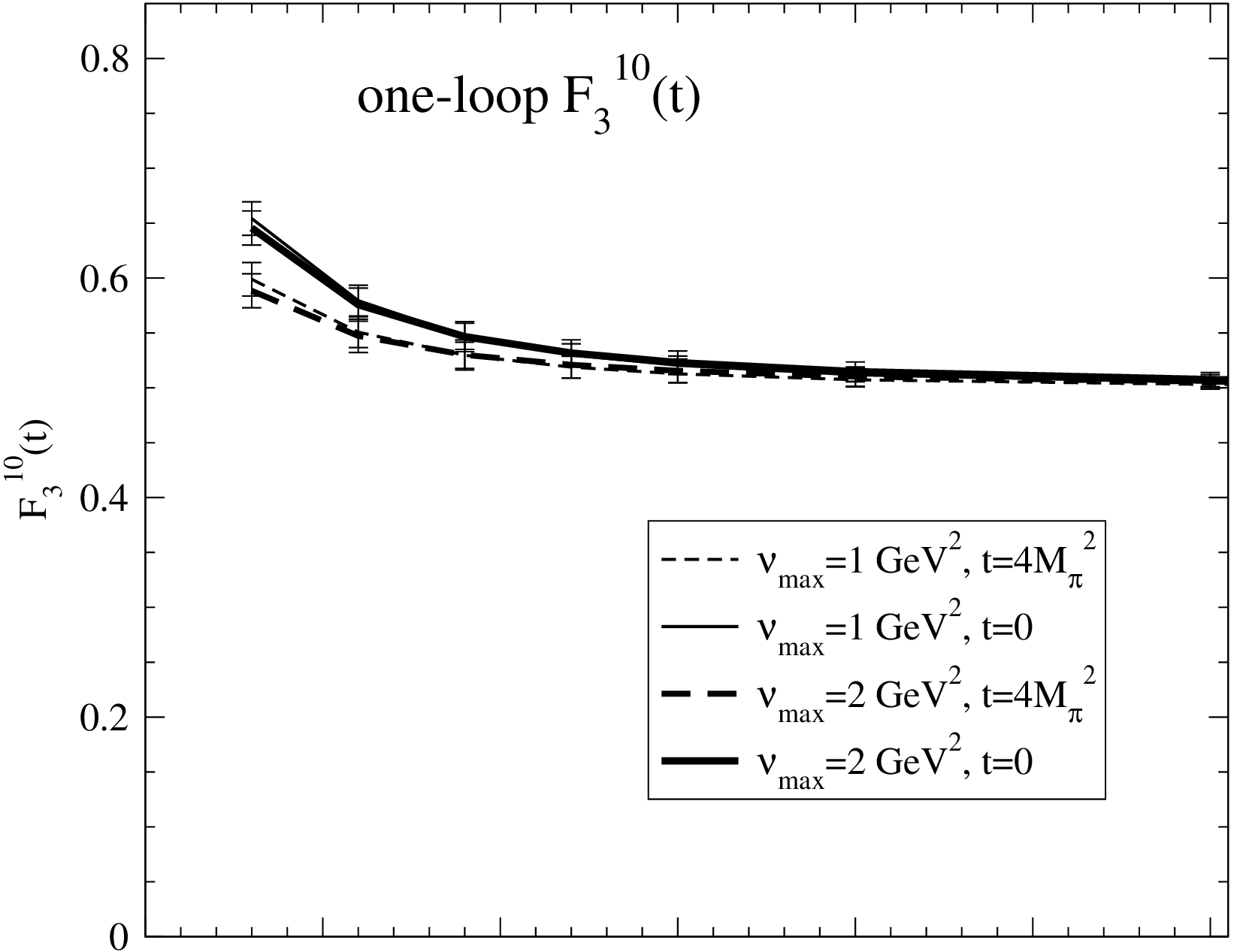}\\
      \includegraphics[scale=0.24]{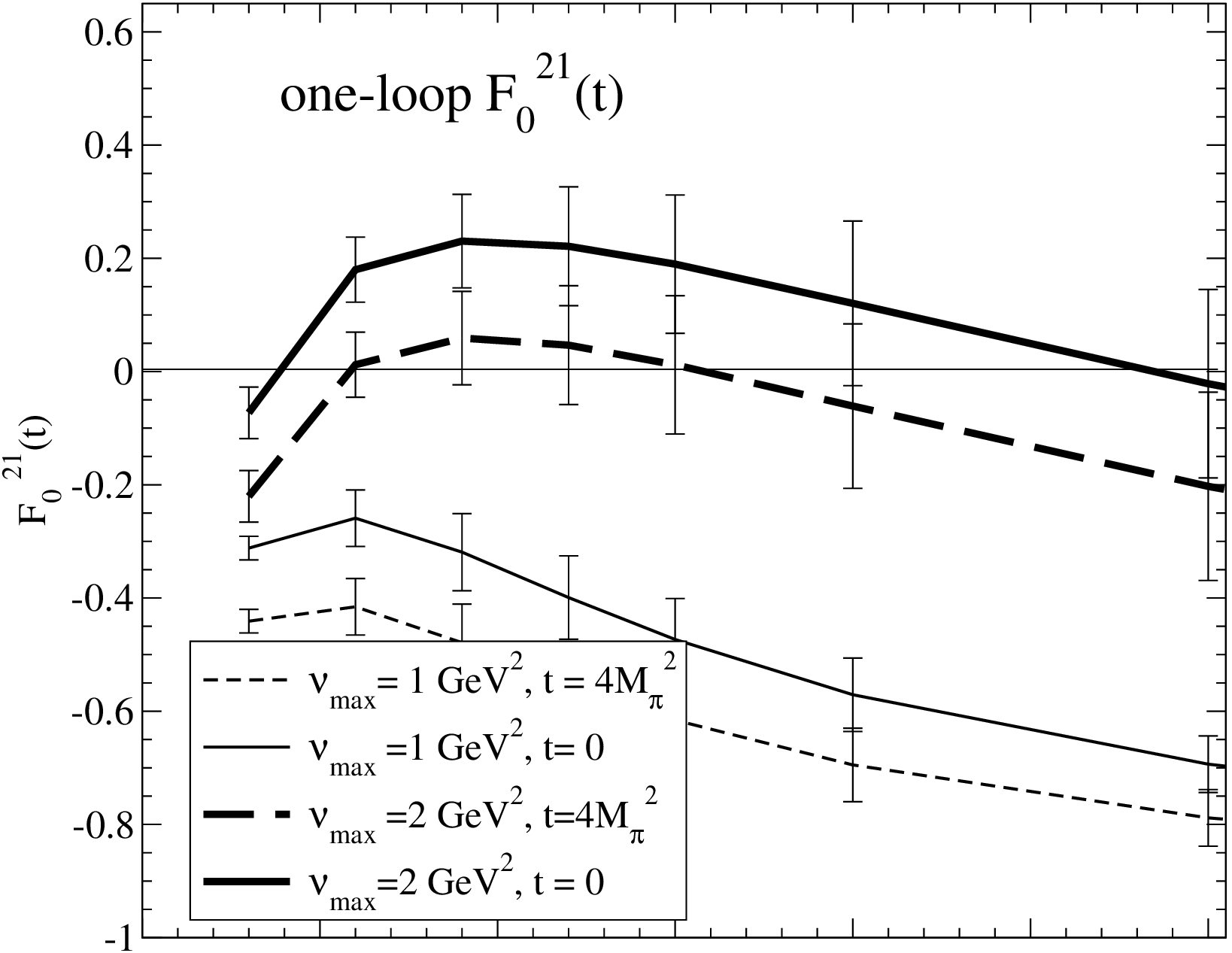}\hspace{4mm}\includegraphics[scale=0.24]{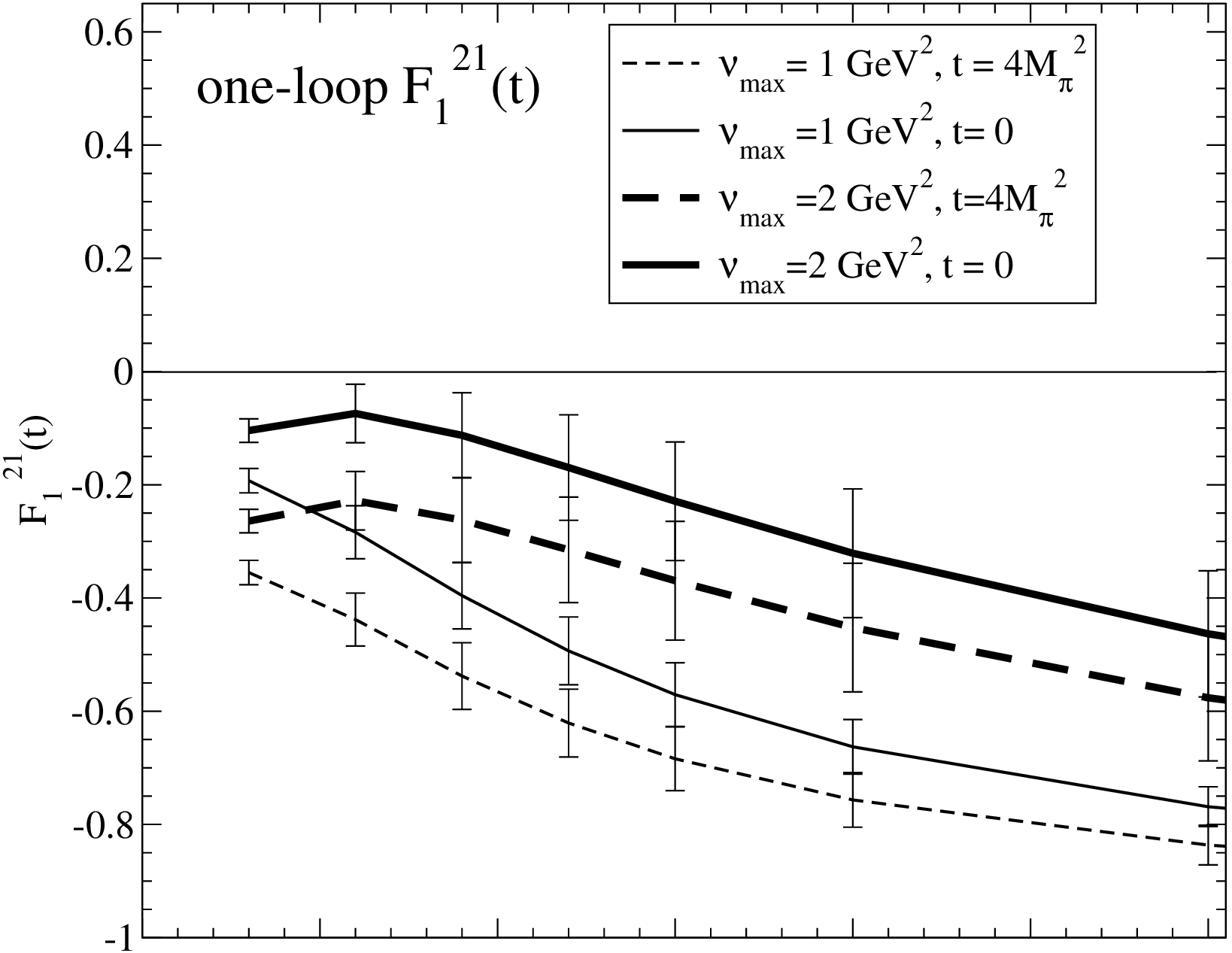}\\
      \includegraphics[scale=0.24]{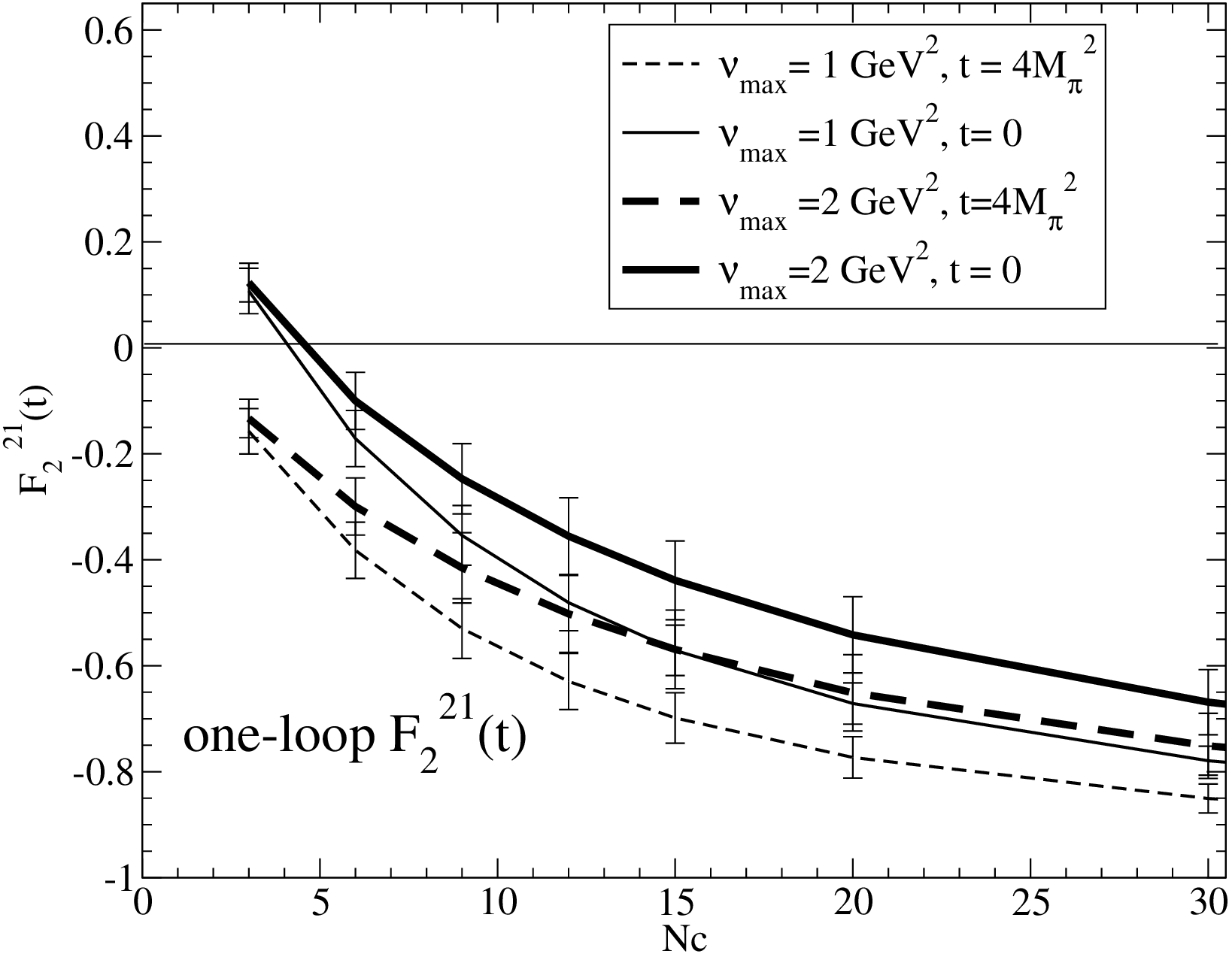}\hspace{4mm}\includegraphics[scale=0.24]{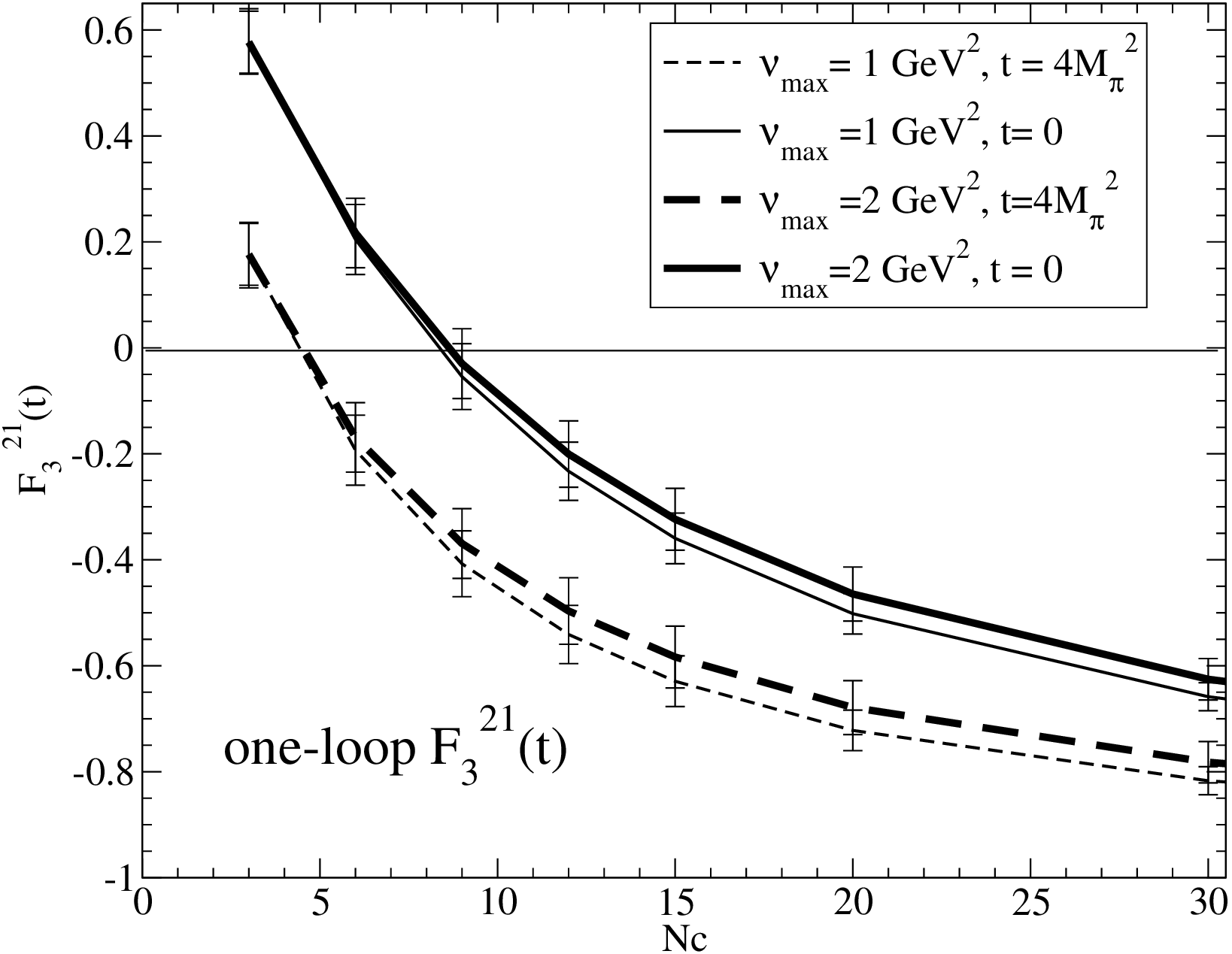}
      \caption{Ratios $F^{I\,I'}_n$ of Eq.~(13) with $n=0-3$. The top four graphs are for $F^{10}$, and the lower four for $F^{21}$. One loop $\chi$PT IAM parameters are from a coupled channel $SU(3)$ fit with $\Nc=\,3$ to data.}\label{fig:ratiossu3}
    \end{center}
  \end{figure*}

  \begin{center}
    \begin{table*}
      \begin{tabular}{|c|c|c||c|c||c|c|}\cline{4-7}
        \multicolumn{3}{c}{} & \multicolumn{4}{|c|}{\textbf{1 loop $SU(3)$ IAM}}\\\cline{4-7}
        \multicolumn{3}{c}{} &\multicolumn{2}{|c||} {$\mathbf{t = t_{\mathrm{th}}}$}&\multicolumn{2}{c|}{\textbf {$t=0$}}\\\hhline{---====}
        & $n$ & $N_\mathrm{c}$ & $\nu_\mathrm{max}$=1 GeV$^2$ & $\nu_\mathrm{max}$=2 GeV$^2$ &$\nu_\mathrm{max}$=1 GeV$^2$ &$\nu_\mathrm{max}$=2 GeV$^2$\\\hline\hline
        \multirow{16}{0.35in}{$\;\mathbf{F^{\;1\,0\;}_n}$~~}& 
        \multirow{4}{0.1in}{0} 
        & 3 &~0.503 $\pm$ 0.008 & ~0.385 $\pm$ 0.023 & ~0.500 $\pm$ 0.010 & ~0.364 $\pm$ 0.027 \\
        & & 6 &~0.527 $\pm$ 0.013 & ~0.475 $\pm$ 0.033 & ~0.534 $\pm$ 0.017 & ~0.468 $\pm$ 0.038\\
        & & 9 &~0.528 $\pm$ 0.015 & ~0.522 $\pm$ 0.039 & ~0.537 $\pm$ 0.020 & ~0.524 $\pm$ 0.046\\
        & & 12 &~0.524 $\pm$ 0.015 & ~0.545 $\pm$ 0.042 & ~0.533 $\pm$ 0.021 & ~0.552 $\pm$ 0.050 \\\Cline{0.8pt}{2-7}
        &\multirow{4}{0.1in}{1} 
        & 3 &~0.521 $\pm$ 0.008 & ~0.457 $\pm$ 0.016 & ~0.526 $\pm$ 0.011 & ~0.452 $\pm$ 0.019 \\
        & & 6 &~0.529 $\pm$ 0.011 & ~0.506 $\pm$ 0.022 & ~0.538 $\pm$ 0.015 & ~0.507 $\pm$ 0.026\\
        & & 9 &~0.525 $\pm$ 0.013 & ~0.525 $\pm$ 0.024 & ~0.532 $\pm$ 0.016 & ~0.530 $\pm$ 0.029\\
        & & 12 &~0.520 $\pm$ 0.012 & ~0.531 $\pm$ 0.027 & ~0.526 $\pm$ 0.016 & ~0.538 $\pm$ 0.030 \\\Cline{0.8pt}{2-7}
        &\multirow{4}{0.1in}{2} 
        & 3 &~0.551 $\pm$ 0.011 & ~0.522 $\pm$ 0.013 & ~0.575 $\pm$ 0.013 & ~0.544 $\pm$ 0.016\\
        & & 6 &~0.536 $\pm$ 0.012 & ~0.526 $\pm$ 0.016 & ~0.550 $\pm$ 0.015 & ~0.538 $\pm$ 0.019\\
        & & 9 &~0.525 $\pm$ 0.011 & ~0.525 $\pm$ 0.016 & ~0.534 $\pm$ 0.015 & ~0.533 $\pm$ 0.020\\
        & & 12 &~0.517 $\pm$ 0.010 & ~0.523 $\pm$ 0.016 & ~0.524 $\pm$ 0.013 & ~0.529 $\pm$ 0.019\\\Cline{0.8pt}{2-7}
        &\multirow{4}{0.1in}{3}
        & 3 & ~0.599 $\pm$ 0.015 & ~0.588 $\pm$ 0.015 & ~0.654 $\pm$ 0.017 & ~0.645 $\pm$ 0.017\\
        & & 6 & ~0.551 $\pm$ 0.014 & ~0.547 $\pm$ 0.015 & ~0.579 $\pm$ 0.017 & ~0.575 $\pm$ 0.018\\
        & & 9 & ~0.530 $\pm$ 0.012 & ~0.530 $\pm$ 0.014 & ~0.547 $\pm$ 0.016 & ~0.547 $\pm$ 0.017\\
        & & 12 & ~0.519 $\pm$ 0.010 & ~0.521 $\pm$ 0.012 & ~0.530 $\pm$ 0.013 & ~0.532 $\pm$ 0.015\\\hline\hline        
        \multirow{16}{0.35in}{$\;\mathbf{F^{\;2\,1\;}_n}$~~}& 
        \multirow{4}{0.1in}{0} 
        & 3 & -0.441 $\pm$ 0.021 & -0.220 $\pm$ 0.045 & -0.312 $\pm$ 0.029 & -0.073 $\pm$ 0.058\\
        & & 6 & -0.415 $\pm$ 0.050 & ~0.012 $\pm$ 0.057 & -0.259 $\pm$ 0.057 & ~0.180 $\pm$ 0.059\\
        & & 9 & -0.479 $\pm$ 0.068 &  ~0.059 $\pm$ 0.083 & -0.319 $\pm$ 0.080 & ~0.230 $\pm$ 0.079\\
        & & 12 & -0.552 $\pm$ 0.074 & ~0.047 $\pm$ 0.105 & -0.399 $\pm$ 0.073 & ~0.221 $\pm$ 0.097\\\Cline{0.8pt}{2-7}
        &\multirow{4}{0.1in}{1} 
        & 3 & -0.355 $\pm$ 0.021 & -0.269 $\pm$ 0.021 & -0.193 $\pm$ 0.022 & -0.104 $\pm$ 0.023\\
        & & 6 & -0.438 $\pm$ 0.047 & -0.228 $\pm$ 0.052 & -0.284 $\pm$ 0.051 & -0.074 $\pm$ 0.052\\
        & & 9 & -0.538 $\pm$ 0.054 & -0.262 $\pm$ 0.077 & -0.396 $\pm$ 0.068 & -0.113 $\pm$ 0.078\\
        & & 12 & -0.621 $\pm$ 0.060 & -0.317 $\pm$ 0.093 & -0.493 $\pm$ 0.073 & -0.170 $\pm$ 0.097\\\Cline{0.8pt}{2-7}
        &\multirow{4}{0.1in}{2} 
        & 3 &-0.157 $\pm$ 0.043 & -0.133 $\pm$ 0.036 & ~0.107 $\pm$ 0.039 & ~0.123 $\pm$ 0.032\\
        & & 6 &-0.382 $\pm$ 0.053 & -0.299 $\pm$ 0.054 & -0.171 $\pm$ 0.054 & -0.100 $\pm$ 0.053\\
        & & 9 &-0.530 $\pm$ 0.056 & -0.415 $\pm$ 0.066 & -0.354 $\pm$ 0.063 & -0.247 $\pm$ 0.069\\
        & & 12 &-0.630 $\pm$ 0.053 & -0.505 $\pm$ 0.072 & -0.481 $\pm$ 0.062 & -0.355 $\pm$ 0.078\\\Cline{0.8pt}{2-7}
        &\multirow{4}{0.1in}{3}
        & 3 & ~0.175 $\pm$ 0.062 & ~0.176 $\pm$ 0.058 & ~0.578 $\pm$ 0.042 & ~0.577 $\pm$ 0.040\\
        & & 6 & -0.193 $\pm$ 0.066 & -0.169 $\pm$ 0.065 & ~0.204 $\pm$ 0.057 & ~0.217 $\pm$ 0.056\\
        & & 9 & -0.407 $\pm$ 0.062 & -0.369 $\pm$ 0.066 & -0.054 $\pm$ 0.061 & -0.030 $\pm$ 0.063\\
        & & 12 & -0.541 $\pm$ 0.055 & -0.497 $\pm$ 0.063 & -0.233 $\pm$ 0.060 & -0.200 $\pm$ 0.064\\\hline
      \end{tabular}  
      \caption{ Ratios for 1 loop UChPT using LECs from a single channel
        fit}\centering\label{Table: SU3IAMA}
    \end{table*}
  \end{center}

We show the results in Table~\ref{Table: SU3IAMA}, and plot
the data in Fig.~\ref{fig:ratiossu3}. If Regge expectations were working at one loop order, we would expect $F^{10}$ to tend to 0.66 and for  $F^{21}$ to be very small in magnitude, just as they are at $\Nc=\,3$, particularly for a cutoff of 2~GeV$^2$, the results for which are shown as the bolder lines. However, as $\Nc$ increases we find that the ratio $F^{10}$ tends to $0.5$, while that for $F^{21}$ tends to $-1$. This is in accord with the $n=1,2$ sum rules becoming increasingly dominated by the $\rho$ with very little scalar contribution. This difference is a consequence of the seeming largely non-${\overline q}q$ nature of the $\sigma$ being incompatible with Regge expectations.  All these results use values for the one loop LECs that accurately fit the low energy $\pi\pi$ phase-shifts up to 1 GeV.

Finally, let us recall that the LECs carry a dependence 
on the regularization scale $\mu$ that cancels with those of the loop
functions to give a finite result order by order. As a consequence, when rescaling
the LECs with $N_c$, a specific choice of $\mu$ has to be made.
In other words, despite the $\chi$PT and IAM amplitudes being scale independent, the $N_c$ evolution is not. 
Intuitively, $\mu$ is related to a heavier scale, which has been integrated out in $\chi$PT
and it is customary to take $\mu$ between 0.5 and 1 GeV \cite{Pelaez:2003dy,Donoghue:1988ed}. This range is confirmed
by the fact that at these scales the measured LECs satisfy 
their leading $1/N_c$ relations fairly well~\cite{Donoghue:1988ed}. 
All the previous considerations about the one-loop IAM have been made
with an $N_c$ scaling at the usual choice of renormalization scale  $\mu=770$ MeV $\simeq M_\rho$, which is the most natural choice given the fact that the values of the LECs are mainly saturated by the first octet of vector resonances, 
with additional contributions from scalars above 1 GeV \cite{Donoghue:1988ed}.

Thus, in 
Fig.~\ref{fig:polessu3} we have also illustrated
the uncertainties in the pole movements for the $\rho$(770) and $f_0$(600) due to the 
choice of $\mu$.
Note that the $\rho$(770)  \qq behaviour is rather stable, since for the LECs in Table~\ref{tab:LECs}
the variation is negligible. Other sets of LECs \cite{Pelaez:2003dy,Pelaez:2010er}, 
which also provide a relatively good description
of the $\rho$(770) , show a bigger variation 
with $\mu$, but they always lead to the expected \qq behaviour. 
In contrast, we observe that the only robust feature of the $f_0$(600) 
is that it does not behave predominantly
as a \qq. Unfortunately, its detailed pole behaviour is not well determined except for the fact 
that it moves away from the 400 to 600 MeV region of the real axis and that at
$N_c$ below 15 its width always increases. However, for $N_c$ around 20 or more  and
for the higher values of the $\mu$ range, the width may start decreasing again and the
pole would start behaving as a \qq. 

In Fig.~\ref{fig:one-loop-uncertainty-in-F} we show how the IAM
uncertainty translates into our calculations of the $F^{21}_n$ ratio for the most
interesting cases $n=2,3$. The thick continuous 
line stands for the central values we have been discussing so far, 
which at larger $N_c$ tend to grow in absolute value and, as already commented, spoil semi-local duality.
The situation is even worse when the $N_c$ scaling of our LECs is performed
at $\mu=500\,$MeV. This is due to the fact,  seen in Fig.~\ref{fig:polessu3}, that, 
with this choice of $\mu$, 
the $\sigma$ pole moves deeper and deeper in to the complex plane 
and its mass even decreases. Let us note that this behavior--- compatible
with our IAM results when the uncertainty in $\mu$ is taken into account---  is also found
when studying the leading $N_c$ behavior within other unitarization schemes, or 
for certain values of the LECs within the one-loop IAM \cite{Guo:2011pa, Sun:2005uk}. We would therefore also expect that in these treatments semi-local duality
would deteriorate very rapidly. In \cite{Guo:2011pa}, there is the
$f_0(980)$, as well as
 other scalar states above 1300 MeV, but all of them seem insufficient
to compensate for
the disappearance of the $\sigma$ pole. As we will discuss in Sect.~\ref{sec:heaviereffects},
this is because 
 the contributions of the $f_0(980)$ resonance and 
the region above 1300 MeV to our $F_n^{IJ}$ ratios are rather small, and in 
\cite{Guo:2011pa} they seem to become even smaller, since all those resonances become narrower
as $N_c$ increases. Of course, as pointed out in \cite{Guo:2011pa} this deserves a detailed calculation within their approach. 

In Fig.~\ref{fig:one-loop-uncertainty-in-F} 
we also find that the $F^{21}_n$ are much smaller 
and may even seem to stabilise if we apply the $N_c$ 
scaling of the LECs at $\mu=1000\,$MeV. In such a case, the $\sigma$ pole, after moving away from the 
real axis, returns back at higher masses, above roughly 1 GeV.
For simplicity we only show  $F^{21}_n$ for 
the $t=0$ case, but a similar pattern is found at $t=4M_\pi^2$: the turning back of the $\sigma$ pole at higher masses helps to keep the $F^{21}_n$ ratios smaller.
This behaviour follows from the existence  of a subdominant \qq component within the $f_0$(600) with a mass which is at least twice that of the original $f_0$(600) pole. However,
at one-loop order such behaviour only occurs at one extreme of the $\mu$ range.
In contrast, as we will see next, it appears in a rather natural way in the two-loop analysis 

\begin{figure}
  \begin{center}    
    \includegraphics[scale=0.26]{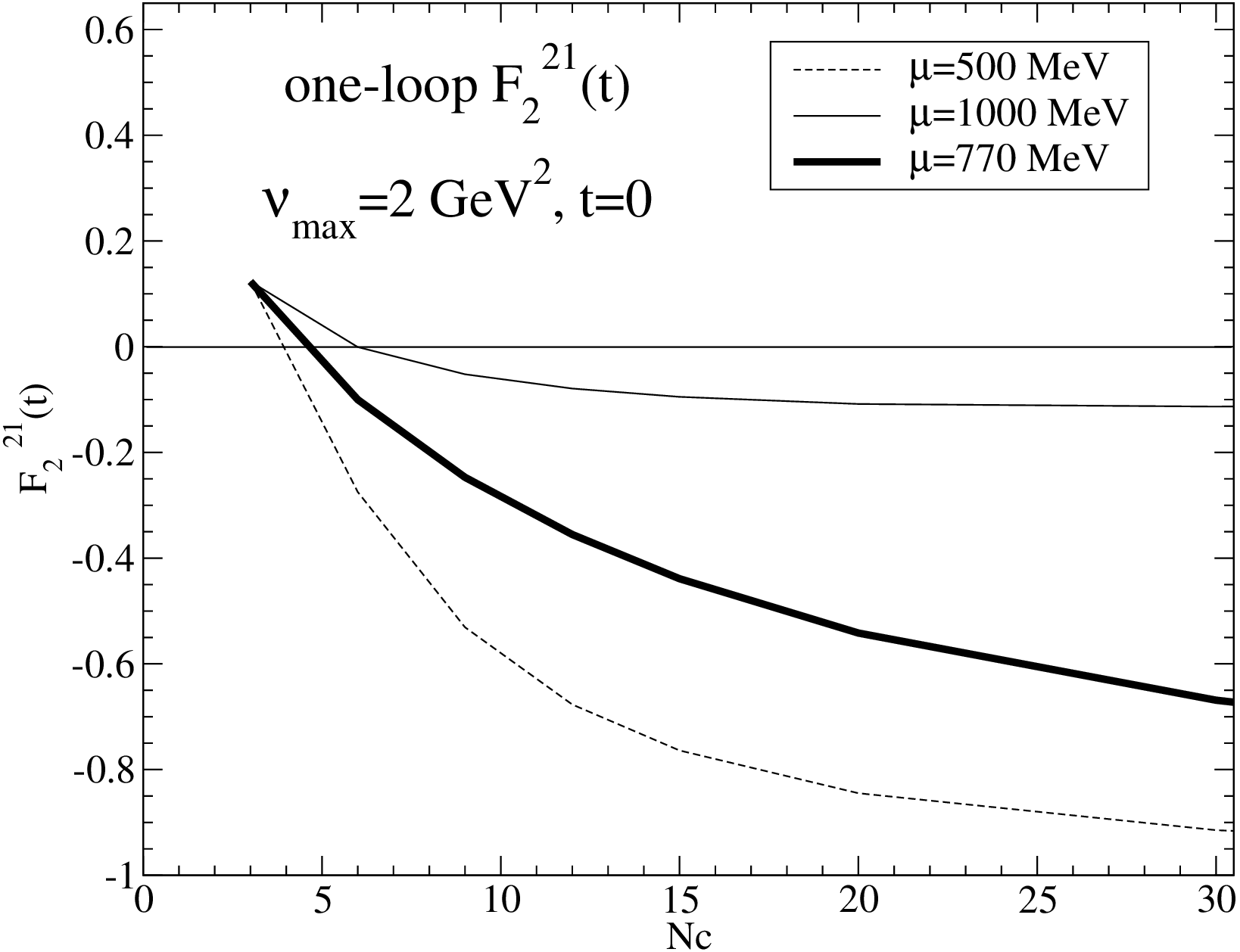}
\hspace{4mm}\includegraphics[scale=0.26]{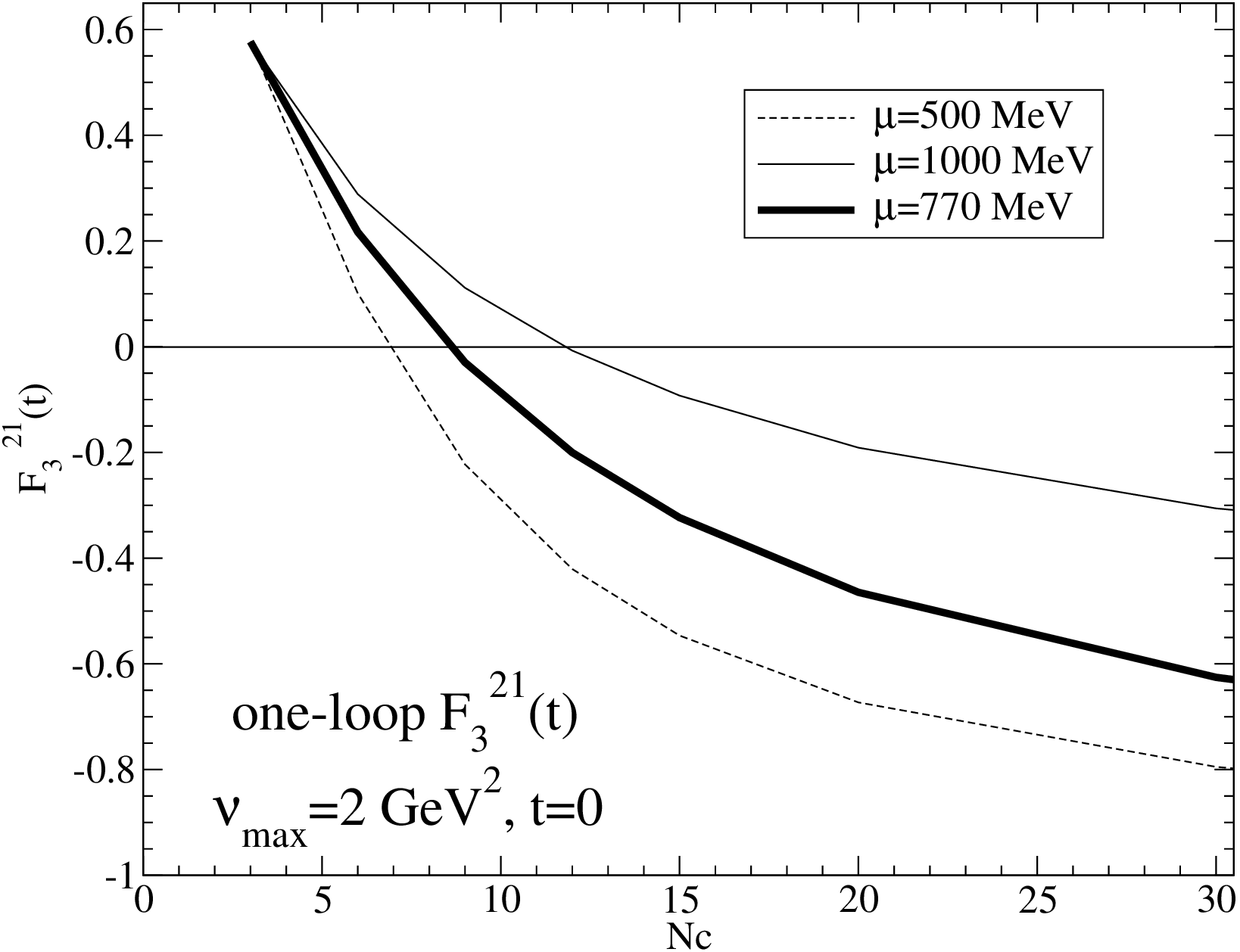}
    \caption{Evolution of the $F^{21}_n$ ratio
calculated with the one-loop IAM when the leading $1/N_c$ behaviour of the LECs is
imposed at different choices of the renormalization scale $\mu$. 
\label{fig:one-loop-uncertainty-in-F}
}
  \end{center}
\end{figure}

\section{$\boldsymbol{N_\mathrm{c}}$ dependence of $\boldsymbol{\pi\pi}$ scattering to two loop 
  UChPT:  Subdominant  \qq component of the $\sigma$}
Now let us move to two loop order in $\chi$PT~\cite{Bijnens:1997vq}  and see if this situation changes. 
The IAM to two loops for pion-pion scattering was first formulated in~\cite{Dobado3}, and first analysed in \cite{Nieves:2001de}.
With a larger number of LECs appearing, we clearly have more freedom. In studying the $1/\Nc$ behaviour,
Pel\'aez and Rios \cite{Pelaez:2006nj} consider three alternatives within single channel $SU(2)$ chiral theory  for fixing these, which we follow here too.  These three cases involve combining agreement with experiment with different underlying structures for the~$\rho$~and~$\sigma$. Agreement with experiment for the $I=0$ and 2 $S$-waves and the $I=1$ $P$-wave is imposed by minimising a suitable  $\chi_{\mathrm{data}}^2$. Our whole approach is one of considering the $1/\Nc$ corrections to the physical $\Nc=\,3$ results. Consequently, to impose an underlying structure for the resonances, we note that
if a resonance is predominantly a $\overline{q}q$ meson, then as a function of $\Nc$, its mass $M\!\sim O(1)$ and width $\varGamma\!\sim O(1/N_{\mathrm{c}})$. Taking into account the subleading orders in $1/\Nc$, it is sufficient to consider a resonance a
$\overline{q}q$ state, if

\begin{eqnarray}
\nonumber\\
  M_{N_\mathrm{c}}^{\overline{q}q}=M_0\,\left(1+\frac{\epsilon_M}{N_\mathrm{c}}\right)\; , \quad \varGamma_{N_{\mathrm{c}}}^{\overline{q}q}=\frac{\varGamma_0}{N_{\mathrm{c}}}\left(1+\frac{\epsilon_\varGamma} {N_{\mathrm{c}}}\right)\; ,
\end{eqnarray} 

\noindent where $M_0$ and $\varGamma_0$ are unknown but $N_{\mathrm{c}}$-independent, with $\epsilon_M$ and $\epsilon_\varGamma$ are naturally taken to be one. Thus for a $\overline{q}q$ state the expected $M_{N_{\mathrm{c}}}$ and
$\varGamma_{N_{\mathrm{c}}}$ can be obtained from those generated by the IAM, 
  \begin{center}
    \begin{table*}
      \begin{tabular}{l c c c c c}\hline\hline
        &&&\\
        IAM Fit & $\chi_{\mathrm{data}}^{\,2}$ & $\chi_{\rho,\overline{q}q}^{\,2}$ & $\chi_{\sigma,\overline{q}q}^{\,2}$ & $\chi_{\sigma,\overline{q}q, N_c=9}^{\,2}$ &
        $\chi_{\sigma,\overline{q}q, N_c=12}^{\,2}$\\
        \hline
        &&&\\
        \textbf{Case A:} $\rho$ as $\overline{q}q$ & ~~1.1~~ & ~~0.9~~ & ~~15.0~~ & ~4.8~ & ~3.4~\\[1.mm]
        \textbf{Case B:} $\rho$ and $\sigma$ as $\overline{q}q$ & ~~1.5~~ & ~~1.3~~ & ~~4.0~~& ~~0.8~~ & ~~0.5~~\\[1.mm]
        \textbf{Case C:} $\sigma$ as $\overline{q}q$ & ~~1.4~~ & ~~2.0~~ & ~~3.5~~& ~~0.6~~ & ~~0.5~~\\[2.mm]
        \hline\hline
      \end{tabular} 
      \caption{Values of the $\chi^2$ for the different $SU(2)$ fits} 
      \centering\label{fits}
    \end{table*}
\vspace{3mm}
  \end{center}

\begin{table}
\begin{center}
  \begin{tabular}{c||c c c}\hline\hline
    LECs & Case A & Case B & Case C\\\hline
    $l_1^r$(x $10^3$) & -5.4 & -5.7 & -5.7  \\
    $l_2^r$(x $10^3$) & ~1.8 & ~2.5& ~2.6 \\
    $l_3^r$(x $10^3$) & ~1.5 & 0.39& -1.7  \\
    $l_4^r$(x $10^3$) & ~9.0 & ~3.5 & ~1.7\\\hline
    $r_1$(x $10^4$)  & -0.6 & -0.58 & -0.6\\
    $r_2$(x $10^4$)  & ~1.5 & ~1.5 & ~1.3\\
    $r_3$(x $10^4$)  & -1.4 &-3.2 & -4.4\\
    $r_4$(x $10^4$)  & ~1.4 & -0.49& -0.03\\
    $r_5$(x $10^4$)  & ~2.4 & ~2.7& ~2.7\\
    $r_6$(x $10^4$)  & -0.6 & -0.62 & -0.7 
  \end{tabular}\caption{Two-loop IAM LECs for the different cases we have used~\cite{Pelaez:2006nj}.}\label{Tab:SU2lecs}
\end{center}
\vspace{-2mm}
\end{table}
\begin{eqnarray}
\nonumber\\
\nonumber
  M_{N_\mathrm{c}}^{\overline{q}q}&\simeq&
  M_{N_{\mathrm{c}}-1}\,\left[1+\epsilon_{M}\left(\frac{1}{N_\mathrm{c}}-\frac{1}{N_\mathrm{c}-1}\right)\right]\\[2mm]
  & =&M_{N_\mathrm{c}-1}\;+\;\Delta
  M_{N_\mathrm{c}}^{\overline{q}q},\\
\nonumber\\
\nonumber 
 \varGamma_{N_\mathrm{c}}^{\overline{q}q}&\simeq&
\frac{N_\mathrm{c}-1}{N_{\mathrm{c}}}\,\varGamma_{N_\mathrm{c}-1}\ \left[1+\epsilon_{\varGamma}\left(\frac{1}{N_{\mathrm{c}}}-\frac{1}{N_{\mathrm{c}}-1}\right)\right]\\[2mm]
 & =&\frac{N_{\mathrm{c}}-1}{N_{\mathrm{c}}}\,\varGamma_{N_\mathrm{c}-1}\;+\;\Delta
  \varGamma_{N_{\mathrm{c}}}^{\overline{q}q}.
\\\nonumber
\end{eqnarray}

We therefore define an averaged $\chi_{\overline{q}q}^{2}$ to measure how
close a resonance is to a $\overline{q}q$ behaviour, using as uncertainty the
$\Delta M_{N_{\mathrm{c}}}^{\overline{q}q}$ and $\Delta \varGamma_{N_{\mathrm{c}}}^{\overline{q}q}$
\begin{eqnarray}
\nonumber\\
\nonumber
  \chi_{\overline{q}q}^{2} = \frac{1}{2
    n}\sum_{N_{\mathrm{c}}=4}^{n}\ \left[\left(\frac{M_{N_{\mathrm{c}}}^{\overline{q}q}-M_{N_{\mathrm{c}}}}{\Delta
    M_{N_{\mathrm{c}}}^{\overline{q}q}}\right)^2\ +\ \left(\frac{\varGamma_{N_{\mathrm{c}}}^{\overline{q}q}-\varGamma_{N_{\mathrm{c}}}}{\Delta
    \varGamma_{N_{\mathrm{c}}}^{\overline{q}q}}\right)^2\right].
\\
\end{eqnarray}
This $\chi^2$ is added to $\chi_{\mathrm{data}}^2$ and the sum is minimised. Case A is where the data are fitted assuming that the $\rho$ is a $\overline{q}q$ meson, while Case B assumes
that both the $\sigma$ and the $\rho$ are $\overline{q}q$ states.
Lastly, Case C is where we minimize $\chi_{\mathrm{data}}^2$ and just
$\chi_{\overline{q}q}^2$ for the $\sigma$. 

We show in Table~\ref{fits} the values of the $\chi^2$ contributions for each case, where we sum over $\Nc$ from 3 to 12.
The two-loop LECs~\cite{Pelaez:2006nj} for each case are shown
in Table~V.
We see from Table~IV that constraining the $\rho$ to be a $\overline{q}q$ state by imposing Eq.~(17) is completely compatible with data at $\Nc=\,3$. In contrast, imposing a $\overline{q}q$ configuration for the $\sigma$ gives much poorer agreement with data and can distort the simple structure for the $\rho$. It is interesting to point out that, the lower energy at which such a sigma's $\overline{q}q$ behaviour
emerges, the
higher energy at which the $\rho$ pole moves with $N_\mathrm{c}$. Therefore, as much as we try to force the
$\sigma$ to behave as a $\overline{q}q$ meson, less the $\rho$ meson does.  However, requiring a $\overline{q}q$ composition for the $\sigma$ for larger $\Nc$ causes no such distortion.

\begin{figure*}
  \begin{center}
    \includegraphics[scale=0.28]{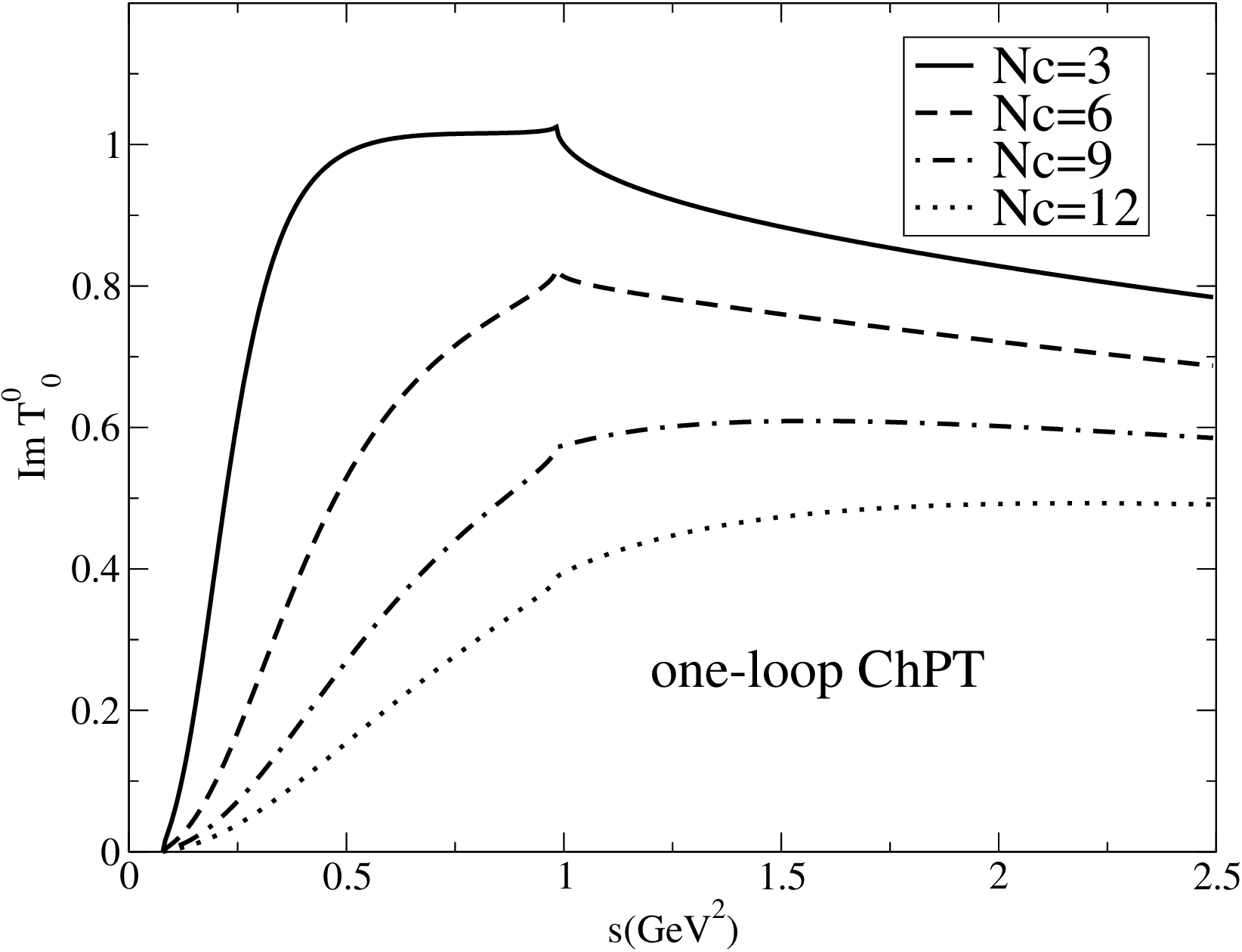}\hspace{4mm}\includegraphics[scale=0.28]{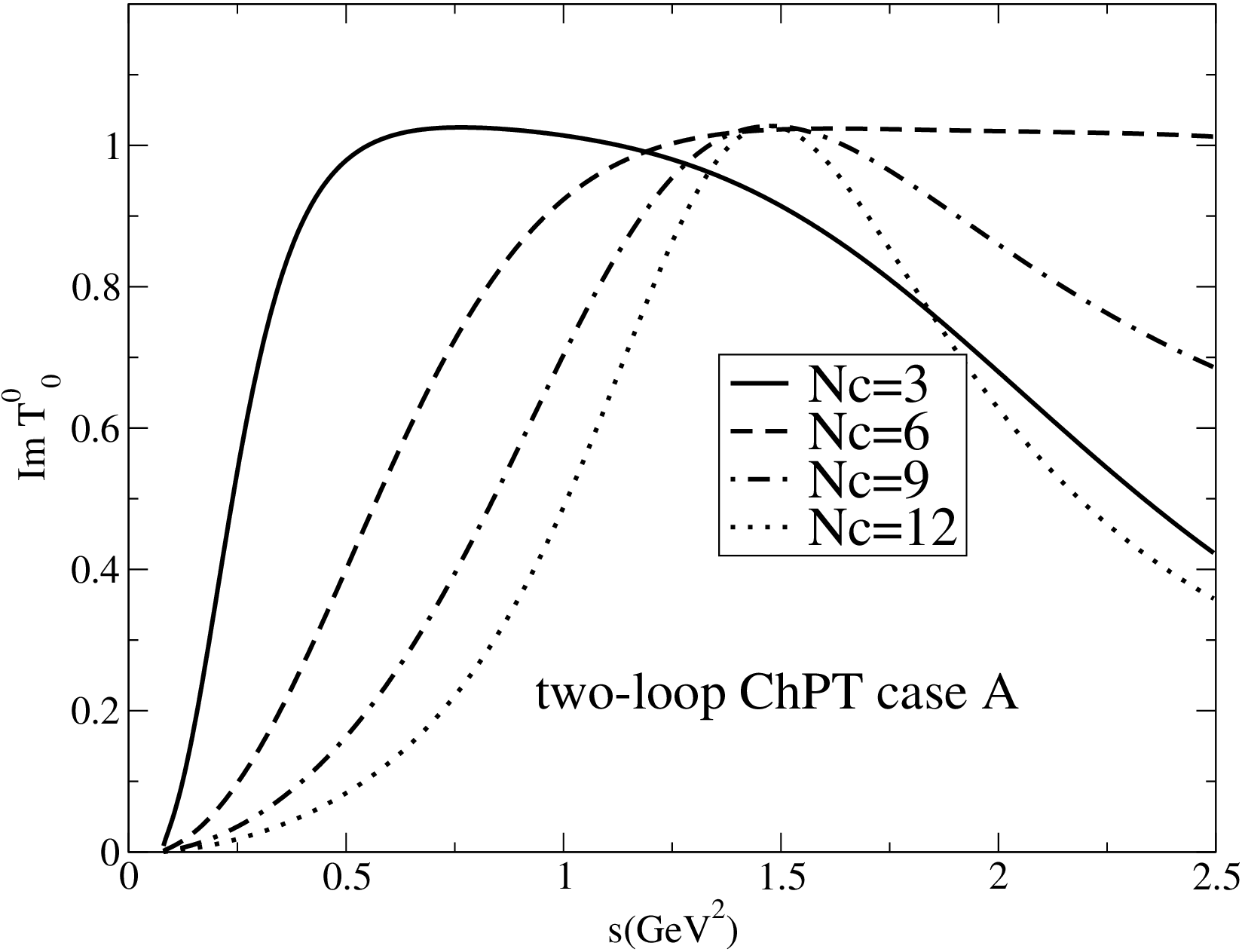}
    \caption{Absorptive parts of the $I=J=0$ partial wave amplitude, ${\mathrm{Im}T^0_0}(s)$, at one loop with the parameters of an $SU(3)$ fit (cf the corresponding coupled channel fit in Fig.~\ref{fig:TIJsu3}) and at two loops an $SU(2)$ fit with $\Nc=\,3$ to data below $0.9$ GeV.These both involve only the $\pi\pi$ channel and so the strong inelastic effects from ${\overline K}K$ threshold are not included, in contrast to Fig.~\ref{fig:TIJsu3}.}\label{fig:TIJsu2}
  \end{center}  
\vspace{3mm}
\end{figure*}

In all parameters sets at two loops, including Case A, which fits the data best and in which the $\rho$ has a clear ${\overline q}q$ structure,
we do see a subleading
$\overline{q}q$ behaviour for the $\sigma$ meson emerge between 1 and 1.5 GeV$^2$. 
This is evident from Fig.~\ref{fig:TIJsu2} where the imaginary part of the $I=J=0$ amplitude is plotted. We see a clear enhancement above 1 GeV emerge
as $\Nc$ increases.
That this enhancement is related to the $\sigma$ at larger $\Nc$ can be seen
by tracking the movement of the $\rho$ and $\sigma$ poles at two loops, and
comparing this with the one loop trajectories in Fig.~\ref{fig:polessu3}.

\begin{figure*}
  \begin{center}
    \includegraphics[scale=0.28]{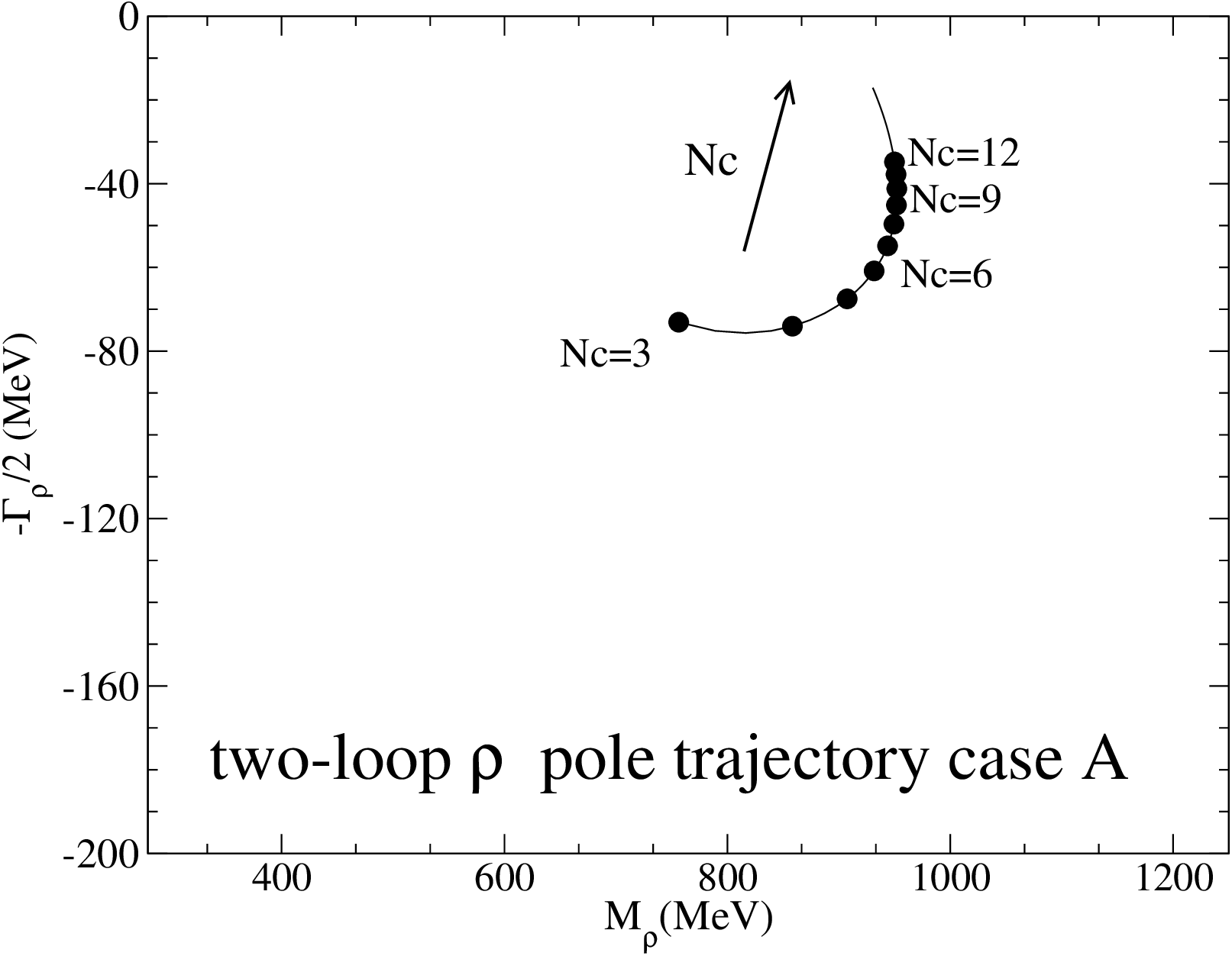} \hspace{4mm}\includegraphics[scale=0.28]{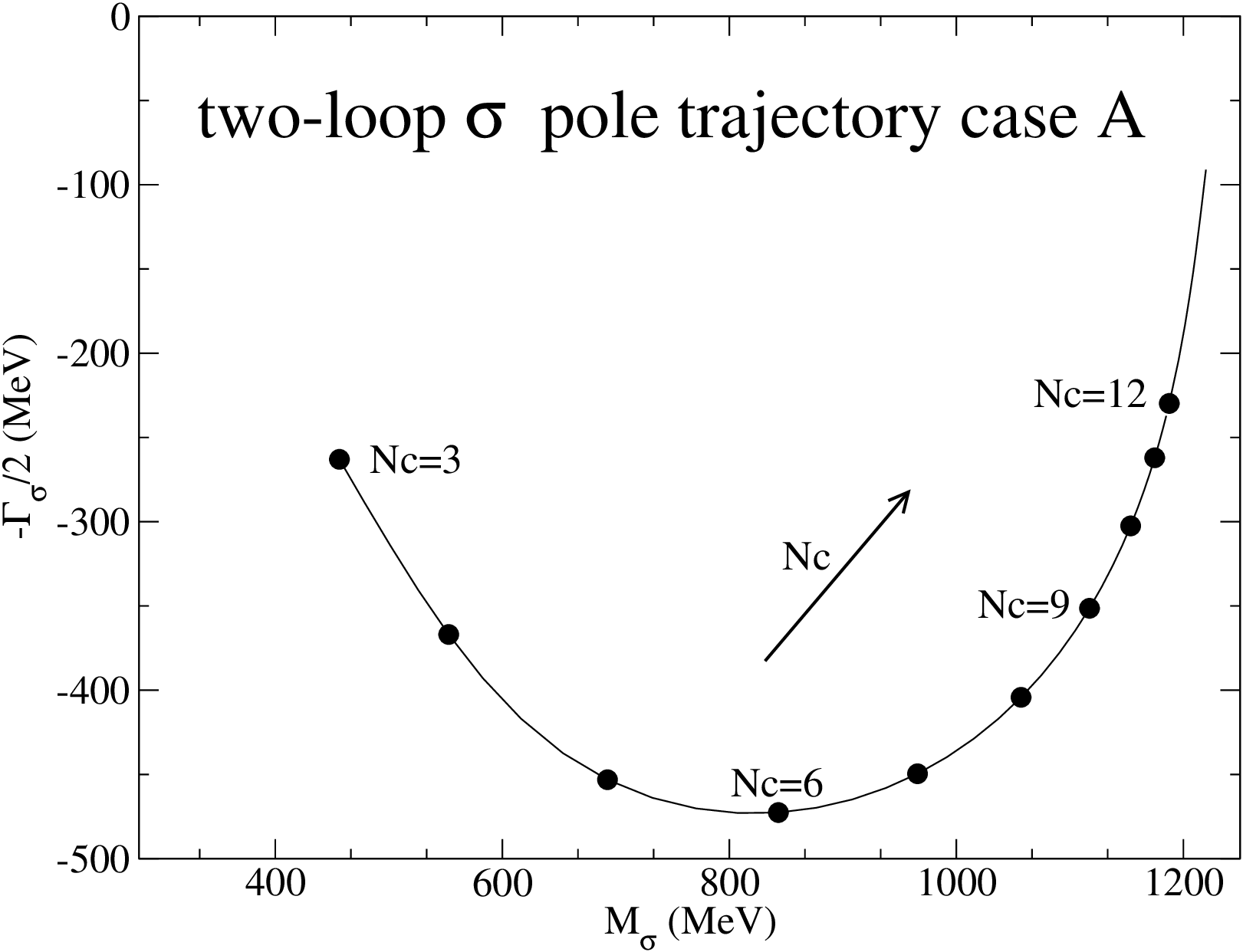}\\
    \caption{Position of the $\rho$ and $\sigma$ poles in the complex energy
      plane as a function of $\Nc$ in two loop $\chi$PT with parameters from
      the $SU(2)$ fit A of Table~IV. This is to be compared with the one loop
      trajectories of Fig.~\ref{fig:polessu3}. 
      Note the different vertical scales for the
      $\rho$ and $\sigma$ plots.}\label{fig:polessu2}
  \end{center}
\end{figure*}

\begin{figure*}
  \begin{center}
    \includegraphics[scale=0.29]{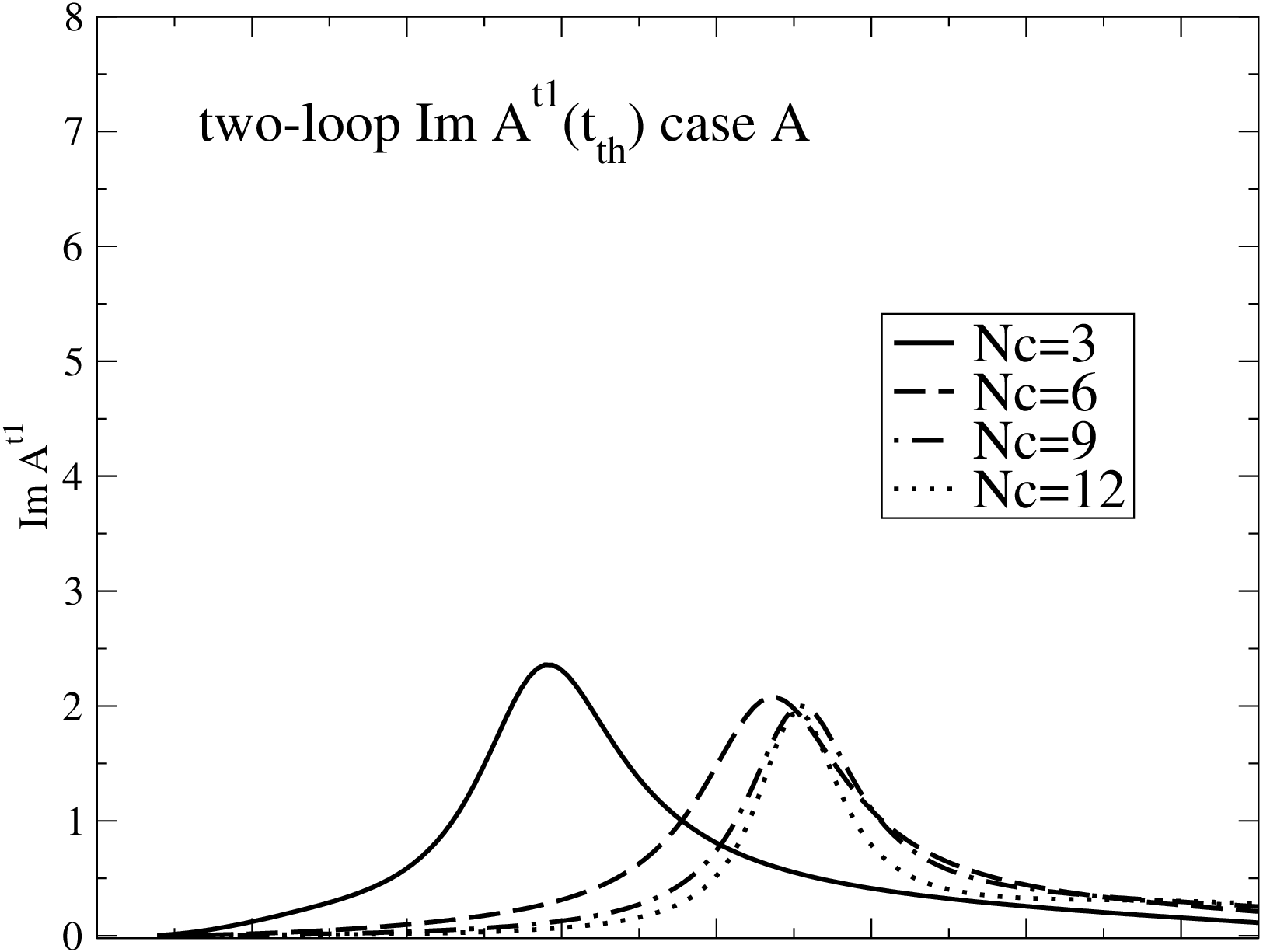}\hspace{4mm}\includegraphics[scale=0.29]{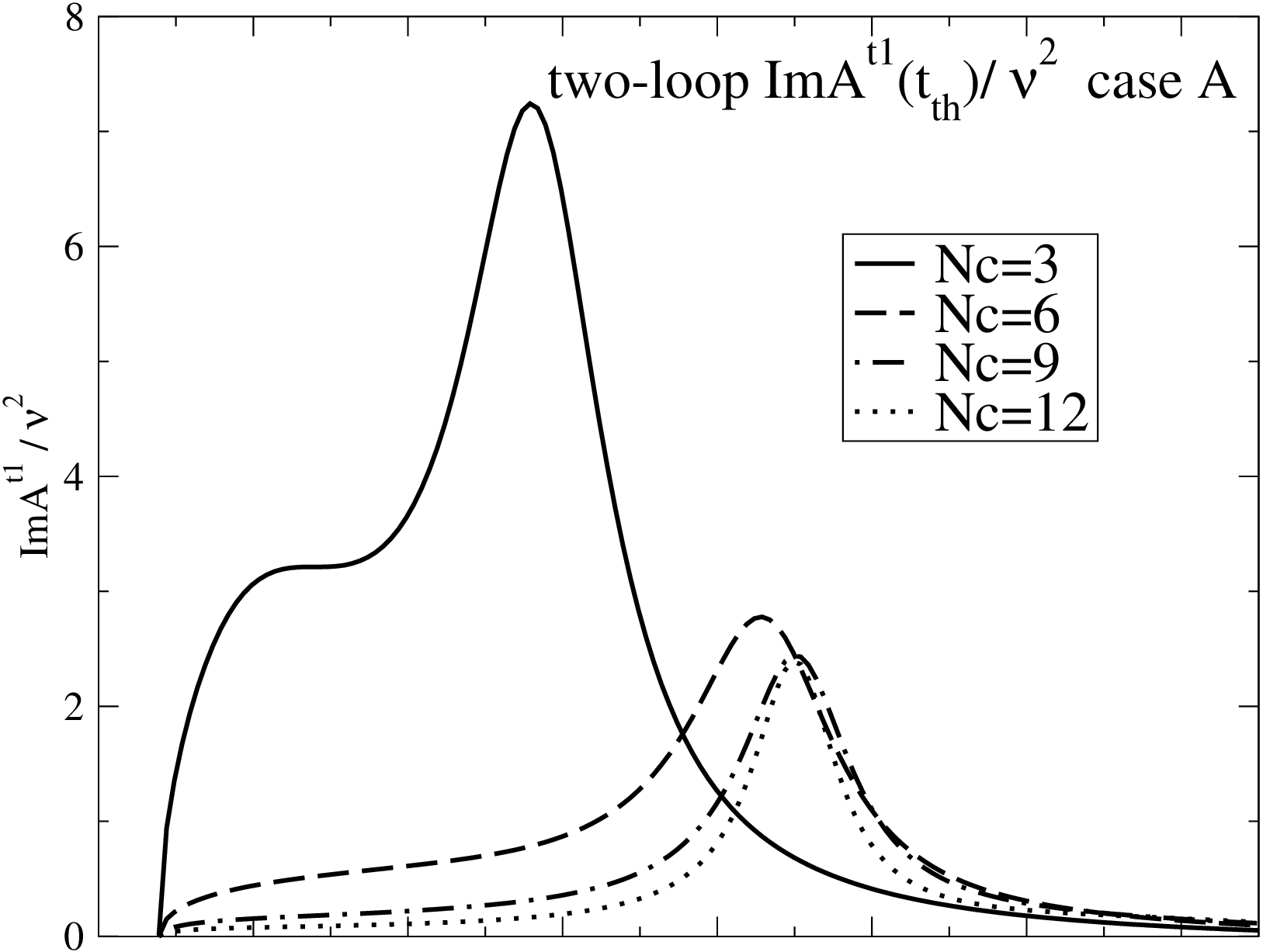}\\
    \includegraphics[scale=0.29]{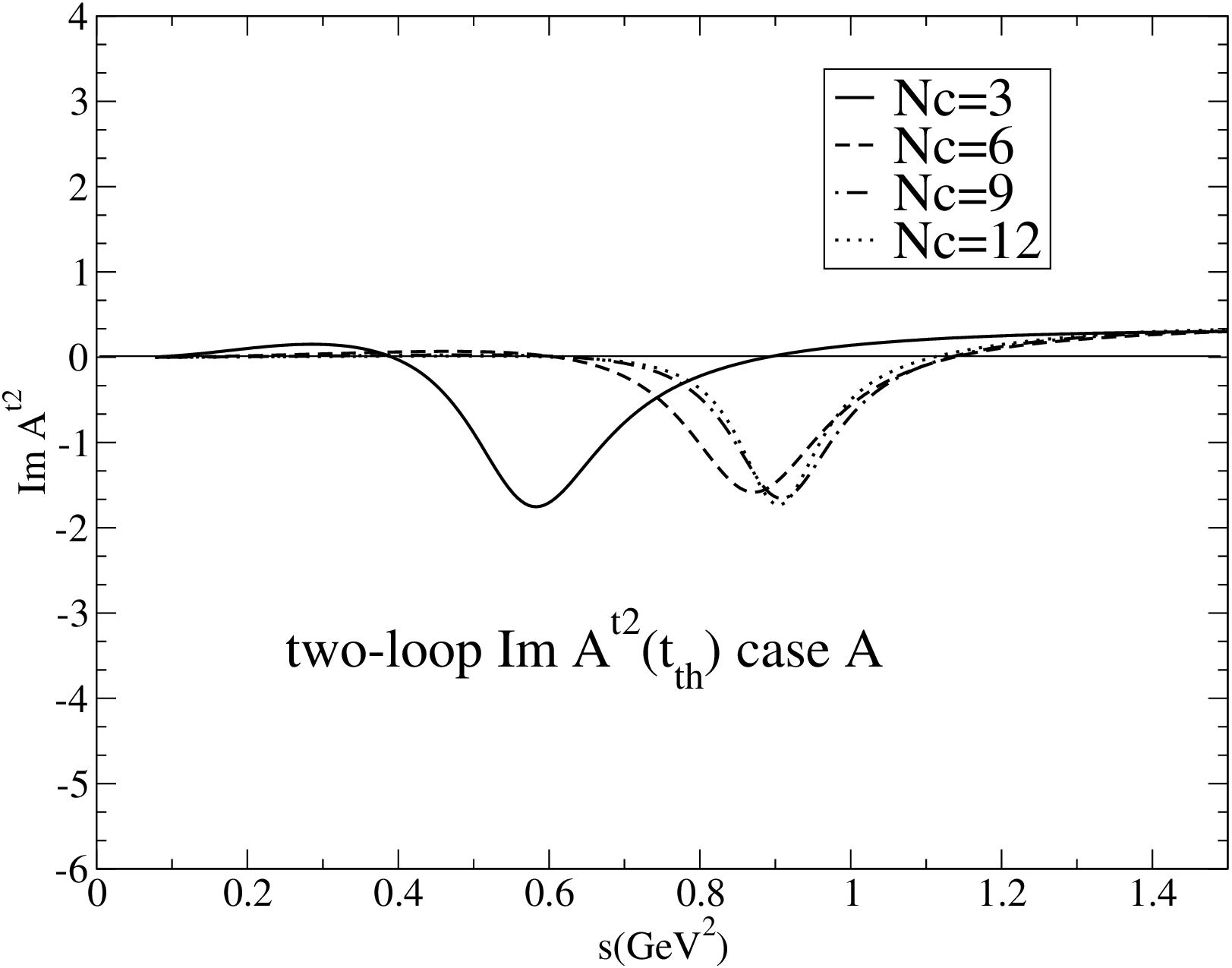}\hspace{4mm}\includegraphics[scale=0.29]{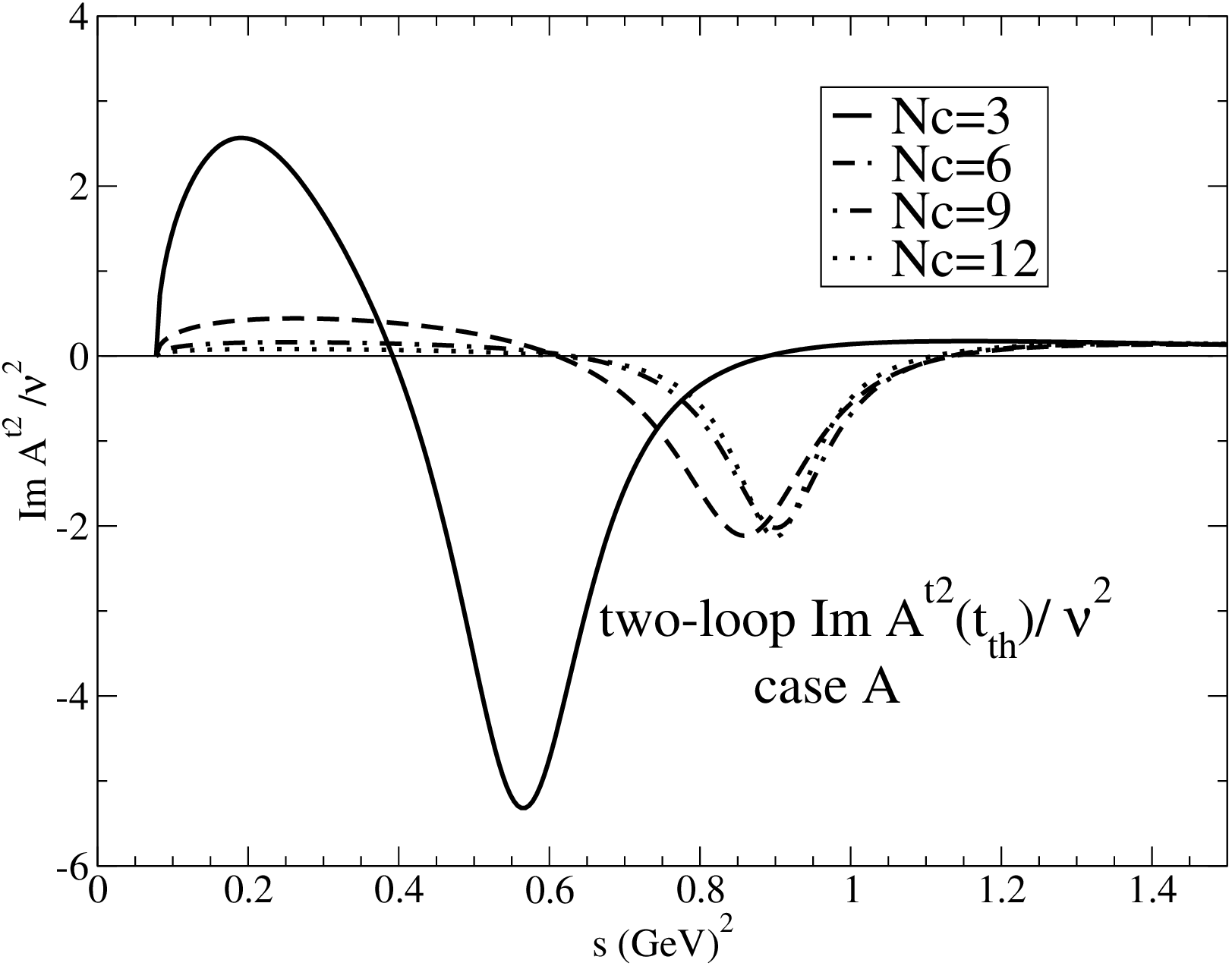}
      \caption{Absorptive parts of amplitudes with definite $t$-channel isospin, ${\mathrm{Im}A^{tI}(s,t_{\mathrm{th}})}/\nu^n$, using the parameters of the two loop $SU(2)$ fit Case A. The top pair of graphs have $I=1$ and the lower with $I=2$, on the left hand $n=0$ and right hand $n=2$. One sees from the lower pair how integrating the curves the positive and negative contributions cancel for all $\Nc$}.\label{fig:Atsu2}
    \end{center}
  \end{figure*}
 \begin{center}
  \begin{table*}
    \begin{tabular}{|c|c|c||c|c||c|c|}\cline{4-7}
      \multicolumn{3}{c}{} & \multicolumn{4}{|c|}{\textbf{2 loops SU2} $\rho$
        \textbf{as} \qq}\\\cline{4-7}
      \multicolumn{3}{c}{} &\multicolumn{2}{|c||}{$\mathbf{t =
          4M}_{\pi}^2$}&\multicolumn{2}{c|}{$\mathbf{t=0}$}\\\hhline{---====}
      & $n$ & $N_\mathrm{c}$ & $\nu_\mathrm{max}$=1 GeV$^2$ & $\nu_\mathrm{max}$=2 GeV$^2$ &$\nu_\mathrm{max}$=1 GeV$^2$ &$\nu_\mathrm{max}$=2 GeV$^2$\\\hline\hline
      \multirow{12}{0.35in}{$\;\mathbf{F^{\;1\:0}_n}$}& 
      \multirow{4}{0.1in}{0} 
        & 3 & 0.493 & 0.359 & 0.488 & 0.334 \\
      & & 6 & 0.494 & 0.370 & 0.492 & 0.349 \\
      & & 9 & 0.491 & 0.395 & 0.490 & 0.376 \\
      & & 12 & 0.489 & 0.422 & 0.488 & 0.404 \\\Cline{0.8pt}{2-7}
      &\multirow{4}{0.1in}{1} 
        & 3 & 0.509 & 0.442 & 0.511 & 0.434 \\
      & & 6 & 0.496 & 0.419 & 0.494 & 0.407 \\
      & & 9 & 0.488 & 0.430 & 0.487 & 0.418 \\
      & & 12 & 0.485 & 0.447 & 0.483 & 0.436 \\\Cline{0.8pt}{2-7}
      &\multirow{4}{0.1in}{2} 
        & 3 & 0.533 & 0.505 & 0.551 & 0.522 \\
      & & 6 & 0.498 & 0.457 & 0.498 & 0.454 \\
      & & 9 & 0.482 & 0.452 & 0.479 & 0.445 \\
      & & 12 & 0.477 & 0.460 & 0.472 & 0.452 \\\Cline{0.8pt}{2-7}
      &\multirow{4}{0.1in}{3}
        & 3 & 0.572 & 0.563 & 0.618 & 0.611 \\
      & & 6 & 0.503 & 0.485 & 0.511 & 0.495 \\
      & & 9 & 0.472 & 0.460 & 0.468 & 0.456 \\
      & & 12 & 0.461 & 0.457 & 0.451 & 0.447 \\\hline\hline
      \multirow{12}{0.35in}{$\;\mathbf{F^{\;2\:1}_n}$}& 
      \multirow{4}{0.1in}{0} 
        & 3 & -0.421 & -0.060 & -0.280 & 0.135 \\
      & & 6 & -0.536 & -0.086 & -0.454 & 0.058 \\
      & & 9 & -0.648 & -0.061 & -0.579 & 0.073 \\
      & & 12 & -0.748& -0.038 & -0.686 & 0.090  \\\Cline{0.8pt}{2-7}
      &\multirow{4}{0.1in}{1} 
        & 3 & -0.351 & -0.202 & -0.183 & -0.028 \\
      & & 6 & -0.438 & -0.196 & -0.335 & -0.069 \\
      & & 9 & -0.578 & -0.215 & -0.497 & -0.102 \\
      & & 12 & -0.699& -0.227 & -0.629 & -0.121  \\\Cline{0.8pt}{2-7}
      &\multirow{4}{0.1in}{2} 
        & 3 & -0.173 & -0.123 & 0.097 & 0.139 \\
      & & 6 & -0.249 & -0.152  & -0.069 & 0.027 \\
      & & 9 & -0.435 & -0.248  & -0.294 & -0.105 \\
      & & 12 & -0.594 & -0.314 & -0.477 & -0.192 \\\Cline{0.8pt}{2-7}
      &\multirow{4}{0.1in}{3}
        & 3 & 0.146 & 0.156 & 0.570 & 0.575 \\
      & & 6 & 0.102 & 0.112 & 0.485 & 0.488 \\
      & & 9 & -0.121 & -0.073 & 0.249 & 0.275 \\
      & & 12 & -0.332 & -0.216 & 0.012 & 0.092 \\\hline     
    \end{tabular}  
    \caption{ Ratios for two-loop UChPT using the LECS of Case A.}
 \centering\label{Table: SU2a}
  \end{table*}
\end{center}

\begin{figure*}
  \begin{center}
    \includegraphics[scale=0.24]{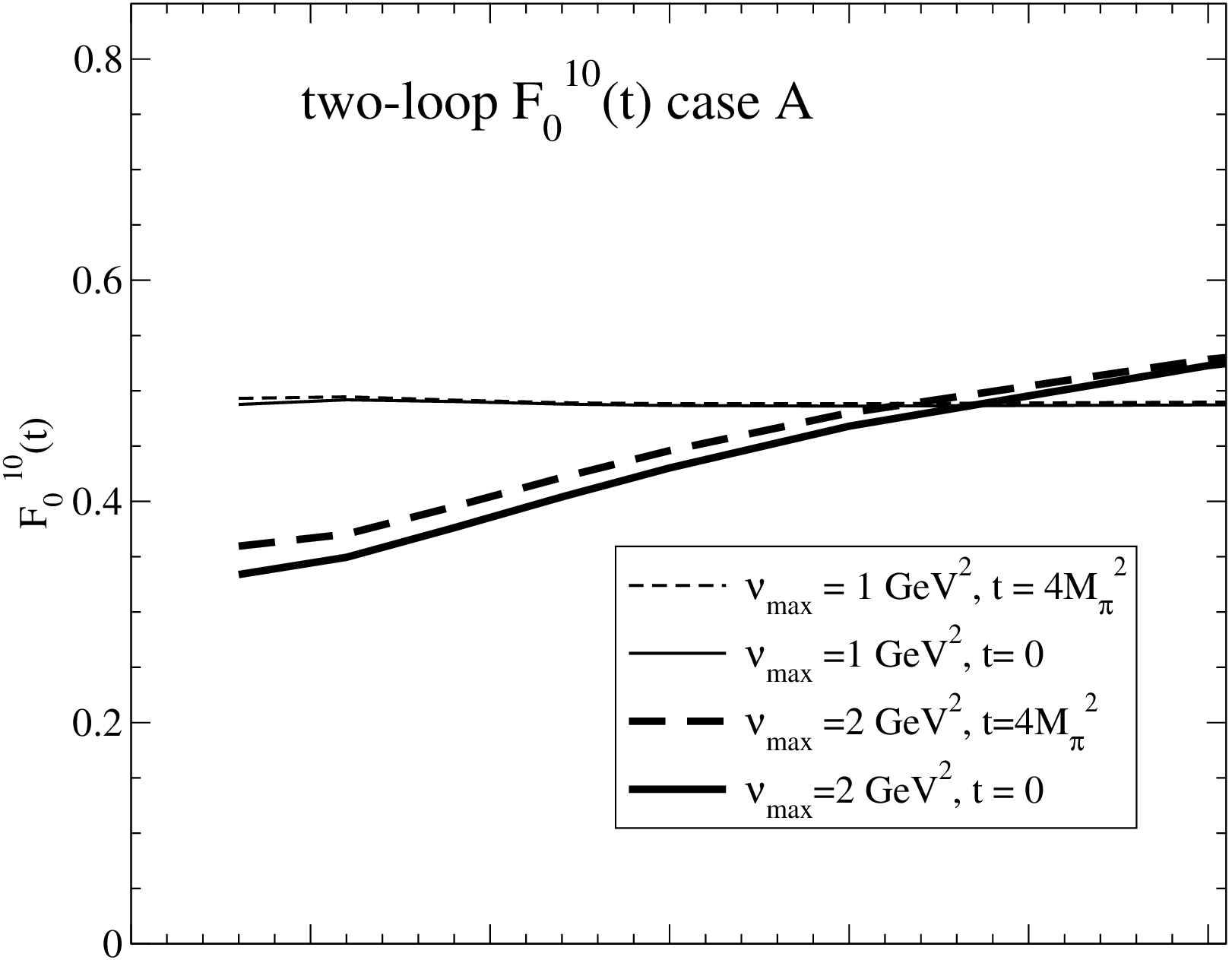}\hspace{4mm}\includegraphics[scale=0.24]{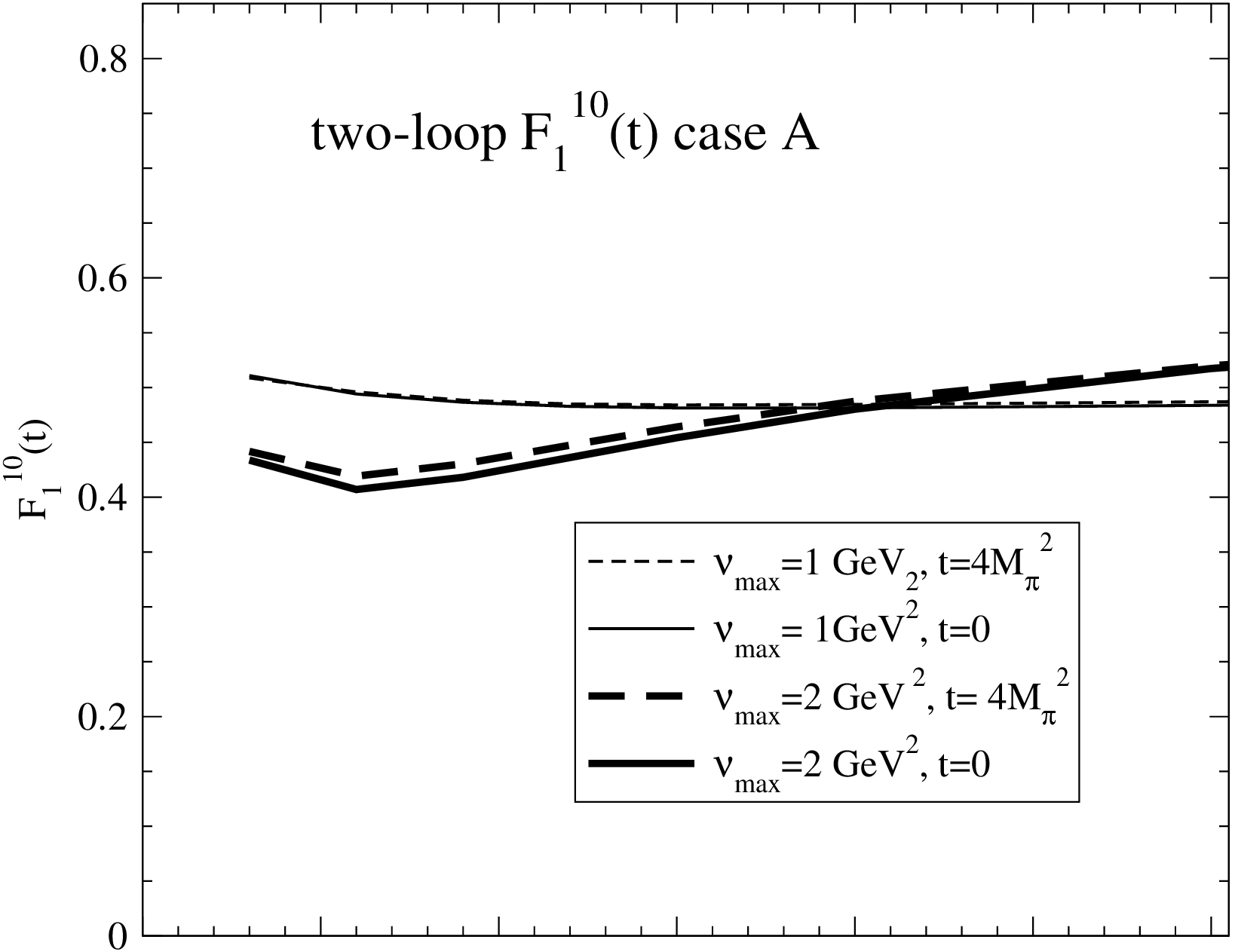}\\
    \includegraphics[scale=0.24]{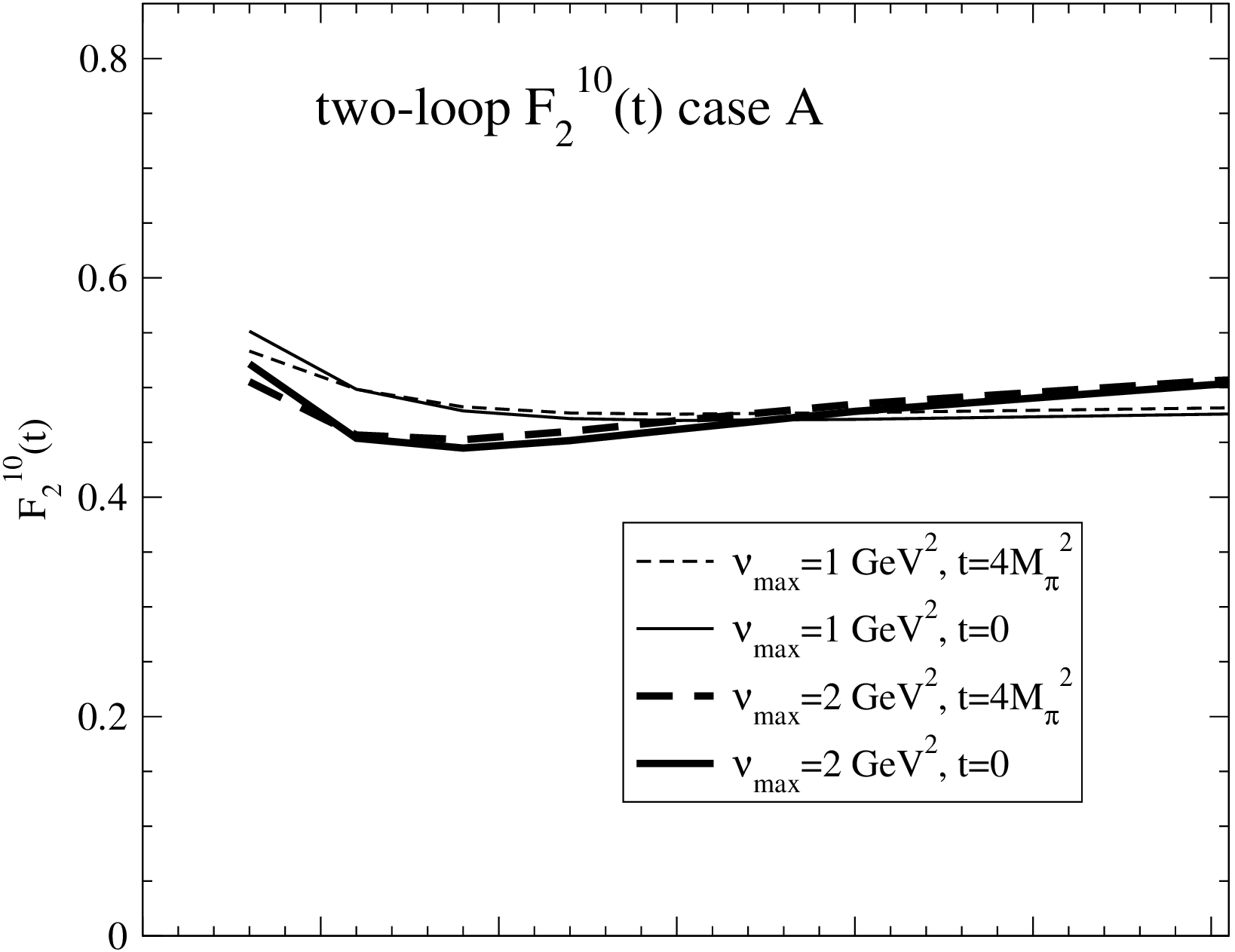}\hspace{4mm}\includegraphics[scale=0.24]{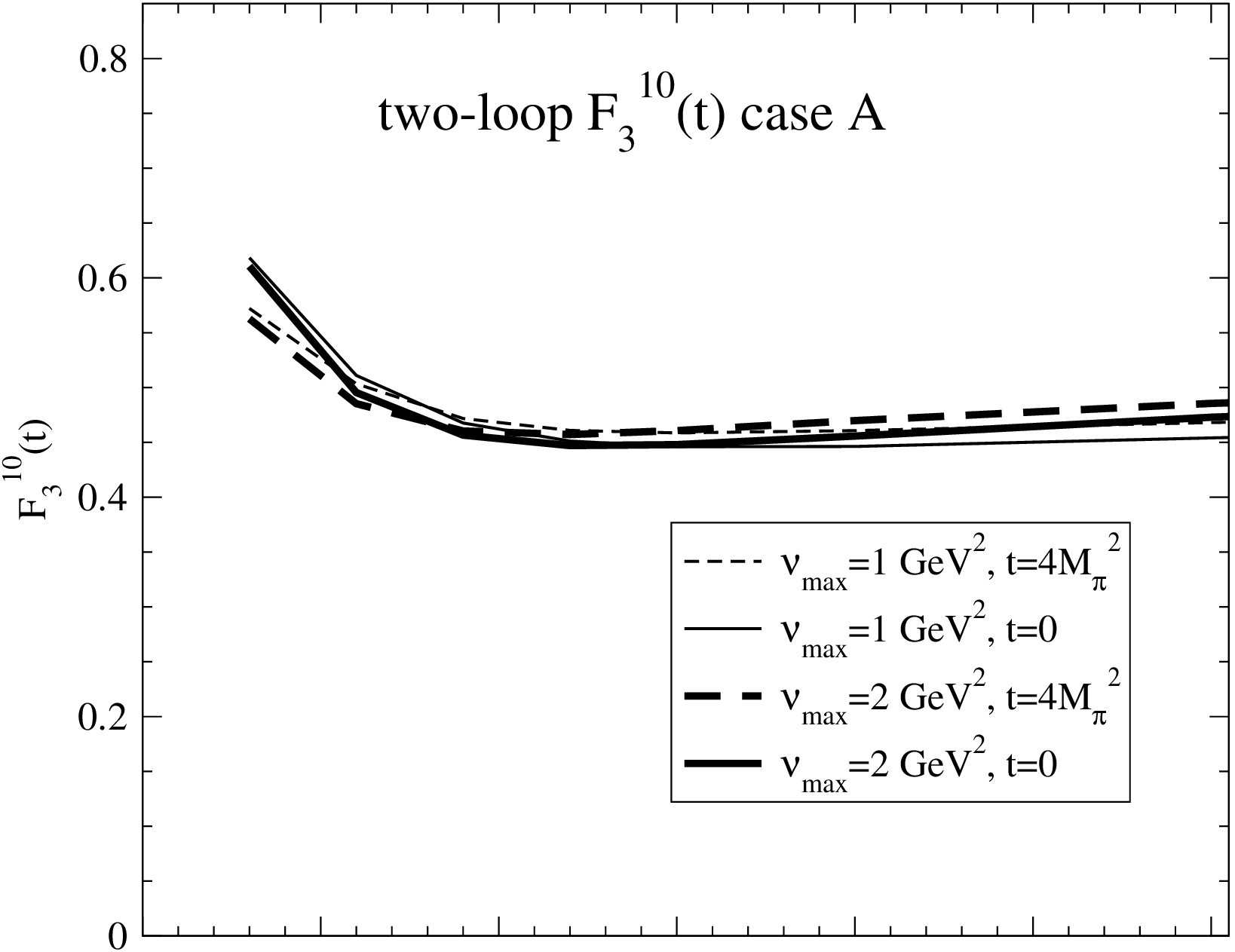}\\
    \includegraphics[scale=0.24]{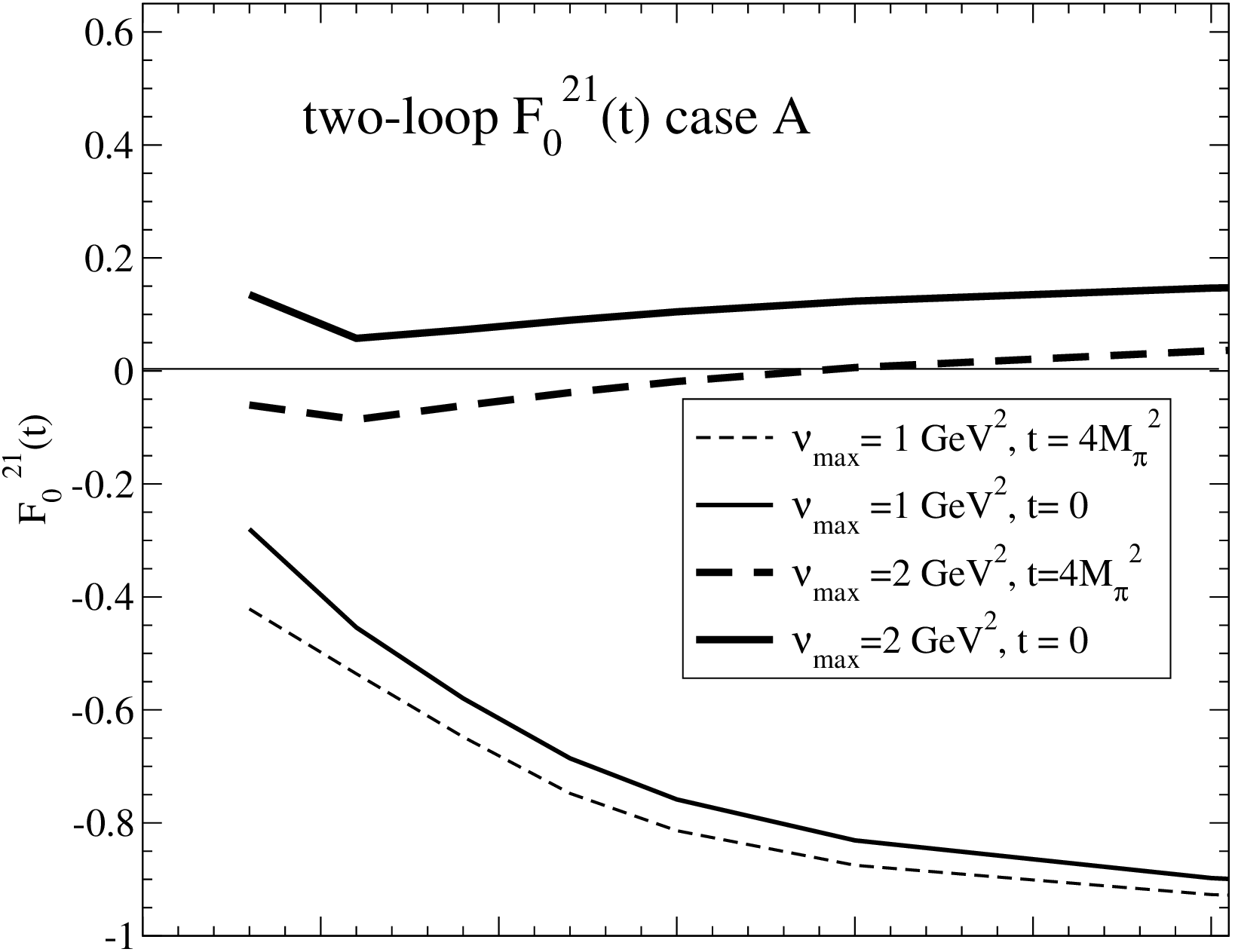}\hspace{4mm}\includegraphics[scale=0.24]{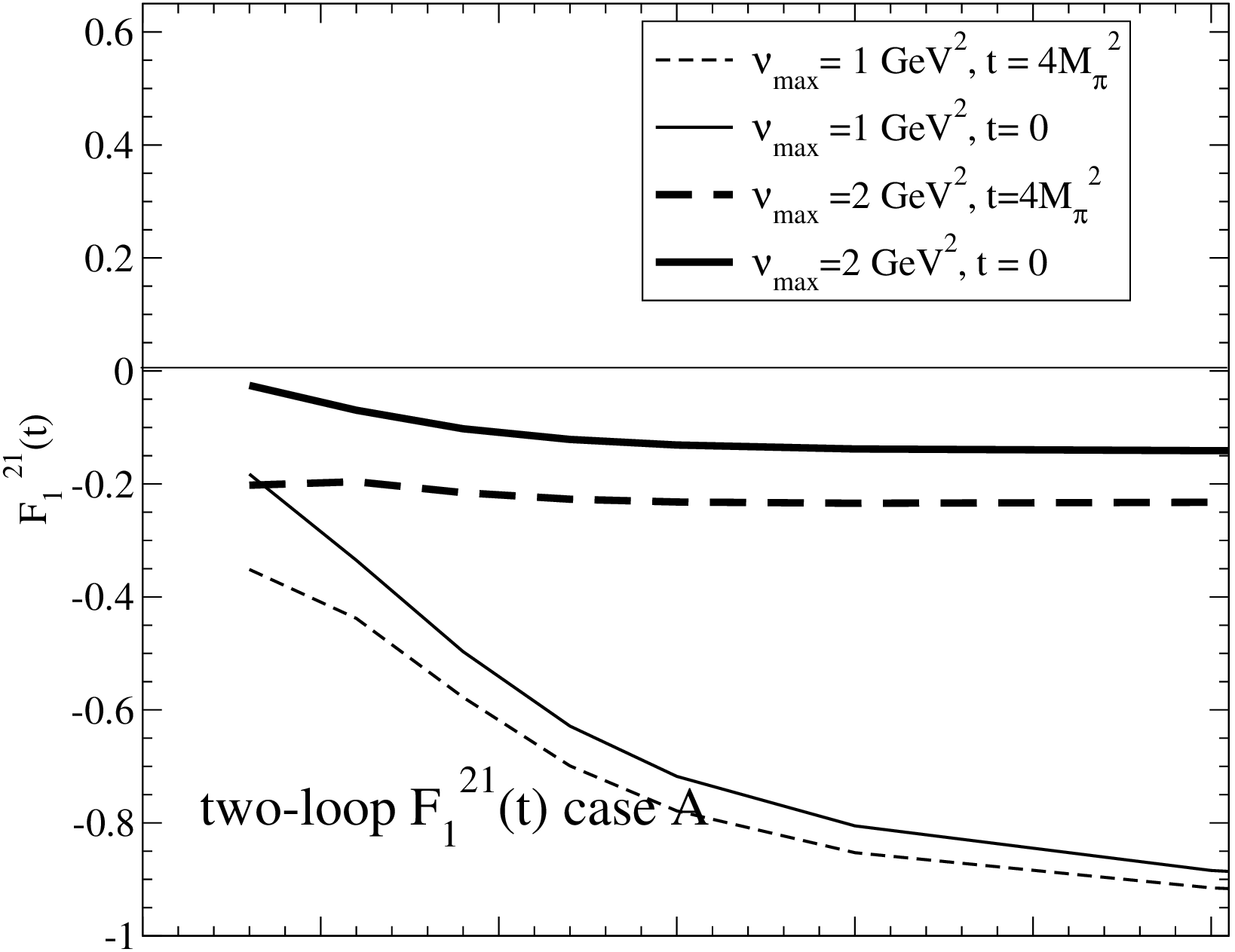}\\
    \includegraphics[scale=0.24]{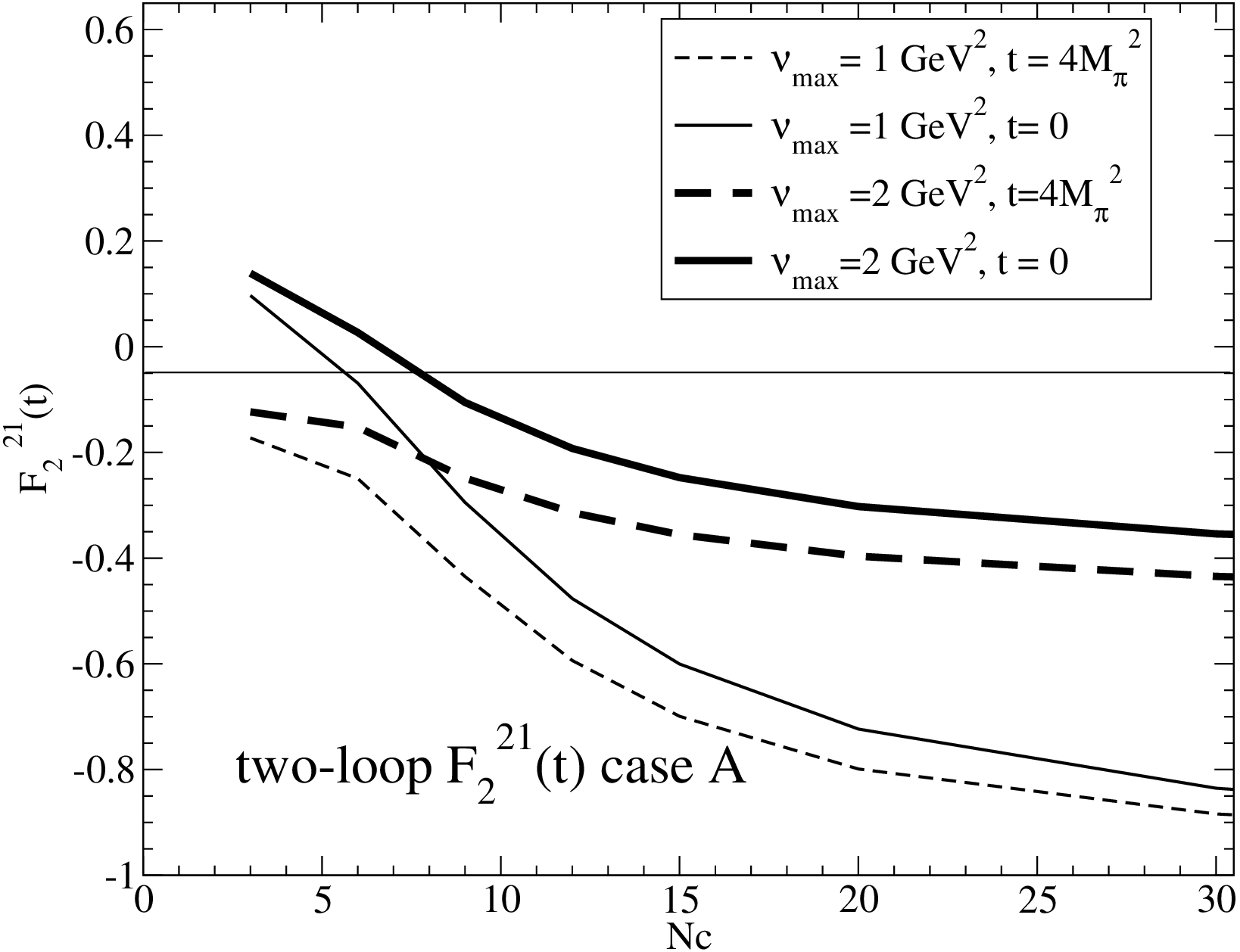}\hspace{4mm}\includegraphics[scale=0.24]{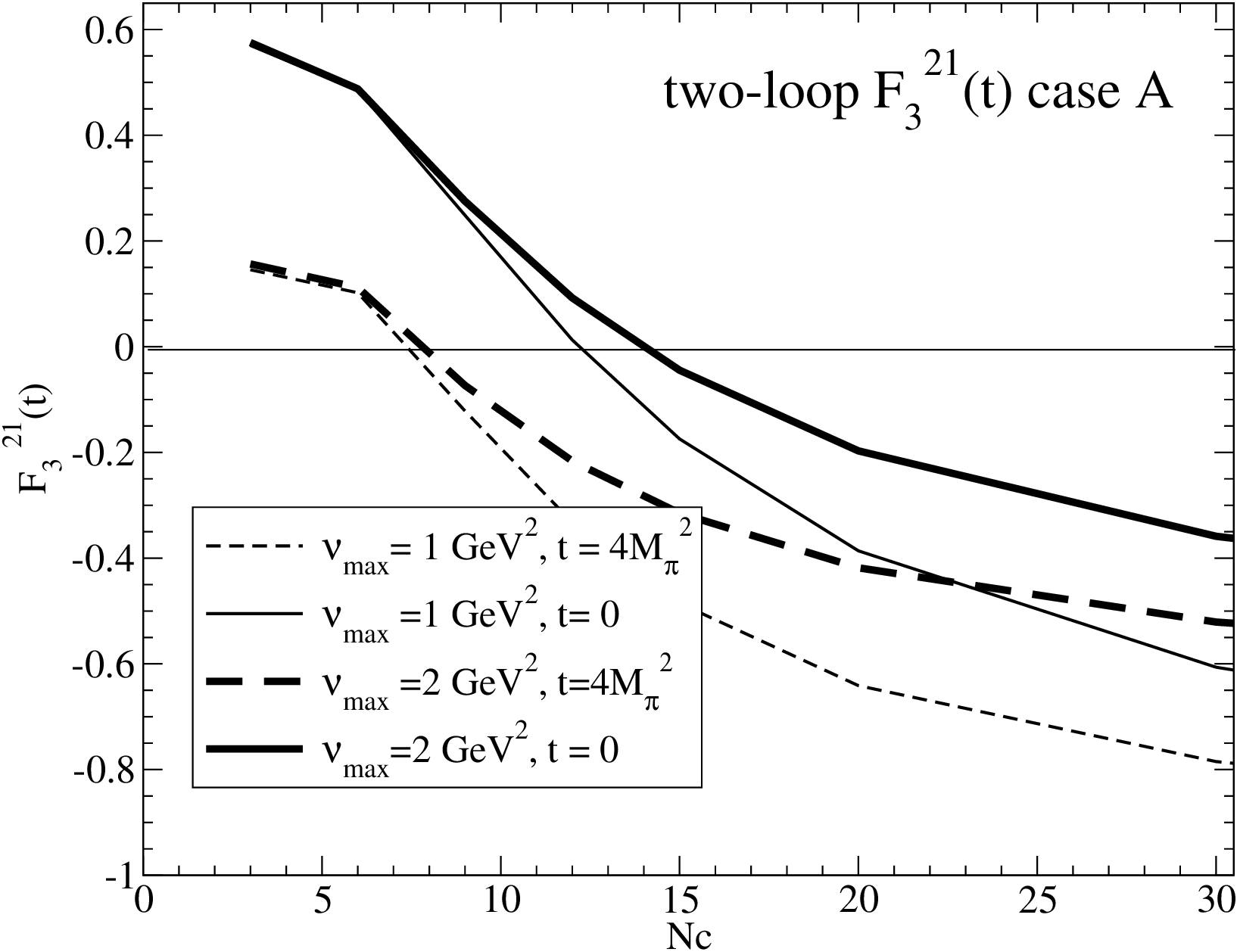}
    \caption{Ratios $F^{I\,I'}_n$ of Eq.~(6) with $n=0-3$. The top four graphs are for $F^{10}$, and the lower four for $F^{21}$. Two loop $\chi$PT IAM parameters are from the $SU(2)$ fit with $\Nc=\,3$ to data: Case A.}\label{fig:ratiossu2}
  \end{center}
\end{figure*} 

We see clearly how the $\sigma$ pole moves away as $\Nc$ increases above 3, just as in the one loop case, but then subleading terms take over as $\Nc$ increases above 6 and the $\sigma$ pole moves back to the real axis close to 1.2 GeV. This clearly indicates dominance of a ${\overline q}q$ component in its Fock space, 
which may well be related to the existence of a scalar \qq nonet above 1 GeV, as suggested in 
\cite{VanBeveren:1986ea,Oller:1998zr,Jaffe,Black:1998wt,Close:2002zu}.
 This is directly correlated with the enhancement seen in Fig.~\ref{fig:TIJsu2} (the pole movement shown in Fig.~\ref{fig:polessu2}) and of course this enhancement makes its presence felt in the amplitudes with definite $t$-channel isospin. Indeed, with $I_t=2$ we see the growth of a positive contribution to the imaginary part that might cancel the negative $\rho$ component as $\Nc$ increases: see Fig.~\ref{fig:Atsu2} and compare with the one loop forms in Fig.~\ref{fig:Atsu3}.

In addition, and though these ratios have only been evaluated at one loop order, as shown in Fig.~\ref{fig:TIJsu3}, to go further one would need to extend this analysis to two or more loops. 
Notwithstanding this caveat, we now compute the finite energy sum rule ratios, $F(t)_n^{\;I\:I'}$ of Eq.~(6) with these same two loop parameters. These ratios are set out in Table~VI.

We should be just a little cautious in recognising the limitations of the single channel approach we use here at two loop in $\chi$PT.  Despite the unitarisation, we are restricted to a region below 1~GeV, where strong coupling inelastic channels are not important. We see in Fig.~\ref{fig:polessu2} (and Fig.~\ref{fig:TIJsu2}) that the sub-dominant \qq components move above 1 GeV as $\Nc$ increases beyond 10 or 12. Consequently, if we take $\Nc$ much beyond 15 without including coupled channels, we do not expect to have a detailed
description of the resonances up to 2 GeV$^2$. 
However, in the scenario where the sigma has a subdominant \qq, it should be interpreted
as a Fock space state that is mixed in all the $f_0$ resonant structures in that region \cite{LlanesEstrada:2011kz}, 
which survives as $N_c$ increases. Then it is easy to see that its contribution
would be dominant in our ratios, and still provide a large cancellation with the $\rho$ contribution.
The reason is that, when this subdominant \qq component approaches the real axis above 1 GeV,
it has a much larger width than any other $f_0$ resonant state in that region.
For instance,  we see in Fig.~\ref{fig:polessu2} that for  $N_c=12$,  the width of the \qq subcomponent in the sigma is 
roughly 450 MeV, whereas the width
of any other \qq component that may exist in that region would have 
already decreased by $3/12=1/4$. Since the other components would be heavier and much narrower their contributions
would be much smaller than that of the \qq state subdominant in the $\sigma$. 
Note that it is also likely that some of the $f_0$'s may have large glueball components (see, for instance, \cite{Close:2002zu}),
which also survive as $N_c$ increase,  but then their widths would decrease even faster---like $1/N_c^2$, and our argument would apply even better. 
For the scenario when we do not see the sigma subdominant component (as in Fig.~\ref{fig:ratiossu3}), we still expect that
the other resonances by themselves will not be able to cancel the $\rho$ contribution, so that the IAM
would still provide a qualitatively good picture of this \lq\lq non-cancellation".
For this reason, although the IAM much beyond $N_c=15$
may not necessarily yield a detailed description
of the resonance structure, we expect the $N_c$ behaviour of the ratios to be
qualitatively correct for both scenarios even at larger $N_c$.

Additional arguments
 to consider the IAM only as a 
qualitative description beyond $N_c=15$ or 30 have been given in 
\cite{Pelaez:2010er}
since the error made in approximating the left cut, as well as the effect of the $\eta'$ may start to become 
numerically relevant around those $N_c$ values.

Remarkably we see with two loop $\chi$PT, that the unitarised amplitudes do reflect semi-local duality with $I=2$ in the $t$-channel suppressed. This is most readily seen from the plots of the ratios $F^{I\,I'}_n$ for the two loop
amplitudes shown in Fig.~\ref{fig:ratiossu2} (to be compared with the one loop ratios of Fig.~\ref{fig:ratiossu3}).
For $F^{21}_{\,n}$, it is clear that, if only considering the integrals up to 1 GeV$^2$, the ratios are still not small in magnitude. Indeed, their absolute value increases with $\Nc$. However, integrating up to 2 GeV$^2$ takes into account the sub-dominant component, and then the ratios stabilise at much smaller values for all $\Nc$, consistent with expectations from semi-local duality.

\section{The effect of heavier resonances}
\label{sec:heaviereffects}

So far we have restricted the analysis of the $\Nc$ behaviour to the $\rho$ and $\sigma$ resonances. 
Of course, one may wonder what is the effect  of heavier resonances on our analysis and conclusions. In particular,
since the 
subdominant \qq component of the $\sigma$ emerges between 1 and 1.5~GeV$^2$,  one might worry
about the $f_0(980)$ and even the $f_0(1370)$ resonance, since the latter has a width of several hundred MeV
and may overlap with the region of interest. (The $f_0(1500)$ and $f_0(1710)$ lie beyond that energy range and
are therefore suppressed by the $1/s^n$ in the denominator).  In addition, we might worry about resonances in higher waves;
in this case the $f_2(1270)$ in the $D$-wave would yield the largest contribution.

Actually, Fig.~\ref{fig:TIJsu3} has been calculated in an $SU(3)$ coupled channel formalism and includes the $f_0(980)$ as a very sharp drop in ${\mathrm{Im}} T^0_0$, which disappears as $\Nc$ increases.
By comparing with Fig.~\ref{fig:TIJsu2}, with no $f_0(980)$ present, it is clear that, by removing the $f_0(980)$ the
 variation in the ${\mathrm{Im}} T^0_0$ integrals, and therefore in the $F_n^{I I'}$ of Eq.~(6), is small compared to the systematic uncertainty that we have estimated as the difference between the $t=0$ and $t=t_{\mathrm{th}}$ calculations. 
 Actually, if the $f_0(980)$  is included in a coupled channel IAM calculation, as in Fig.~\ref{fig:TIJsu3}, the new $F_n^{\,21}$ 
values would all lie between 
our $t=0$ and $t=t_{\mathrm{th}}$ 
results listed in Table III without the $f_0(980)$.
The error we make 
by ignoring the $f_0(980)$ is,
 at most 30$\%$ of the estimated systematic uncertainty.
For sure the $f_0(980)$ will not be able to compensate the $\rho$ contribution.
Still, one might wonder whether this is also the case at two-loops if the $f_0(980)$
 or $f_0(1370)$ have a \qq component around 1 to 1.5~GeV$^2$ that survives when $\Nc$ increases.
 However, at least the lightest such component would be precisely the same \qq state that we already see in the $f_0(600)$.
 Actually the interpretation of the IAM results is that all these scalars are a combination of all possible
states from Fock space \cite{LlanesEstrada:2011kz}, namely, \qq, tetraquarks, molecules, glueballs, etc...., but as $\Nc$ grows
only the \qq survives between 1 and  1.5 GeV$^2$, whereas the other components are either more massive or disappear in the deep complex plane. And it is precisely that component, which we already have in our calculation, the one compensating the $\rho$ contributions, as we have just seen above.

In the very preliminary interpretation of ~\cite{LlanesEstrada:2011kz}, the ${\overline q}q$ subdominant component of the $f_0(600)$ within the IAM naturally accounts for 20-30\% of its total composition. This is in fairly good agreement  with the 40\% estimated in \cite{Fariborz:2009cq}. Indeed, given the two caveats raised by the authors of \cite{Fariborz:2009cq}, their 40\% may be considered an upper bound. Firstly, this 40\% refers to the \lq\lq tree level masses'' of the scalar states. These mesons, of course, only acquire their physical mass and width after unitarization, which is essentially generated by  $\pi\pi$ final state interactions. Intuitively we would expect these to enhance the non-${\overline q}q$ component, and so bring the ${\overline q}q$ fraction below the \lq\lq bare'' 40\%. Secondly in \cite{Fariborz:2009cq} the authors also suggest that ``\lq\lq a possible glueball state is another relevant effect'' not included in their analysis. In \cite{LlanesEstrada:2011kz}, the glueball component is of the order of 10\%.
Consequently, the results of \cite{Fariborz:2009cq}, those presented here and in \cite{LlanesEstrada:2011kz}, are all quite consistent.
  
Finally, we will show that the contribution of the $f_2(1270)$ to the FESR cancellation, even assuming it follows exactly a ${\overline q}q$ leading $\Nc$ behaviour,
is rather small and does not alter our conclusions. All other resonances coupling to $\pi\pi$ are more massive and therefore less relevant.

In order to describe the $I=0$ $J=2$ channel we will again use  the parametrization of KPY in terms of the $I=0$, $J=2$ phase-shift $\delta^0_2$, namely
\begin{equation}\label{pwconformal}
\mathcal{A}_{2}^{\,0}=\frac{1}{\sigma(s)}\frac{1}{\cot\delta_{(2)}^{0}-\imath}
\end{equation}
where $\cot \delta_{2}^{0}$, which is proportional to $s-M_{f_2}^2$, is given in detail on the Appendix of KPY~\cite{Kaminski:2006qe}. 
Now, by replacing
\begin{equation}\label{cotconformal}
\cot \delta_{2}^{0}\rightarrow \frac{\Nc}{3}\cot \delta_{2}^{0}.
\end{equation}
we ensure that the amplitude itself scales as $1/\Nc$. This also ensures that the resonance mass
$M_{f_2}$ is constant, and its width scales as $1/\Nc$. We require the $f_2(1270)$ to behave
as a perfect ${\overline q}q$ at leading order in $1/\Nc$, while reproducing the KPY fit to the $D$-wave at $\Nc=3$.
As can be noticed in Fig.~\ref{Fig:Dwave}, 
for $F^{21}_2$ and $F^{21}_3$, which are the most 
relevant ratios for our arguments, the difference between  
adding this $D$-wave contribution 
to our previous results is smaller than 
the effect of the sigma \qq component around 1 to 1.5 GeV$^2$.
For the  ratio $F^{21}_3$, the effect of the $D$-wave contribution is larger,
but it is the effect of the sigma \qq sub-component the one that
makes the curves flatter and bounded between -0.2 and 0.2, 
whereas the slope is clearly negative without such a contribution and the absolute value of
the ratio can be as large as 0.5 and still growing.
Note that in Fig.~\ref{Fig:Dwave}
we compare  
our previous one- and two-loop $F_n^{21}$ calculations (bolder line)
to those which include the $f_2(1270)$ 
resonance as a pure ${\overline q}q$ (thin lines).
Therefore the effect of including 
the $f_2(1270)$ does not
modify our conclusions. 
The main FESR cancellation at $\Nc$ larger than 3 is between the $\rho(770)$
and the subdominant \qq component of the $f_0(600)$ resonance, which appears around 1 to 1.5 GeV$^2$.

This is even more evident if we extrapolate our results to even higher $N_c$, 
as in Fig.~\ref{nc100}, where all curves include the effect of the $f_2(1275)$.
As already explained above, for such high $N_c$ the IAM cannot be trusted as a precise 
description, but just as a qualitative model of the effect of a \qq state
around 1 to 1.5 GeV$^2$, which has a width much larger than the states seen there at $N_c=3$
and will dominate the integrals in $F_n^{21}$.
It is clearly seen that the effect of such a state will compensate the $\rho(770)$ contribution
and preserve semi-local duality. Other states that survive the $N_c$ limit in that
 region---which would be heavier and much narrower--- 
would only provide smaller corrections to this qualitative picture.
 Nevertheless, it would be desirable to extend this study to  a
more ambitious
treatment of the higher mass states in future work.

\begin{figure*}
\includegraphics[scale=0.25]{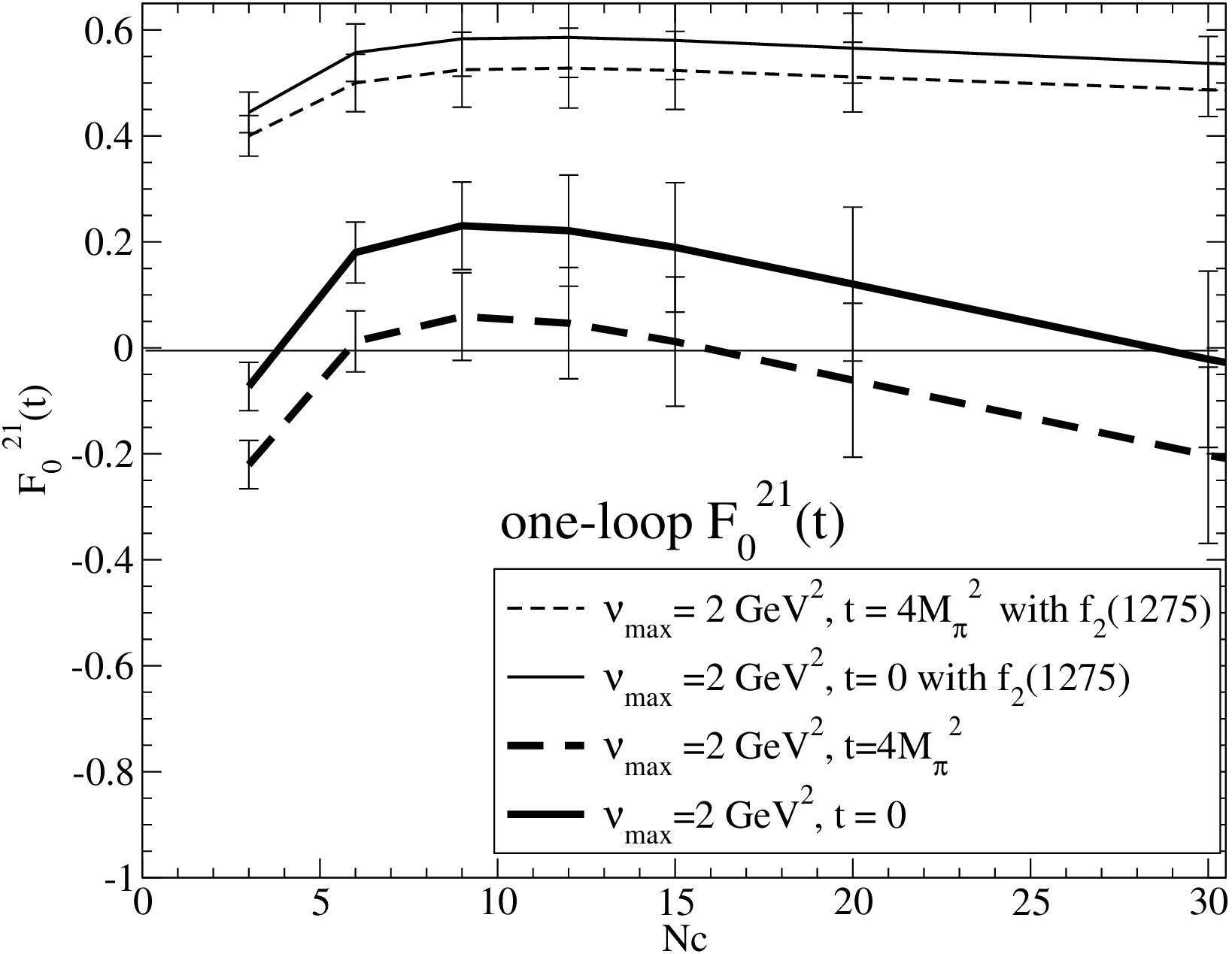}\includegraphics[scale=0.25]{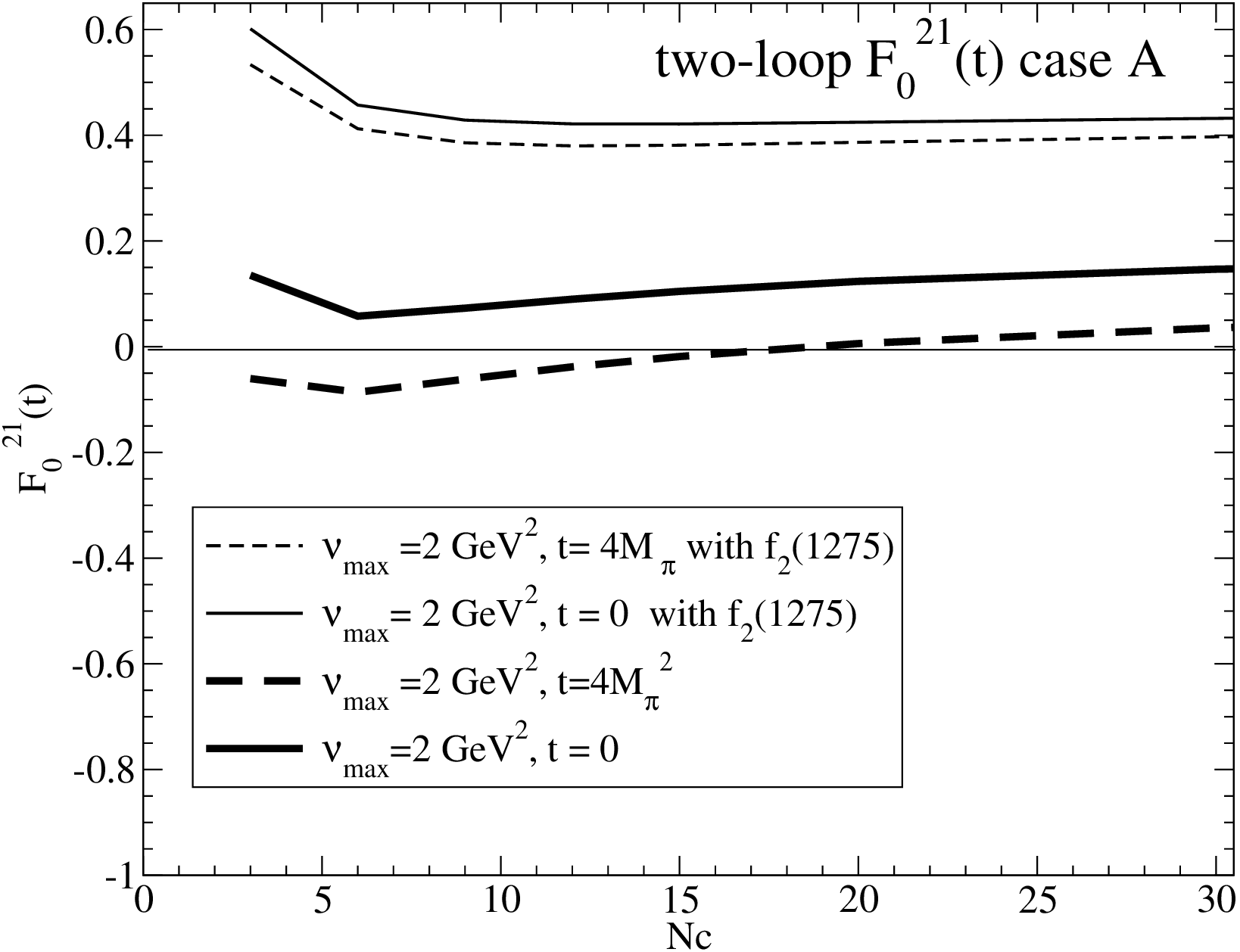}
\includegraphics[scale=0.25]{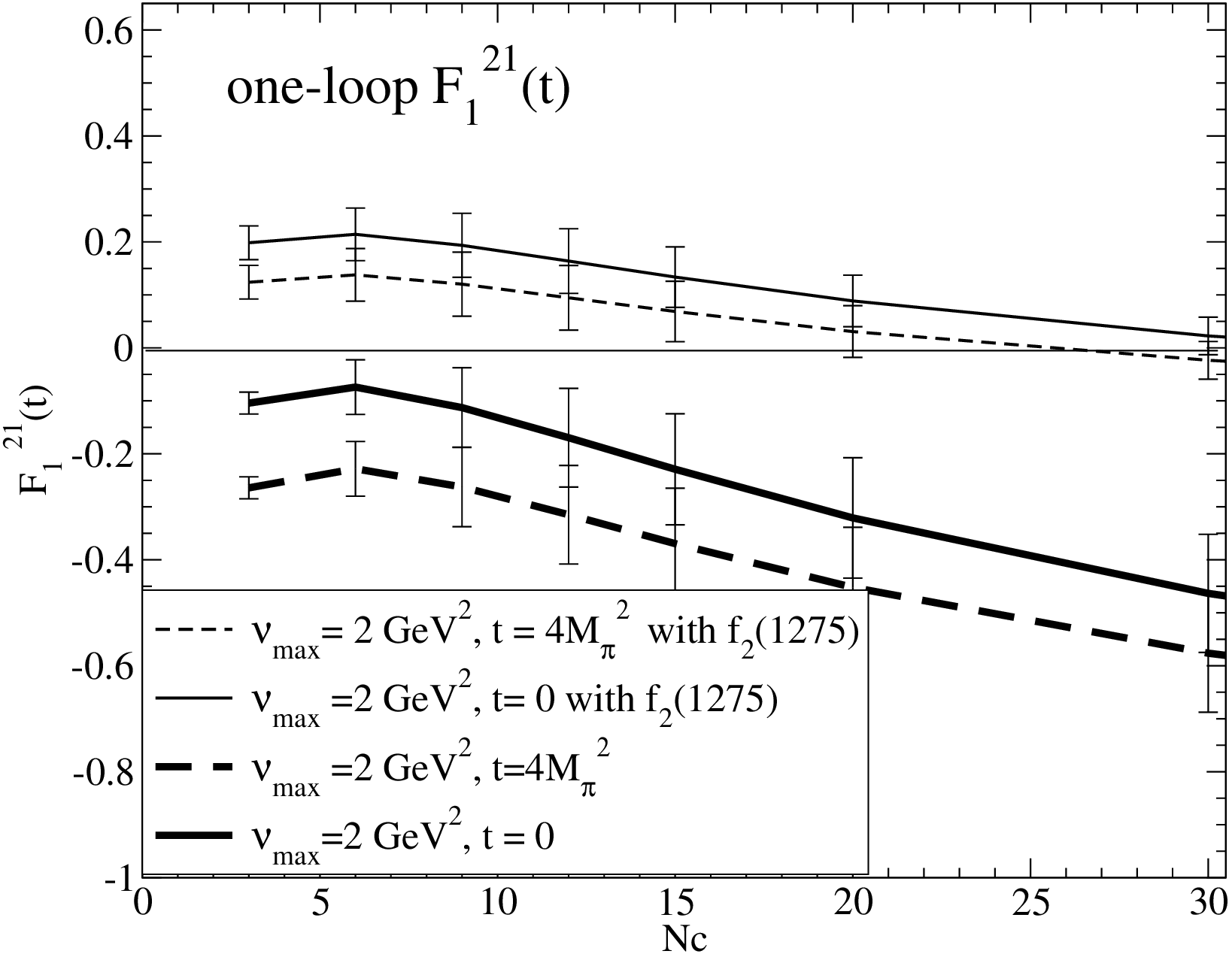}\includegraphics[scale=0.25]{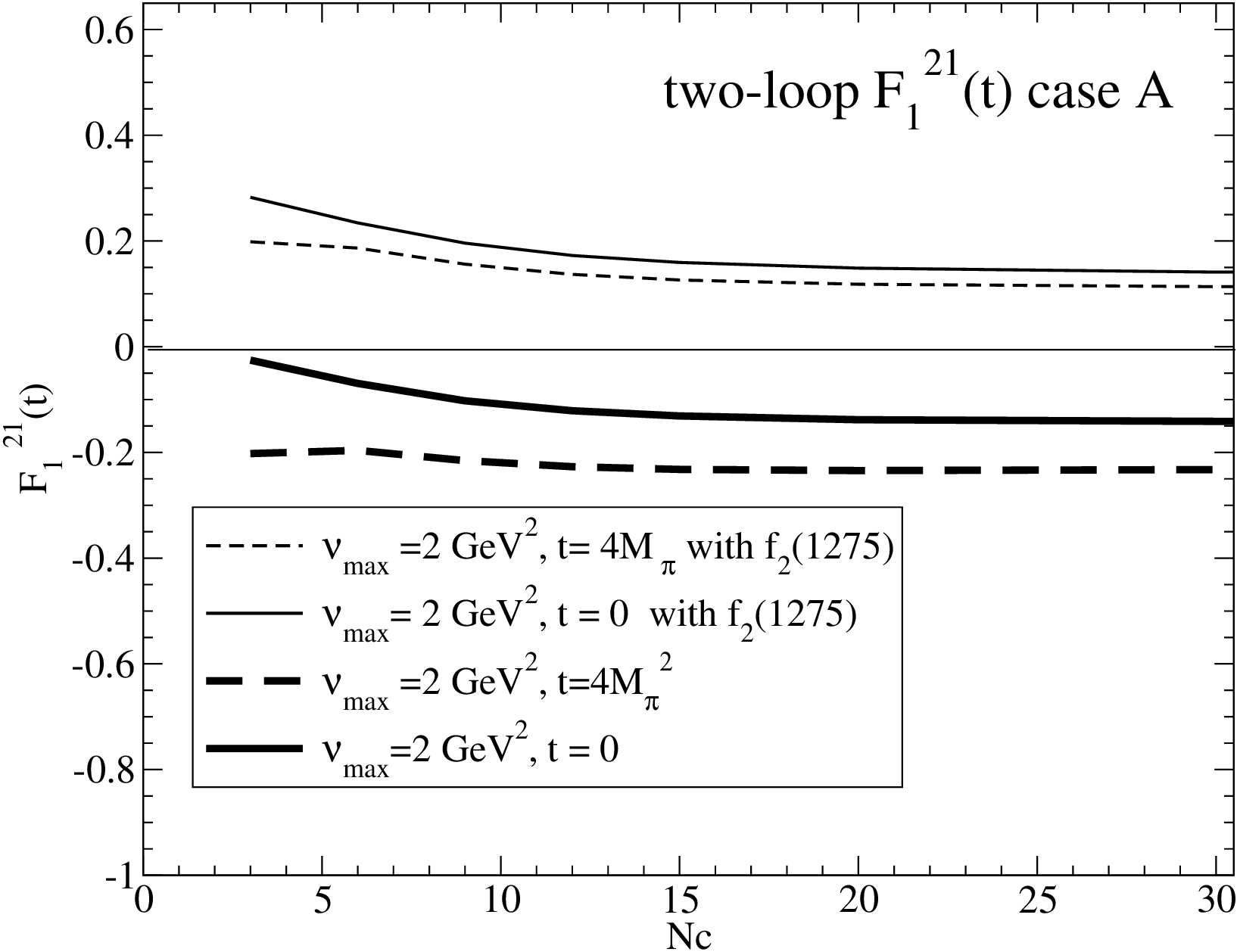}
\includegraphics[scale=0.25]{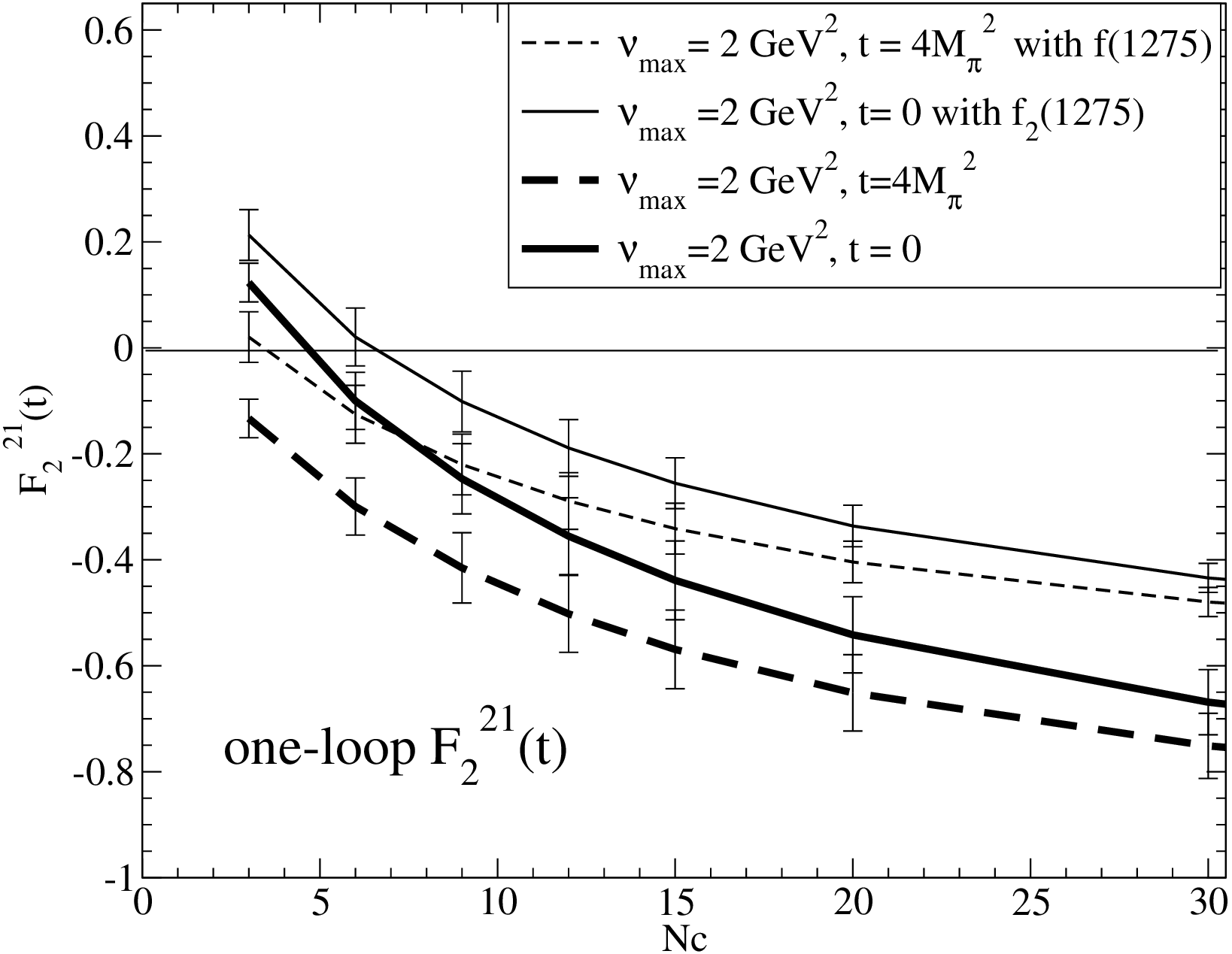}\includegraphics[scale=0.25]{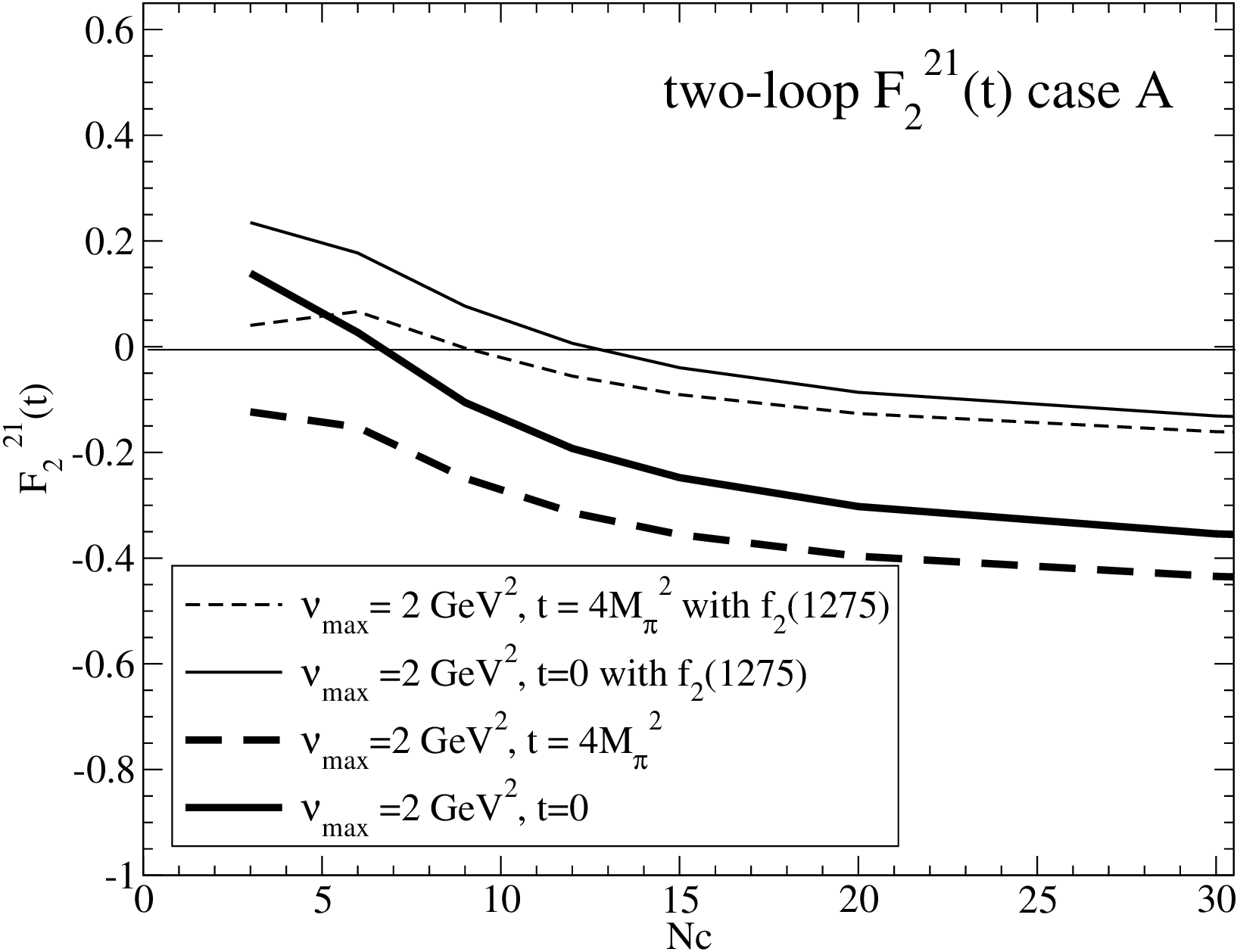}
\includegraphics[scale=0.25]{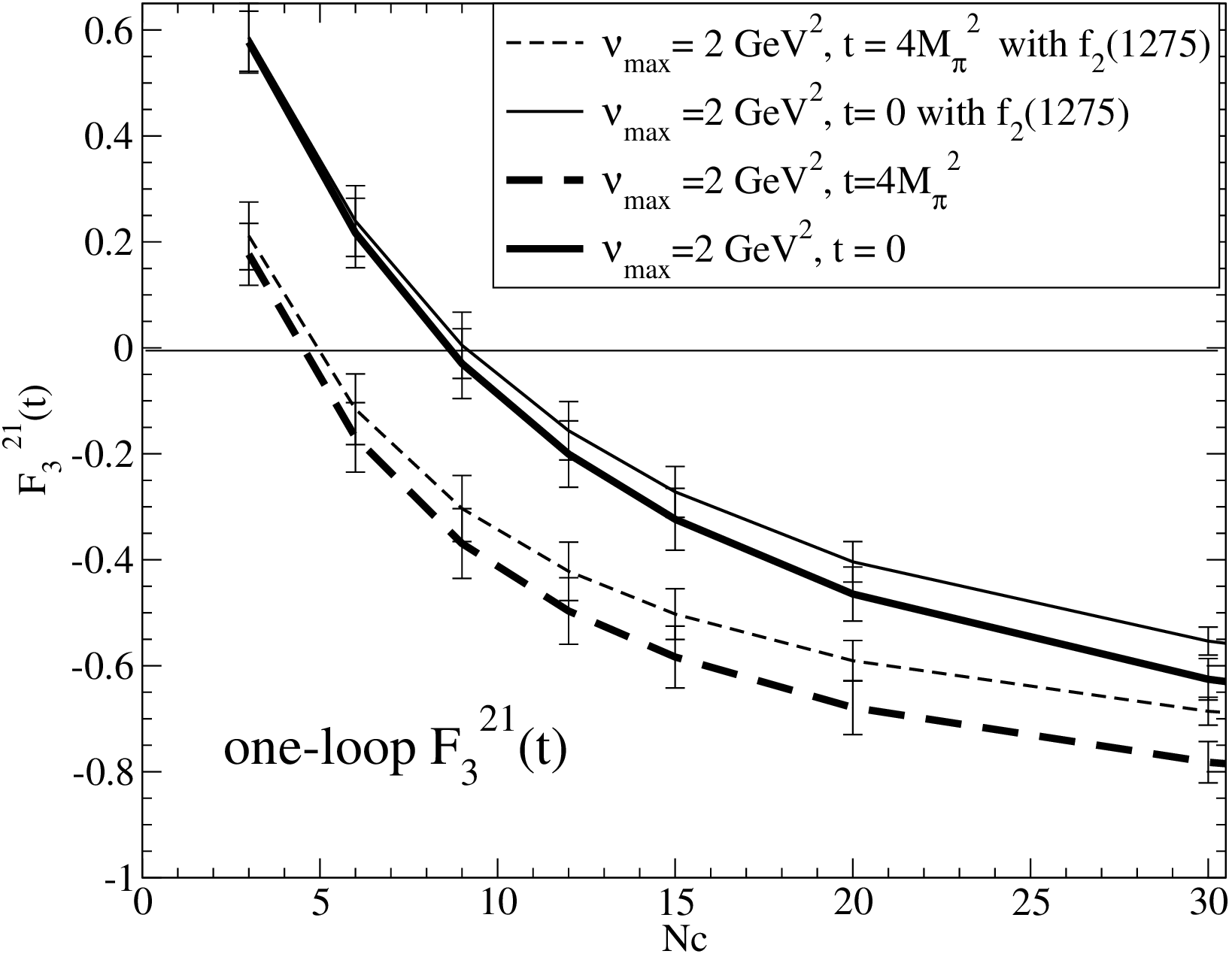}\includegraphics[scale=0.25]{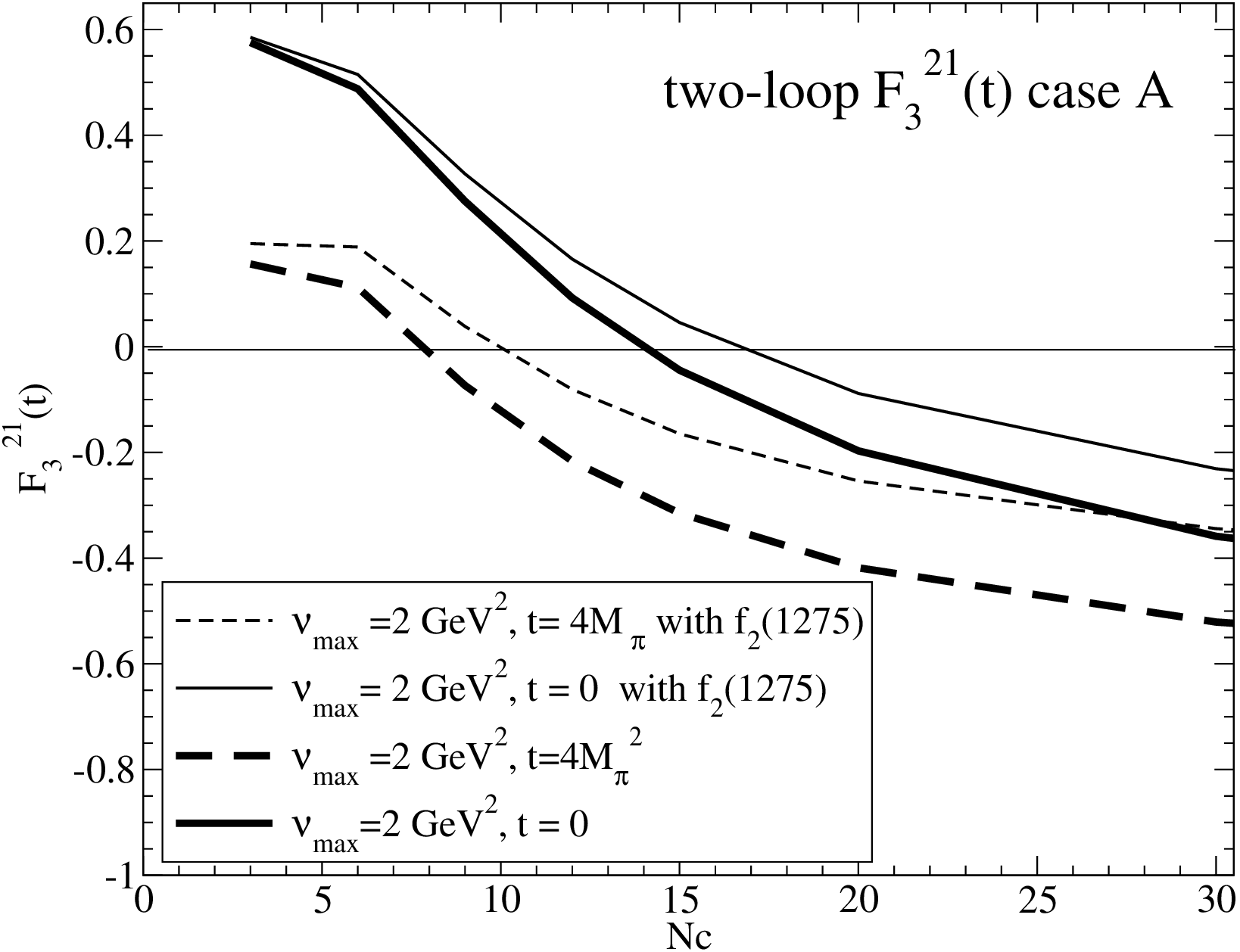}
\caption{ Results for $F_n^{21}$ with and without the $f_2(1270)$ resonance scaled as a pure \qq (thin and bolder lines, respectively).
The left panels are for one-loop IAM results, and the right 
ones for the two-loop results.
The latter contain a subdominant
 \qq component of the $f_0(600)$ around 1 to 1.5 GeV$^2$ whose
effect is relevant for the cancellation of  
$F_n^{21}$.}
\label{Fig:Dwave}
\end{figure*}

\section{Discussion}
It is a remarkable fact that hadronic scattering amplitudes from threshold upwards build their high energy Regge behaviour. This was learnt from detailed studies of meson-nucleon interactions more than forty years ago. This property is embodied in semi-local duality, expressed through finite energy sum-rules. Perhaps just as remarkably we have shown here that the Regge parameters fixed from high energy $NN$ and $\pi N$ scattering yield the correct $\pi\pi$ $P$ and $D$-wave scattering lengths, {\it cf.} Eqs.~(11,12). Indeed, there is probably no closer link between amplitudes with definite $t$-channel quantum numbers and their low energy behaviour in the $s$-channel physics region than that shown here. What is more, such a relationship should hold at all values of $\Nc$. At low energy the scattering amplitudes of pseudo-Goldstone bosons are known to be well described by their chiral dynamics, and their contribution to
finite energy sum-rules is dominated by the $\rho$(770) and $f_0$(600) contributions. However,
there are many proposals in the literature, including the 
$\Nc$ dependence of the unitarised chiral amplitudes,
suggesting that the $f_0$(600), contrary to the $\rho$(770), 
may not be an ordinary \qq meson. 
This is a potential problem for 
the concept of semi-local duality between resonances and Regge exchanges. The reason is that
for $I=2$
$t$-channel exchange it requires a cancellation between the $\rho$(770) and $f_0$(600) resonances,
which may no longer occur if the $f_0$(600) contribution becomes comparatively
smaller and smaller as $N_c$ increases.

This conflict actually  occurs for the most part of
one-loop unitarised chiral perturbation theory
parameter space.  
In contrast, for a small part of the one-loop parameter space and in a very natural way
at higher order in the chiral expansion, the $\sigma$ may 
have a ${\overline q}q$ component in its Fock space, which though sub-dominant at $\Nc=\,3$, 
becomes increasingly important as $\Nc$ increases. This is critical, 
as we have shown here, in ensuring semi-local duality for $I=2$ exchanges is 
fulfilled as $\Nc$ increases. As we show in Fig.~\ref{nc100} this better 
fulfillment of semi-local duality keeps improving 
even at much larger $N_c$, where the IAM can only be interpreted as a 
very qualitative average description.  

Thus the chiral expansion contains the solution to the seeming paradox of how a  distinctive nature for the $\rho, \,\sigma$ at $\Nc=\,3$  
is reconciled with semi-local duality at larger values of $\Nc$. Indeed, despite the additional freedom brought about by the extra low energy constants at two loop order, fixing these from experiment at
$\Nc=\,3$ automatically brings this compatibility with semi-local duality as $\Nc$ increases. This is a most satisfying result.

The $P$ and $D$-wave scattering lengths evaluated using Eqs.~(7,~8) that agree so well with local duality at $\Nc=3$ can, of course, be computed at larger $\Nc$  by inputting chiral amplitudes on each side of the defining equations. The scattering lengths themselves involve only the real parts, while the Froissart-Gribov integrals require the imaginary parts that are determined by the unitarization procedure. Explicit calculation shows that these  agree as $\Nc$ increases. While the agreement at one loop order is straightforward, at two loops there is a subtle interplay of dominant and sub-dominant terms placing constraints on the precise values of the LECs. As this takes us beyond the scope of the present work we leave this for a separate study.
 
  
Though beyond the scope of this work, we can then ask what does this tell us about the nature of the enigmatic scalars~\cite{Pennington:2010dc}?
 At $\Nc=\,3$, the behaviour of the $\sigma$ is controlled by its coupling to $\pi\pi$. Its Fock space is dominated by this non-${\overline q}q$ component~\cite{Jaffe,Black:1998wt,LlanesEstrada:2011kz}. In dynamical calculations of resonances and their propagators, like that of van Beveren, Rupp and their collaborators~\cite{VanBeveren:1986ea} and of Tornqvist~\cite{torn}, 
the seeds for the lightest scalars are an ideally mixed \qq multiplet of higher mass. 
\begin{figure*}
  \begin{center}
    \includegraphics[scale=0.25]{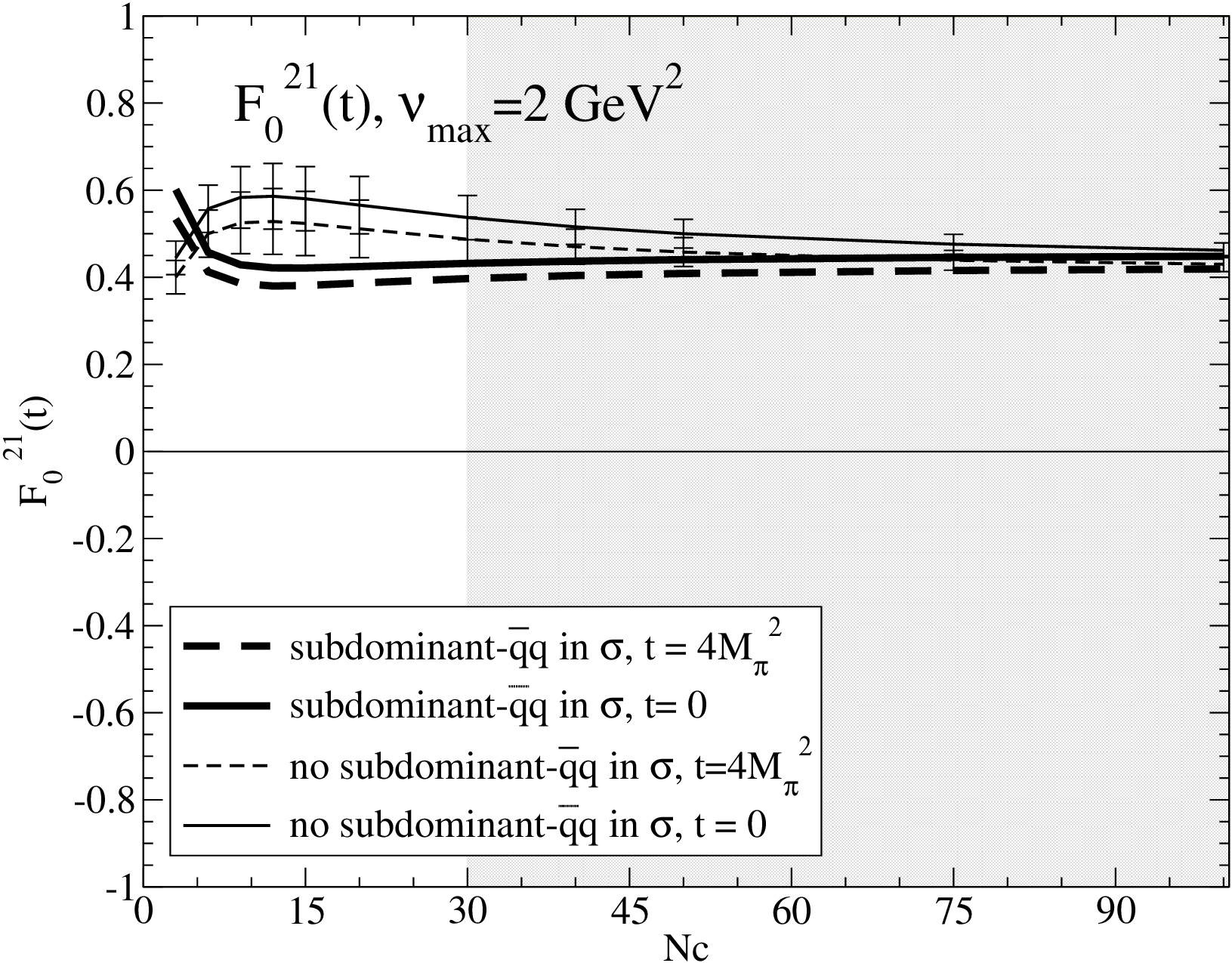}\hspace{4mm}\includegraphics[scale=0.25]{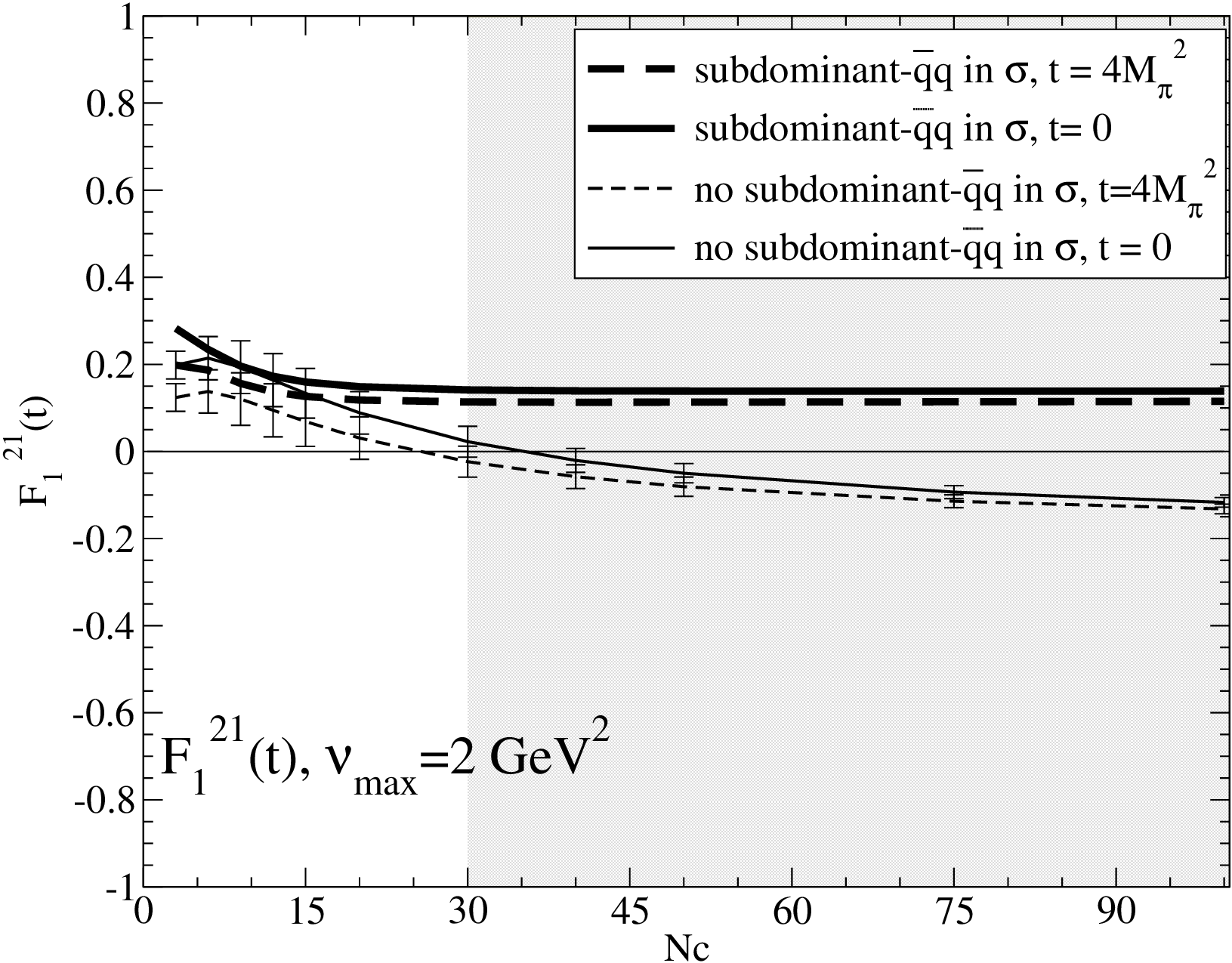}\\
    \includegraphics[scale=0.25]{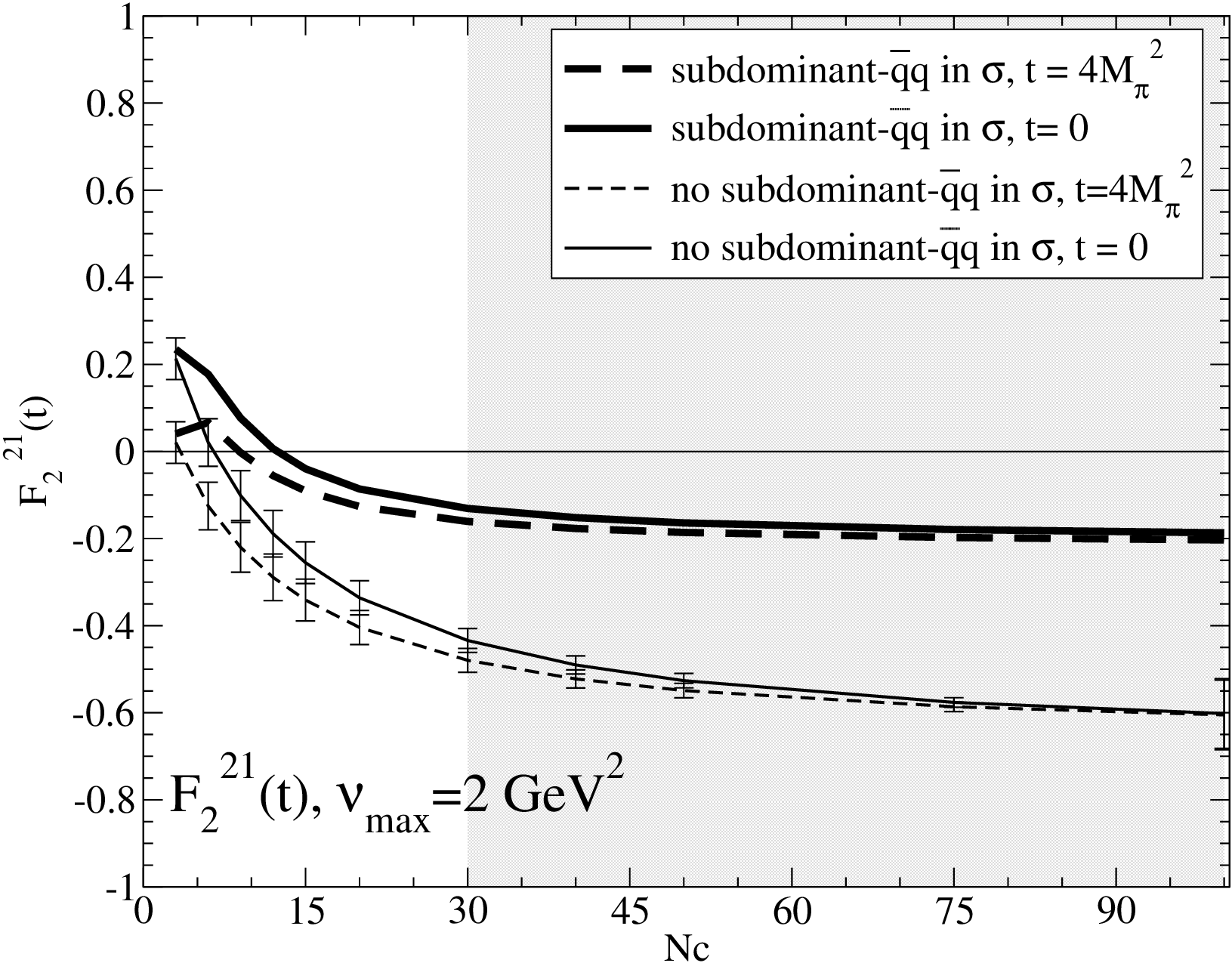}\hspace{4mm}\includegraphics[scale=0.25]{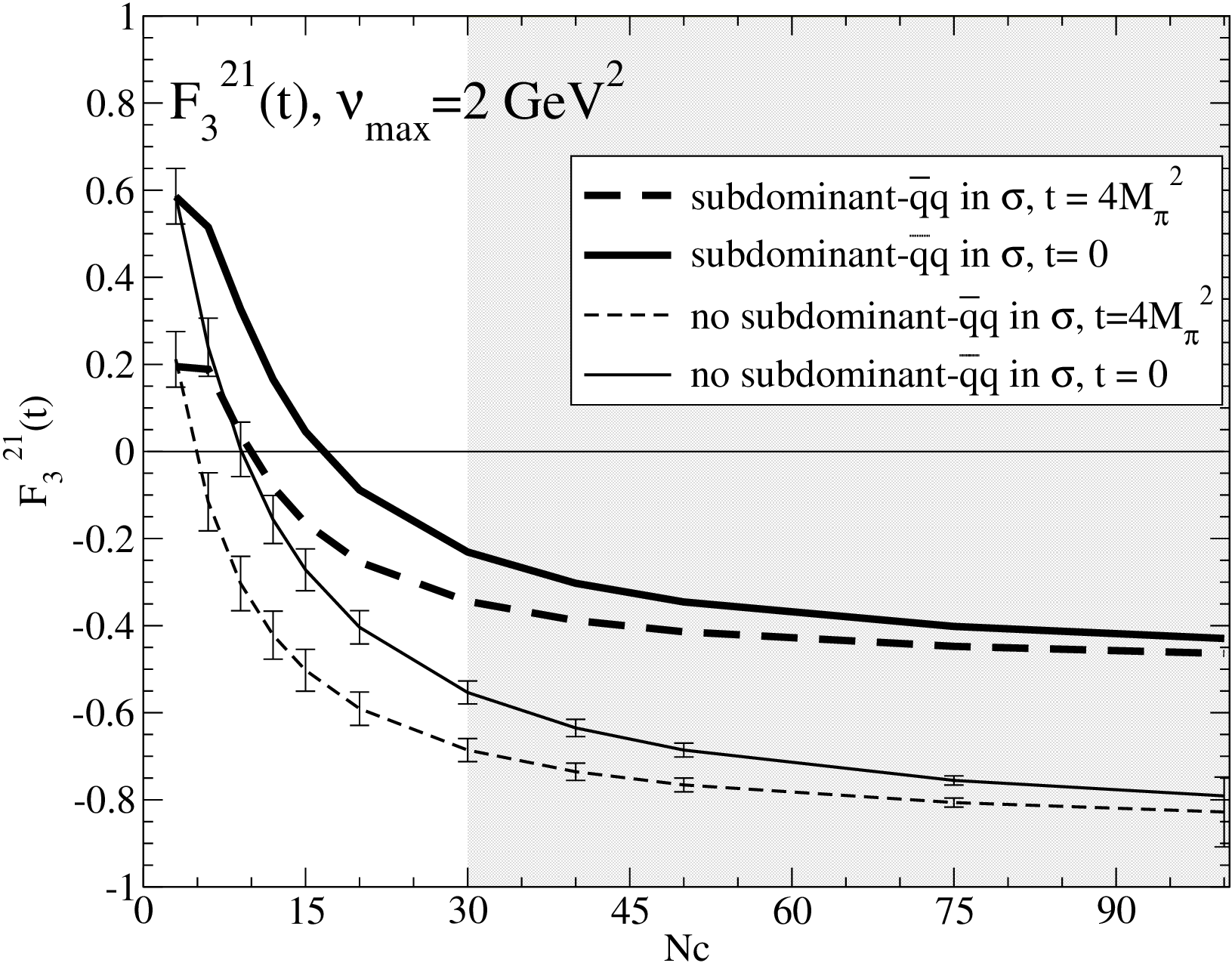}
    \caption{ Results for the $F_n^{21}$ ratios including the $f_2(1275)$ model 
to the $\chi$PT unitarised S and P waves. 
The bolder lines correspond to our two loop calculation that yields a subdominant \qq component
around 1 to 2 GeV$^2$, whereas the thin lines are the one loop 
unitarised calculation that does not contain such a component.
As explained in the text, beyond $N_c=15$ or 30 (gray area)
we consider the unitarised amplitudes 
to provide just a qualitative description of the dominant 
\qq state in the 1 to 1.5 GeV$^2$ region.
It is nevertheless clear that the effect of such a \qq component 
brings a large cancellation in the ratios, improving the fulfillment of
semi-local duality.}
\label{nc100}
  \end{center}
\end{figure*} 
These seeds may leave a conventional \qq nonet near 1.4~GeV~\cite{VanBeveren:1986ea,Oller:1998zr,Black:1998wt,Close:2002zu}, while the dressing by hadron loops dynamically generates a second set of states, whose decay channels  dominate their behaviour at $\Nc=\,3$ and pull their masses close to the threshold of their major decay:
the $\sigma$ down towards $\pi\pi$ threshold, and the $f_0(980)$ and $a_0(980)$ to ${\overline{K}}K$ threshold. The leading order in the $1/\Nc$ expansion  discussed here may
be regarded {\it a posteriori} as providing a quantitative basis for this. The scalars are at $\Nc$ larger than 3 controlled by \qq seeds of mass well above 1 GeV (1.2 GeV for the intrinsically non-strange scalar). Switching on decay channels, as one does as $\Nc$ decreases, changes their nature dramatically, inevitably producing non-\qq or di-meson components in their Fock space at $\Nc=\,3$~\cite{Pennington:2010dc}.
We see here that the $\sigma$ having a sub-dominant \qq component with a mass above 1 GeV is essential for semi-local duality, that suppresses $I=2$ amplitudes, to hold.


\noindent {\bf{Acknowledgments}}\\
MRP is grateful to Bob Jaffe for discussions about the issues raised at the start of this study.
The authors (JRdeE, MRP and DJW) acknowledge partial support of the EU-RTN Programme, Contract No. MRTN--CT-2006-035482, \lq\lq Flavianet'' for this work, while at the IPPP in Durham.
 DJW is grateful to the UK STFC for the award of a postgraduate studentship and to Jefferson Laboratory for hospitality while this work was completed.

\end{document}